\documentclass[12pt]{article}

\usepackage{amsmath,amsfonts,enumerate,array}
\allowdisplaybreaks
\usepackage{graphicx}
\usepackage[usenames]{color}
\usepackage[english]{babel}
\usepackage{fancybox}
\usepackage{chngpage} 

\usepackage{tikz-cd}

\newcommand{\captionfonts}{\small}

\makeatletter  
\long\def\@makecaption#1#2{%
  \vskip\abovecaptionskip
  \sbox\@tempboxa{{\captionfonts #1: #2}}%
 \ifdim \wd\@tempboxa >\hsize
    {\captionfonts #1: #2\par}
  \else
    \hbox to\hsize{\hfil\box\@tempboxa\hfil}%
  \fi
  \vskip\belowcaptionskip}
\makeatother   

\usepackage{mciteplus}

\usepackage[colorlinks=true,      linkcolor=blue,      urlcolor=blue,      filecolor=blue,      citecolor=blue,
      pdfstartview=FitH,     pdfpagemode=UseNone,      bookmarksopen=true]{hyperref}  
      
\usepackage[all]{hypcap}     

\usepackage{cite}

\begin{document}

\numberwithin{equation}{section}


\mathchardef\mhyphen="2D


\newcommand{\be}{\begin{equation}} 
\newcommand{\ee}{\end{equation}} 
\newcommand{\bea}{\begin{eqnarray}\displaystyle}
\newcommand{\eea}{\end{eqnarray}}
\newcommand{\bt}{\begin{tabular}}
\newcommand{\et}{\end{tabular}}
\newcommand{\bs}{\begin{split}}
\newcommand{\es}{\end{split}}

\newcommand{\I}{\text{I}}
\newcommand{\II}{\text{II}}

\renewcommand{\a}{\alpha}	
\renewcommand{\b}{\beta}
\newcommand{\g}{\gamma}		
\newcommand{\G}{\Gamma}
\renewcommand{\d}{\delta}
\newcommand{\D}{\Delta}
\renewcommand{\c}{\chi}			
\newcommand{\C}{\Chi}
\newcommand{\p}{\psi}			
\renewcommand{\P}{\Psi}
\newcommand{\s}{\sigma}		
\renewcommand{\S}{\Sigma}
\renewcommand{\t}{\tau}		
\newcommand{\e}{\epsilon}
\newcommand{\n}{\nu}
\newcommand{\m}{\mu}
\renewcommand{\r}{\rho}
\renewcommand{\l}{\lambda}

\newcommand{\nn}{\nonumber\\} 		
\newcommand{\newotimes}{}  				
\newcommand{\diff}{\,\text{d}}		
\newcommand{\h}{{1\over2}}				
\newcommand{\Gf}[1]{\G \Big{(} #1 \Big{)}}	
\newcommand{\floor}[1]{\left\lfloor #1 \right\rfloor}
\newcommand{\ceil}[1]{\left\lceil #1 \right\rceil}

\def\cA{{\cal A}} \def\cB{{\cal B}} \def\cC{{\cal C}}
\def\cD{{\cal D}} \def\cE{{\cal E}} \def\cF{{\cal F}}
\def\cG{{\cal G}} \def\cH{{\cal H}} \def\cI{{\cal I}}
\def\cJ{{\cal J}} \def\cK{{\cal K}} \def\cL{{\cal L}}
\def\cM{{\cal M}} \def\cN{{\cal N}} \def\cO{{\cal O}}
\def\cP{{\cal P}} \def\cQ{{\cal Q}} \def\cR{{\cal R}}
\def\cS{{\cal S}} \def\cT{{\cal T}} \def\cU{{\cal U}}
\def\cV{{\cal V}} \def\cW{{\cal W}} \def\cX{{\cal X}}
\def\cY{{\cal Y}} \def\cZ{{\cal Z}}

\def\mC{\mathbb{C}} \def\mP{\mathbb{P}}  
\def\mR{\mathbb{R}} \def\mZ{\mathbb{Z}} 
\def\mT{\mathbb{T}} \def\mN{\mathbb{N}}
\def\mH{\mathbb{H}} \def\mX{\mathbb{X}}
\def\CP{\mathbb{CP}}
\def\RP{\mathbb{RP}}
\def\Z{\mathbb{Z}}
\def\N{\mathbb{N}}
\def\H{\mathbb{H}}

\newcommand{\bin}[1]{{\bf {\color{blue} BG:}} {{\color{blue}\it#1}}}
\newcommand{\shaun}[1]{{\bf {\color{red} SH:}} {{\color{red}\it#1}}}


\addtolength{\skip\footins}{0pc minus 5pt}

\def\b{\bigskip}

\begin{flushright}
\end{flushright}
\vspace{15mm}
\begin{center}
{\LARGE Transitions of three-charge black hole microstates\\ 
\vspace{1mm}
in the D1D5 CFT}
\\
\vspace{18mm}
\textbf{Shaun}~ \textbf{D.}~ \textbf{Hampton}
\\
\vspace{8mm}
\text{School of Physics},\\ \text{Korea Institute for Advanced Study},\\ \text{Seoul 02455, Korea}\footnote{sdh2023@kias.re.kr}
\vspace{3mm}
\end{center}

\vspace{2mm}

\thispagestyle{empty}

\begin{abstract}
Using the D1D5 CFT we investigate transitions involving a member of a certain class of states called superstrata states, which are holographically dual to certain smooth, horizonless, $1/8$-BPS, three-charge black hole microstates known as superstrata. We study these transitions by deforming the CFT away from the free orbifold point using a marginal deformation which contains a twist operator and a supercharge operator. We apply two marginal deformations to an initial state containing a graviton acting on a superstratum state. We compute amplitudes capturing transitions from this state to a graviton acting on a microstratum state, a member of a class of states which are holographically dual to certain smooth, horizonless, non-BPS, three-charge black hole microstates known as microstrata, non-BPS analogues of superstrata. We compare the resulting amplitude for various initial and final state energies to determine the preferred transition process. This may give hints as to how the dual superstratum geometry may preferentially back-react in this setting.
\end{abstract}
\newpage

\setcounter{page}{1}

\numberwithin{equation}{section} 

\tableofcontents

\newpage
\section{Introduction}
The D1D5 system has yielded much fruit in the endeavor to understand the microscopic nature of black holes \cite{Callan:1996dv,Das:1996wn,Das:1996ug,Maldacena:1996ix}. The work of \cite{Strominger:1996sh,Maldacena:1999bp} showed that microstates of the three charge black hole, which account for the black hole entropy, could be computed in a field theoretic way.
Since then, many new insights have been made, relying critically on the holographic principle and AdS/CFT \cite{Maldacena:1997re,Gubser:1998bc,Witten:1998qj}. For an early review of the D1D5 system and black holes see \cite{David:1999ec}. The fuzzball \cite{Lunin:2001fv,Lunin:2001jy} and microstate geometries \cite{Bena:2007kg,Bena:2016ypk} programs have sought to describe the details of black hole microstucture through explicit constructions and precision tests of the corresponding microstates \cite{Kanitscheider:2006zf,Kanitscheider:2007wq,Taylor:2007hs,Giusto:2015dfa,GarciaTormo:2019inl,Giusto:2019qig,Ganchev:2021ewa}. See \cite{Bena:2022rna,Bena:2022ldq,Shigemori:2020yuo,Mathur:2005zp} for nice reviews of the subject and various references therein. These microstates can be obtained, choosing a particular frame, by wrapping collections of D1 and D5 branes around compact dimensions and giving them motion which is perpendicular to the direction along with they are wrapped. The two charge configurations, corresponding to the microstates of the two charge black hole, were identified holographically as the number of Ramond sector ground states in the D1D5 CFT \cite{Lunin:2001jy}. 

More recently, smooth, $1/8$-BPS three charge microstate geometries have been constructed where the third charge, a momentum wave, was added to the compact directions of the D1D5 system. These configurations are known as superstrata \cite{Bena:2016ypk,Bena:2015bea,Bena:2017xbt,Ceplak:2018pws,Heidmann:2019zws,Bena:2016agb,Ganchev:2022exf,Ceplak:2022pep}. Their CFT duals have also been identified and correspond to coherent superpositions of supergraviton states \cite{Shigemori:2020yuo}. A different class of three charge microstates obtained from spectral flow transformations of two charge configurations were obtained in \cite{Giusto:2012yz}.

The D1D5 CFT at the free point in moduli space is given by the symmetric product orbifold $\mathcal{M}^{N_1N_5}/ S_{N_1N_5}$ where the target space is $\mathcal M=T^4\text{ or }K3$ \cite{Seiberg:1999xz,Larsen:1999uk}. This theory contains twisted sectors and untwisted sectors. The twisted sectors are created by twist operators which act on copies of the CFT. Significant progress has been made in understanding aspects of the twist operator at the free point in a variety of contexts \cite{Jevicki:1998bm,deBoer:1998kjm,Pakman:2009zz,Pakman:2009ab,Pakman:2009mi,Burrington:2012yq,Burrington:2012yn,Carson:2017byr,Dei:2019iym,Burrington:2018upk,DeBeer:2019oxm,Belin:2020nmp,Eberhardt:2018ouy,Eberhardt:2019ywk,Lunin:2000yv,Lunin:2001pw} with earlier works on orbifold CFTs being investigated in \cite{Dixon:1986qv,Atick:1987kd,Arutyunov:1997gt}.

Since the supergravity regime, where the vast majority of constructions have been performed, is far from the free point in moduli space, in the CFT, to connect the different regions, one must add an interaction, a marginal deformation to the theory which drives it towards the supergravity regime. This operator includes both a twist operator and a supercharge operator \cite{Avery:2010er,Avery:2010hs,Avery:2010qw,Burrington:2014yia,Carson:2014xwa,Carson:2017byr,Carson:2014yxa,Carson:2014xwa,Carson:2014ena,Carson:2016cjj,Carson:2015ohj,Carson:2016uwf,Carson:2015ohj,Guo:2022zpn}. This operator causes the copies of the branes wrapping the circle to interact with one another, i.e. to join or split as a function time. This introduces dynamical changes within the CFT and seems quite essential if one desires to understand the dynamics of black hole microstates such as the scrambling of  information, formation of microstates, transitions between microstates, etc. from the dual field theory perspective.

Along these lines several works have been explored by computing quantities at various orders in twist and deformation operators with various processes in mind. For example, in \cite{Hampton:2018ygz,Guo:2019pzk,Guo:2019ady,Benjamin:2021zkn,Keller:2019yrr,Lima:2020boh,Lima:2020nnx,Lima:2020urq} the lifting of states from action of the marginal operator has been studied rather extensively. This helps identify which states are protected from corrections and which of those become stringy as you deform away from the free theory. In \cite{Hampton:2019csz,Hampton:2019hya} the authors sought to understand aspects of thermalization of initial excitations in the D1D5 CFT into more complicated final states. This is a very complex phenomena whose full understanding should be obtained in the regime of strongly coupled field theory, however one must identify a setting where computations can be carried out explicitly. In \cite{Chang:2025rqy} the authors utilize recently developed techniques of supercharge cohomology to construct and classify BPS states away from the orbifold point and \cite{Chen:2024oqv} investigates scrambling dynamics of $1/4$-BPS states dual to the two charge black hole. The field theory dual of probes moving in AdS were explored in \cite{Guo:2021ybz} and the dual of tidal excitations of probes in superstrata geometries \cite{Martinec:2020cml,Ceplak:2021kgl,Guo:2024pvv} were explored \cite{Guo:2021gqd}. There the background was considered to be fixed in both the initial and final states, corresponding to a fixed background geometry in the gravity dual.

In the current work we aim to investigate how a graviton acting on an initial superstratum state interacts to produce a final state in which the superstratum state itself has been \textit{changed}, qualitatively corresponding to the back-reaction of the superstratum geometry in the bulk through interaction with an in-falling graviton. This provides a setting to investigate back-reaction dynamics of heavy states from the perspective of the D1D5 CFT addressing fundamental questions within the fuzzball and microstrate geometry programs about how black hole microstates evolve in their phase space. For work addressing various aspects of this evolution see \cite{Mathur:2008kg,Marolf:2016nwu,Guo:2022and}.
The approach taken in this paper involves computing transition amplitudes between the initial state mentioned above and a final state corresponding to a specified background change at second order in the marginal perturbation. The final state we consider will be a combination of a graviton acting on what we call a microstratum state, a particular member of the CFT dual states of microstrata, a family of three-charge, smooth, horizonless non-BPS microstate geometries \cite{Ganchev:2021ewa,Ganchev:2021pgs,Ganchev:2023sth}, non-BPS analogues of superstrata. The first construction of non-BPS microstate geometries was given in \cite{Jejjala:2005yu}. By analyzing the resulting amplitude that we compute, we find that it contains two main components: 1) an oscillatory term in $t$ and II) a term which grows as $t^2$. Similar oscillatory behavior alone has been shown to appear elsewhere such as
when studying the CFT dual of a graviton moving in vacuum AdS \cite{Guo:2021ybz}. The combination of both effects have been shown to appear elsewhere such as when studying the CFT dual of interacting gravitons in vacuum AdS \cite{Hampton:2019csz} and the CFT dual of tidal excitations of a graviton probe moving within a superstratum geometry \cite{Guo:2021gqd}. The $t^2$ growth explored in this work, for long time scales, suggests a significant change of the initial superstratum state through interaction with a graviton into a microstratum state with a graviton which, in the bulk, corresponds to a graviton in a microstratum geometry. We compare the behavior of the computed transition amplitude for various quantum numbers to determine which set yields the largest growth. We also consider a large $N$ limit and extrapolate the second order results to a strong coupling region, yielding a corresponding transition timescale. This may provide some understanding about the back-reaction of bulk, three-charge geometries from the perspective of the dual CFT. Though our computations are far from the supergravity point, they do provide hints and suggestions in field theory of how three charge states may explore some of their phase space through transitions due to interactions between the geometry and in-falling matter.

The plan of the paper is as follows, in section \ref{review} we review the D1D5 CFT. In section \ref{cylinder modes} we record the modes on the base space that will enter the amplitude computation. In section \ref{transitions of ss} we discuss initial and final states involved in the transition amplitude we plan to compute. In \ref{amplitudes} we write down the corresponding transition amplitudes. In \ref{covering space} we use the covering space method to compute the unintegrated amplitudes and in section \ref{integration} we integrate those amplitudes to obtain the full result. We perform a numerical analysis of these amplitudes in section \ref{analysis} and discuss the large $N$ limit in \ref{large N}. In section \ref{discussion conclusion} we discuss and conclude.
\section{Review of D1D5 CFT}\label{review}
Consider Type IIB string theory compactified on 
\begin{equation}
    M_{4,1}\times S^1\times T^4
\end{equation} 
with $N_1$ D1 branes wrapping $S^1$ and $N_5$ D5 branes wrapping $S_1\times T^4$. The low energy limit of this configuration (where $T^4$ is taken to be much smaller than $S^1$) flows to an $\mathcal N=(4,4)$ SCFT in $1+1$ dimensions wrapping $S^1$. 
In moduli space, it is conjectured that the free point of this theory, also know as the orbifold point, is a symmetric product orbifold theory of $N=N_1N_5$ copies of a seed CFT with target $\mathcal M$ of central  $c=6$  by the permutation group $S_N$ denoted $\mathcal M^N/S_N$. The target space can be chosen to be $\mathcal M = T^4$ or $K3$. In our case we consider $\mathcal M=T^4$ and theory is the symmetric product orbifold
\begin{equation}
(T^4)^{N_1N_5}/S_{N_1N_5}
\end{equation}
Each copy of the torus yields four bosons $X_1,X_2,X_3,X_4$ and four left and four right moving fermions $\psi_1,\psi_2,\psi_3,\psi_4$ and $\bar\psi_1,\bar\psi_2,\bar\psi_3,\bar\psi_4$ respectively. The total central charge of the theory is then $c_{\text{tot}} =cN_1N_5= 6N_1N_5$
\subsection{Symmetries of the CFT}
The D1D5 CFT includes an $\mathcal N = (4,4)$ superconformal symmetry. We give the corresponding generators, OPEs and algebra in appendix \ref{algebra}.
Each copy of the algebra includes in $SU(2)_L\times SU(2)_R$ R-symmetry with charges
\begin{equation}
    SU(2)_L : (j,m),\quad SU(2)_R : (\bar j, \bar m)
\end{equation}
There is also an internal R-symmetry coming from the directions of the torus $T^4$ given by $SO(4)\equiv SU(2)_1\times SU(2)_2$. This symmetry is broken by the compactification of the torus but at the free point it still provides a useful organizing principle. 
The indices $\alpha,\dot\alpha=\pm$ denote charges under $SU(2)_L\times SU(2)_R$ and $A,\dot A=\pm$ indices denote charges under $SU(2)_1\times SU(2)_2$. The four real left moving fermions are grouped into a complex left-moving fermion $\psi^{\alpha A}$ and complex a right moving fermion $\bar\psi^{\dot\alpha \dot A}$. 
The collection of four bosons $X^i$ form a vector in $T^4$ and are thus singlets under $SU(2)_L\times SU(2)_R$. They can be organized into a complex field $X_{A\dot A}$ charged unto $SU(2)_1\times SU(2)_2$. More details are given in appendix \ref{algebra}.

\subsection{Coordinates}
The base space of the CFT is a cylinder defined in terms a complex coordinate $w$ with
\begin{equation}\label{w}
w = \tau + i\sigma, \quad -\infty<\tau<\infty,\quad 0\leq \sigma \leq 2\pi
\end{equation}
where the coordinate $\tau$ is the Wick rotated and rescaled physical time coordinate $t$ and $\sigma$ is the rescaled compact coordinate $y$ given as 
\begin{equation}
    \tau = i {t\over R_y},\quad \sigma ={y\over R_y}
\end{equation}
with $R_y$ the radius of the compact physical $y$ circle. In this computation we will map are result to the plane and then a covering surface. The map from the cylinder to the plane is 
\begin{equation}
z=e^w
\end{equation}
We will show the covering space computation in a later section. In order to simplify expressions which come later we define a convenient choice of coordinates in terms of the center of mass and difference of the two twist operators which we record below
\begin{equation}\label{s dw}
s={w_1+w_2\over2},\quad \Delta w = w_2-w_1
\end{equation}
Inverting these coordinates
\begin{equation}
    w_1=s-{\Delta w\over2}, \quad w_2= s+{\Delta w\over2}
\end{equation}
In terms of the $z$ coordinate
we find that 
\begin{equation}\label{wz}
    z_1 = e^{w_1}=e^{s-{\Delta w\over2}},\quad z_2 = e^{w_2}=e^{s+{\Delta w\over2}}
\end{equation}
\subsection{NS and R vacua}\label{NS R}
Let's take, for the moment, $c=6$. The lowest energy state is then the NS vacua given by the dimension and charge
\begin{equation}
    |0_{NS}\rangle, \quad h=m=0
\end{equation}
where we've only written the left moving sector. However, there is also the R sector, which is given by
\begin{align}
    |0_R^{\pm}\rangle,\quad h = {1\over4}, m = \pm{1\over2}
\end{align}
These sectors can be related by spectral flow \cite{Schwimmer:1986mf,Sevrin:1988ew} denoted by the parameter $\alpha$ (not to be confused with the R-charge under $SU(2)_L\times SU(2)_R$). This is an automorphism of the $\mathcal N=(4,4)$ superconformal algebra in which the charges and dimensions change in the following way for central charge $c$ 
\begin{align}\label{sf}
    h' &= h + \alpha m + {c\alpha^2\over24}\nn
    m' &= m + {\alpha c\over 12}
\end{align}
which for $c=6$ yields
\begin{align}\label{sf}
    h' &= h + \alpha m + {\alpha^2\over4}\nn
    m' &= m + {\alpha\over 2}
\end{align}
The vacua transform in the following ways under the indicated spectral flow 
\begin{align}\label{vacuum sf}
    \alpha = 1&:\quad |0^-_R\rangle \to |0_{NS}\rangle,\quad |0_{NS}\rangle\to|0^+_R\rangle\nn
    \alpha = -1&:\quad |0^+_R\rangle \to |0_{NS}\rangle,\quad |0_{NS}\rangle\to|0^-_R\rangle
\end{align}
Similar relations also hold for the right moving sector. Combining the two sectors, in this paper we will utilize, when necessary, the notation that 
\begin{equation}\label{Ramond}
    |0^{\alpha\dot\alpha}_R\rangle \equiv|\alpha\dot\alpha\rangle\equiv|0^{\alpha\dot\alpha}_R\rangle|\bar0^{\alpha\dot\alpha}_R\rangle
\end{equation}
with the hermitian conjugate denoted as 
\begin{equation}\label{Ramond conj}
    \langle 0_{R,\alpha\dot\alpha}| \equiv \langle \alpha\dot\alpha|\equiv \langle0_{R,\alpha}|\langle\bar0_{R,\dot\alpha}|,\quad \alpha,\dot\alpha = \pm
\end{equation}
Furthermore, any operator $\mathcal O(z)$ carrying R-charge $m$ will transform under spectral by $\alpha$ units at a location $z_0$ in the following way
\begin{equation}
    \mathcal O(z)\to  (z-z_0)^{\alpha m}\mathcal O(z)
\end{equation}
\subsection{Deformation operator}
In this work we will consider deforming away from the free orbifold theory by applying a marginal deformation. This deformation defined in the following way
\bea \label{S}
S=S_0 + \lambda \int d^2z D(z,\bar z)
\eea
with $S_0$ the free action and where the deformation operator $D$ is defined as
\bea\label{D}
D(z,\bar z)\equiv \e^{\dot A\dot B}G^-_{-{1\over2},\dot A}\bar G^-_{-{1\over2},\dot B}\sigma_{2}^{++}(z,\bar z)=\e^{\dot A\dot B}G^+_{-{1\over2},\dot A}\bar G^+_{-{1\over2},\dot B}\sigma_{2}^{--}(z,\bar z)
\eea
We see that this operator contains two primary components. 
One component of the deformation is called the twist operator which is given as 
\begin{equation}\label{bare plus sf}
\sigma_2^{\alpha\dot\alpha}(z,\bar z) =  S^{\alpha}(z)\bar S^{\dot\alpha}(\bar z)\sigma_2(z,\bar z),\quad \alpha,\dot\alpha=\pm
\end{equation}
Here $\sigma_2(z,\bar z)$ is known as the bare twist and $S^{\alpha}(z),\bar S^{\bar\alpha}(\bar z)$ are called spin fields. The bare twist takes two copies of the CFT and joins them together into a doubly wound copy by changing their boundary conditions. For example given two copies of a free boson $X^{(1)}, X^{(2)}$ (where we have momentarily omitted the charge indices for brevity). The bare twist joins these two singly wound copies into a doubly wound copy. On the cylinder, for example, a schematic representation of the twist action as a permutation of copies can be written as
\begin{equation}
    X^{(1)}\to X^{(2)}, \quad X^{(2)}\to X^{(1)}
\end{equation}
More formally under the change in boundary conditions we have
\begin{equation}
    X^{(1)}(\tau,\sigma + 2\pi)=X^{(2)}(\tau,\sigma),\quad X^{(1)}(\tau,\sigma + 2\pi)=X^{(2)}(\tau,\sigma)
\end{equation}
One should really think of the resulting bosonic field as a single field, $X$, where each segment is equivalent to a different copy 
\begin{align}
    X(\tau,\sigma) &= X^{(1)}(\sigma,\tau),\quad 0\leq \sigma < 2\pi\nn
    X(\tau,\sigma) &= X^{(2)}(\sigma,\tau),  \quad 2\pi\leq \sigma < 4\pi
\end{align}
The dimension of the bare twist operator, which is of order $n=2$, for central charge $c=6$ is
\begin{equation}
h ={c\over24}(n-{1\over n})={6\over24}(2-{1\over2}) = {3\over8} 
\end{equation}
The bare twist is sufficient to join and split bosonic fields. To join and split fermions we must also consider the spin field $S^{\alpha}(z)$ (the right moving spin field is similarly given by $\bar S^{\dot\alpha}(\bar z)$). It's dimension for a twist-2 operator is
\begin{equation}
    h={1\over4n}={1\over8}
\end{equation}
Since fermions contain $R$-charge their behavior is more subtle. As seen in the previous subsection, there are two types of vacua for fermions. The NS vacuum which is anti-periodic and the R vacuum which is periodic. The spin field is the operator which acts on the NS sector and yields the R sector. In order maintain the correct boundary conditions for fermions under twisting and untwisting, the bare twist operator must be accompanied by a spin field such that the combined action of the twist operator gives
\begin{equation}
\psi^{(1)}\to \psi^{(2)}, \quad \psi^{(2)}\to \psi^{(1)}
\end{equation}
where the signs of the fermions are maintained under twisting procedures. 
If multiple copies are involved in the twist process then we write
\bea 
\sigma^{\alpha\dot\alpha}_2(z,\bar z) = \sum_{i>j}\sigma^{\alpha\dot\alpha,(i)(j)}_2(z,\bar z)
\eea
where $i,j$ correspond to the respective copies on which the twist operator acts. Thus the full twist operator, including the bare twist and spin field, have left and right moving dimension and charge
\begin{equation}
    h= j = \alpha m = {1\over2},\quad  \bar h= \bar j = \dot \alpha m = {1\over2}
\end{equation}
Where $\alpha,\dot\alpha=\pm$. The twist operator has an OPE with its conjugate given by the following
\begin{equation}
    \s^{\alpha\dot\alpha}(z,\bar z)\s_{\alpha\dot\alpha}(z',\bar z')\sim {1\over |z -z'|^2}
\end{equation}
which has been normalized to the identity. We note that $\alpha,\dot\alpha$ are not summed over.
The other component of the deformation operator includes a supercharge operator which is defined as a contour integral acting on the twist operator as follows
\begin{equation}
 G^{\alpha}_{\dot A,-{1\over2}} \bar G^{\dot\alpha}_{\dot B,-{1\over2}}\s^{\beta\dot\beta}_2(w_i,\bar w_i) = \bigg[{1\over2\pi i}\oint_{w_i}dw G^{\alpha}_{\dot A}(w)\bigg]\bigg[{1\over2\pi i}\oint_{\bar w_i}d\bar w \bar G^{\dot \alpha}_{\dot B}(w)\bigg]\s^+_2(w_i,\bar w_i)
\end{equation}

In this paper we will eventually study the large $N$ limit of the amplitudes we obtain where $N=N_1N_5$. We can thus define the following coupling
\begin{equation}\label{g}
    g=\lambda \sqrt N
\end{equation}
where $g$ is like the t'Hooft coupling in this scenario and should not to be confused with the string coupling, $g_s$. By matching the string spectrum in the PP-wave limit \cite{Gava:2002xb}, the coupling $\lambda$ can be identified as the six-dimensional string coupling, $g_6$, where $g_6 = g_s\sqrt{Q_1/ Q_5}$. In the gravity dual, $g_6$ is related to the radius of AdS$_3$ and $S^3$ as $(R_{AdS}/l_s)^2=g_6\sqrt N$.
Assuming that $N_1\sim N_5$, we can describe the perturbative CFT in the limit $g\ll 1$ while the D1D5 supergravity solution corresponds to the parameter regime where
\begin{equation}
    1\ll g\ll \sqrt N
\end{equation}
The first inequality satisfies the condition that the radius of AdS be much larger than the string length $(R_{AdS}\gg l_s)$ whereas the second inequality satisfies the condition that the string coupling remains small $(g_s\ll1)$.

In this work we compute transition processes in the CFT in the perturbative regime, $g\ll1$ and then extrapolate our results to strong coupling.
Next we define the bosonic and fermionic modes which use to compute the necessary amplitudes.
\section{Cylinder modes}\label{cylinder modes}
In this section we define the mode expansion of the bosonic and fermionic fields on the cylinder which we utilize in the explicit computation of the amplitude we investigate in this work. We consider configurations where the twist operator joins two singly wound copies into a doubly wound copy and the second twist splits the doubly wound copy back into two singly wound copies. Thus the modes in both the initial and final states will defined on singly wound copies of the CFT.  We also consider the modes to be in the Ramond sector where fermions are integer moded. We start with modes defined before the action of the twist operators
\bea\label{w modes before}
\a^{(j)}_{A\dot A,m} &=& {1\over 2\pi }\int_{\tau<\tau_i,\sigma=0}^{2\pi}dw e^{mw}\partial X^{(j)}_{A\dot A}(w)\cr
d^{\a A(j)}_{n} &=&{1\over 2\pi i}\int_{\tau<\tau_i,\sigma=0}^{2\pi}dw  e^{nw}\psi^{\a A,(j)}(w)
\eea
where we choose a constant time slice $\tau$ with the integration along the circle direction $\sigma$
The index $i=1,2$ labels the twist locations. The label $j=1,2$ labels the copy of the CFT. 
The modes after the action of the two twist operators, at constant $\tau$, are given by 
 \bea\label{w modes after twist}
\a^{(j)}_{A\dot A,m} &=& {1\over 2\pi }\int_{\tau>\tau_i,\sigma=0}^{2\pi}dw e^{mw}\partial X^{(i)}_{A\dot A}(w)\cr
d^{\a A(j)}_{n} &=&{1\over 2\pi i}\int_{\tau>\tau_i,\sigma=0}^{2\pi}dw  e^{nw}\psi^{\a A,(i)}(w)
\eea
Since both the initial and final state modes are defined in the singly wound sector, the commutation relations for both cases are given by
\begin{align}
[\alpha^{(i)}_{A\dot A,m},\alpha^{(j)}_{B\dot B,n}] &=-m\e_{ A B}\e_{\dot A\dot B}\d^{(i)(j)}\d_{m+n,0} \nn
\lbrace d^{(i)\a A}_{r} , d^{(j)\beta B}_{s}\rbrace&= - \e^{\a\beta}\e^{AB}\d^{(i)(j)}\d_{r+s,0}
\end{align}
The deformation operator is given by the following supercharge contours circling their respective twist operators
\bea\label{w def}
G^{-}_{\dot A,-{1\over2}}\s^+_2(w_1) &=& {1\over2\pi i}\oint_{w_i}dw G^{-}_{\dot A}(w)\s^+_2(w_1)\cr
G^{+}_{\dot A,-{1\over2}}\s^-_2(w_1) &=& {1\over2\pi i}\oint_{w_i}dw G^{+}_{\dot A}(w)\s^-_2(w_2)
\eea
Next we discuss the initial and final states involved in the transition we desire to compute through the action of the deformation operators.

\section{The transition}\label{transitions of ss}
In this paper we are interested in studying, within the holographical field theory, how an initial mode within gravity could begin to back-react with certain three charge microstate geometries called superstrata. In general, from the gravitational side, studying this back-reaction is a highly nontrivial and complex process. In field theory it is also nontrivial as it corresponds to understanding non-equilibrium dynamics in strongly coupled systems. With this in mind, we hope to make steps toward this understanding by considering a simple scenario that allows us to perform concrete computations which in this case, comes from considering deformations away from a free CFT by a marginal operator. With the understanding that this is a perturbative computation, we hope to at least gain some hints about this back-reaction on the field theory side.
In this section we set up the computation of calculating the CFT transition amplitude of an initial state dual to a graviton mode in the bulk interacting with a superstratum state, a state in the CFT, which is holographically dual to a superstratum geometry in the bulk. 
Let's begin by defining the initial state we are interested in.

\subsection{$(1,0,n)$ Superstratum state}
The CFT dual of the set of ${1\over8}$-BPS smooth microstate geometries called superstrata have been identified at the free point in field theory moduli space \cite{Bena:2016ypk}. In general each bulk geometry with a specified set of charges is given holographically as a coherent superposition of states in the D1D5 CFT. We will assume, for simplicity, that this state is highly peaked around its average value whose quantum numbers correspond to supergravity charges in the bulk. We consider the CFT dual of one of the simpler such configurations, the $(1,0,n)$ superstratum geometry, which we write in the Ramond sector as
\bea
|\psi\rangle &=& \bigg({1\over n!}(L_{-1}-J^3_{-1})^n|00\rangle\bigg)^{N_{00}}\bigg(|++\rangle\bigg)^{N_{++}}
\label{ss state}
\eea
We see that this state is composed of two main ingredients, 1) a combination of generators, $L_{-1}-J^3_{-1}$, raised to the power of $n$, acting on the state $|00\rangle$ which all together are called 00 strands, and 2) another set of states called $++$ strands, both which are in the singly wound sector. The $00$ strands are neutral under R-charge and thus contain no net rotation along $S^3$. The state $(L_{-1}-J^3_{-1})^n|00\rangle$ corresponds to a supergraviton mode with momentum along the $S^1$ direction. The $++$ strands carry R-charge under $SU(2)_L\times SU(2)_R$ and thus correspond to angular momentum along $S^3$ in the bulk theory. Here $N_{00}$ is the number of $00$ strands and $N_{++}$ is the number of $++$ strands where their combined total should add up to the effective winding number
\begin{equation}
    N_1N_5 = N_{00} + N_{++}
\end{equation}
The state $|00\rangle$ is defined as

\bea
|00\rangle &=&{1\over\sqrt2}\e_{AB}|AB\rangle = {1\over\sqrt2}\e_{AB} d^{-A}_0 \bar d^{-B}_0 |++\rangle 
\eea
with $|++\rangle$ defined in (\ref{Ramond}). The state $|++\rangle$ has quantum numbers 
\begin{equation}
   j=\bar j= m = \bar m = {1\over2}
\end{equation}
The superstratum state $|\psi\rangle$ has quantum numbers 
\begin{equation}
    h = nN_{00}+{N_{00}+N_{++}\over4},\quad j =m= {N_{++}\over2}, \qquad \bar h = {N_{00}+N_{++}\over4},\quad \bar j =\bar m= {N_{++}\over2} 
\end{equation}
The relation between the CFT charges and parameters of the $(1,0,n)$ superstratum geometry are given by
\begin{equation}
 N_{++} = \mathcal N a^2,\qquad N_{00} = \mathcal N {b^2\over2}   
\end{equation}
with 
\begin{equation}
    \mathcal N = {V_4 R_y^2\over(2\pi)^4g_s\alpha'^4}
\end{equation}
Here $V_4$ is the volume of $T^4$, $R_y$ is the radius of $S^1$, and $g_s$ is the 10-D string coupling.
\subsection{Initial state}
We are interested in the scenario of how an in-falling graviton mode into the superstratum geometry can interact with and even alter it. In \cite{Guo:2021gqd} the authors considered the CFT dual of tidal excitations of an in-falling graviton in a superstratum geometry. There the superstratum state was maintained as a fixed background, whose quantum numbers did not change from the initial state to the final state. Thus the initial graviton probe time involved into an excited string state identified as the splitting of the initial graviton mode identified with the CFT dual of the probe, into several modes in the final state. The difference in this current work is that the corresponding superstratum state will now be allowed to change due to the interaction with the graviton mode. Roughly, this will correspond to a back-reacted process in the bulk where the in-falling graviton is no longer treated as just a probe. Let's discuss the initial state we will be interested in. First we consider a single $N_{00}$ strand which can be simplified in the following way using the algebra in appendix \ref{algebra}
\bea\label{LJ on zero modes}
{1\over n!}(L^{(i)}_{-1} - J^{3(i)}_{-1})^n|00\rangle^{(i)} = {1\over\sqrt2}\e_{AB}d^{-A(i)}_{-n}\bar d^{-B(i)}_0| ++\rangle^{(i)}
\eea
where $i$ gives the copy label.
We can therefore write a single superstratum state, combining the a $00$ strand and a $++$ strand in the following way
\bea\label{ss state ij}
|\psi_{i,j}\rangle &=& {1\over n!}(L^{(i)}_{-1}-J^{3(i)}_{-1})^n|00\rangle^{(i)}|++\rangle^{(j)}\nn
&=& {1\over\sqrt2}\e_{AB}d^{-A(i)}_{-n}\bar d^{-B(i)}_0|++\rangle^{(i)} |++\rangle^{(j)}
\eea
We can write our initial state as a graviton mode acting on the superstratum state given in (\ref{ss state ij})
\begin{align}\label{psi init}
|\Psi_0\rangle &\equiv
{1\over \sqrt2m}\a^{(2)}_{++,-m}\bar\a^{(2)}_{--,-m} |\psi_{1,2}\rangle + {1\over \sqrt2m}\a^{(1)}_{++,-m}\bar\a^{(1)}_{--,-m} |\psi_{2,1}\rangle
\nn
 &={1\over2m}\a^{(2)}_{++,-m}\bar\a^{(2)}_{--,-m} \e_{AB}d^{-A(1)}_{-n}\bar d^{-B(1)}_0|++\rangle^{(1)}|++\rangle^{(2)}  \cr
& +  {1\over2m}\a^{(1)}_{++,-m}\bar\a^{(1)}_{--,-m} \e_{AB}d^{-A(2)}_{-n}\bar d^{-B(2)}_0|++\rangle^{(2)}|++\rangle^{(1)}
\end{align}
The above state is copy symmetric as is required by the symmetric product condition and has been normalized to $1$. 

Since we are considering the CFT dual of a back-reaction process in the bulk we should choose which final state to consider, there being many to choose from. This consideration, from the CFT perspective, really concerns how microstate geometries/fuzzballs time evolve, exploring their phase space upon interaction with in-falling matter. We consider a choice of final state, which corresponds to a nontrivial change of the background, while being simple enough to compute, and also providing some insight into the transition process. 
The aim of this paper is to study, perhaps, the simplest nontrivial transition in this twist configuration for which the initial superstratum state itself changes,
\begin{align}\label{transitions}
     \text{superstratum state and graviton} &\to \text{microstratum state and graviton'}
\end{align}
The prime on the right hand side denotes a graviton in the final state which has different quantum numbers than its initial state counterpart. 
We have introduced a terminology called a microstratum state which we label as the CFT dual state, found in \cite{Ganchev:2021ewa}, of one member of the more recently constructed microstrata, a class of smooth back-reacted non-BPS, three-charge microstate geometries. We will write their explicit form in the next section where we construct the final state we are interested in.  
\subsection{Final state}
Let us now explicitly construct the final state which will enter into our transition amplitude. 
The CFT dual of a microstratum geometry is given in \cite{Ganchev:2021ewa} and can be written in the NS sector as the following
\footnote{Note here that we've chosen the $R$ - charges in the left and right moving sectors to be $j=\bar j=-m=-\bar m={1\over2}$, which is opposite to those in \cite{Ganchev:2021ewa} where $j=\bar j=m=\bar m={1\over2}$. This corresponds to angular momentum along $S^3$ moving in the opposite direction. We make this choice to align with previous computations involving superstrata states in the D1D5 CFT.}$^{,\thinspace}$\footnote{The state recorded above is perhaps the simplest microstratum state with more intricate ones being identified in \cite{Ganchev:2023sth}. Furthermore, in \cite{Ganchev:2021ewa} the quantum numbers for which the state is identified with the corresponding gravitational solution is for $m=1$. With this in mind we will keep $m$ general throughout the computation and numerically compute the amplitude for various values, including $m=1$ (in the actual amplitude $m$ here will be changed to $p$). Lastly, since (\ref{micro}) is non-BPS it should receive higher order corrections when flowing from the free theory to the supergravity regime. We also keep this in mind while performing computations.}
\begin{align} \label{micro}
 &\bigg({1\over(m+n)!m!}(L_{-1})^{m+n}(\bar L_{-1})^m\big|O_{-{1\over2},-{1\over2}}\rangle\bigg)^{N_1}\big(|0\rangle_{NS}\big)^{N_{++}}
 \end{align}
 where, using the notation of \cite{Ganchev:2021ewa}, we have
 \begin{align}
 \big|O_{-{1\over2},-{1\over2}}\rangle&\equiv {1\over\sqrt2}\e_{AB}d^{-A}_{-{1\over2}}\bar d^{-B}_{-{1\over2}}|0_{NS}\rangle\nn
\end{align}
and
\begin{align}\label{ms}
 &\bigg({1\over(m+n)!m!}(L_{-1})^{m+n}(\bar L_{-1})^m\big|O_{-{1\over2},-{1\over2}}\rangle\bigg)^{N_1}\big(|0\rangle_{NS}\big)^{N_{++}}\nn 
 &= \bigg({1\over(m+n)!m!}\big(L_{-1})^{m+n}(\bar L_{-1})^m{1\over\sqrt2}\e_{AB}d^{-A}_{-{1\over2}}\bar d^{-B}_{-{1\over2}}|0_{NS}\rangle\bigg)^{N_1}\big(|0_{NS}\rangle\big)^{N_{++}}
\end{align}
Since our computations will be performed in the Ramond sector, let us spectral the above state by $\alpha=1$ which, using (\ref{sf}), leads to the following changes in the generators and states
\bea 
L_{-1}&\to& L_{-1}- J^{3}_{-1}\nn
|0_{NS}\rangle&\to&|++\rangle\nn 
{1\over\sqrt2}\e_{AB}d^{-A}_{-{1\over2}}\bar d^{-B}_{-{1\over2}}|0_{NS}\rangle&\to&{1\over\sqrt2}\e_{AB}d^{-A}_{0}\bar d^{-B}_{0}|++\rangle\nn
\eea
Inserting these changes into (\ref{ms}) gives the state
\begin{align}
\bigg({1\over(m+n)!m!}(L_{-1}-J^{3}_{-1})^{m+n}(\bar L_{-1}-\bar J^{3}_{-1})^m{1\over\sqrt2}\e_{AB}d^{-A}_{0}\bar d^{-B}_{0}|++\rangle_{R}\bigg)^{N_1}\big(|++\rangle_{R}\big)^{N_{++}}
\end{align}
where every copy is in the singly wound sector.
Using the identity (\ref{LJ on zero modes}) we find that 
\begin{align}
   & \bigg({1\over(m+n)!m!}(L_{-1}-J^{3}_{-1})^{m+n}(\bar L_{-1}-\bar J^{3}_{-1})^m{1\over\sqrt2}\e_{AB}d^{-A}_{0}\bar d^{-B}_{0}|++\rangle_{R}\bigg)^{N_1}\big(|++\rangle_{R}\big)^{N_{++}}
   \nn
&=\bigg({1\over\sqrt2}\e_{AB}d^{-A}_{-(m+n)}\bar d^{-B}_{-m}|++\rangle_{R}\bigg)^{N_1}\big(|++\rangle_{R}\big)^{N_{++}}
\end{align}
Let us now define a microstratum state with a single graviton mode for $N_1=N_{++}=1$ where we appropriately symmetrize over the two copies involved
\begin{align}\label{ms plus gr ket}
|\Psi_f\rangle 
&= {1\over2r}\a^{(2)}_{++,-r}\bar\a^{(2)}_{--,-r}\e_{AB}d^{-A(1)}_{-(p+q)}\bar d^{-B(1)}_{-p}|++\rangle^{(1)}_{R}|++\rangle^{(2)}_{R}
\nn
&+ {1\over2r}\a^{(1)}_{++,-r}\bar\a^{(1)}_{--,-r}\e_{AB}d^{-A(2)}_{-(p+q)}\bar d^{-B(2)}_{-p}|++\rangle^{(2)}_{R}|++\rangle^{(1)}_{R}
\end{align}
Since we will be computing amplitudes with these final states let's write the above in \text{bra} form 
\begin{align}\label{ms+gr}
\langle\Psi_f|&={1\over2r}{}^{(2)}_R\langle++|{}^{(1)}_R\langle++|\e_{AB}\bar d^{+B(1)}_{p}d^{+A(1)}_{p+q}\bar\alpha^{(2)}_{++,r}\alpha^{(2)}_{--,r}\nn
&+{1\over2r}{}^{(1)}_R\langle++|{}^{(2)}_R\langle++|\e_{AB}\bar d^{+B(2)}_{p}d^{+A(2)}_{p+q}\bar\alpha^{(1)}_{++,r}\alpha^{(1)}_{--,r}
\end{align}
where this state has also been normalized to $1$.
Next we discuss the amplitude to be computed.

\section{Transition amplitude to be computed}\label{amplitudes}
Let us now compute the actual amplitudes which correspond to the physical transition process we would like to analyze. Considering second order in the deformation (\ref{S}), utilizing (\ref{D}), (\ref{psi init}), (\ref{ms+gr}) we define the following integrated amplitude
\begin{align}\label{gen int}
A^{0\to f}(\tau) &= 
{1\over2}\lambda^2\int d^2w_2d^2w_1\mathcal{A}^{0 \to f}(w_1,w_2,\bar w_1,\bar w_2)
\end{align}
with the unintegrated amplitude defined as
\begin{align}\label{Asec5}
\mathcal{A}^{0 \to f}(w_1,w_2,\bar w_1,\bar w_2) &\equiv \langle \Psi_f|DD|\Psi_0\rangle = \mathcal{A}^{0\to f}_{1221} + \mathcal{A}^{0\to f}_{1212} + \mathcal{A}^{0\to f}_{2121} + \mathcal{A}^{0\to f}_{2112}
\end{align}
where the individual amplitudes are defined as  
\begin{equation}\label{subsub ms plus gr}
\mathcal{A}^{0\to f}_{ijkl} \equiv{1\over 4mr}\e_{AB}\e^{\dot A_2\dot B_2}\e^{\dot A_1\dot B_1}\e_{A'B'}\big[\mathcal{A}^{0\to f}_{ijkl}\big]^{AA'}_{\dot A_2\dot A_1}\big[\bar{\mathcal{A}}^{0\to f}_{ijkl}\big]^{BB'}_{\dot B_2\dot B_1}
\end{equation}
where
\begin{align}\label{subsub ms plus gr lr}
&\big[\mathcal{A}^{0\to f}_{ijkl}\big]^{AA'}_{\dot A_2\dot A_1}\nn
&\equiv{}^{(j)}\langle0_{R,+}|{}^{(i)}\langle0_{R,+}| d^{+A(i)}_{p+q}\alpha^{(j)}_{--,r}G^+_{-{1\over2},\dot A_2}\sigma_{2}^{-}(w_2)G^-_{-{1\over2},\dot A_1}\sigma_{2}^{+}(w_1)\a^{(k)}_{++,-m}d^{-A'(l)}_{-n}|0^+_R\rangle^{(l)}|0^+_R\rangle^{(k)}
\nn
\nn
&\big[\bar{\mathcal{A}}^{0\to f}_{ijkl}\big]^{BB'}_{\dot B_2\dot B_1}\nn
&\equiv{}^{(j)}\langle\bar0_{R,+}|{}^{(i)}\langle\bar0_{R,+}|\bar d^{+B(i)}_p \bar\alpha^{(j)}_{++,r}\bar G^+_{-{1\over2},\dot B_2}\sigma_{2}^{-}(\bar w_2)\bar G^-_{-{1\over2},\dot B_1}\sigma_{2}^{+}(\bar w_1)\bar\a^{(k)}_{--,-m}\bar d^{-B'(l)}_0|\bar0^+_R\rangle^{(l)}|\bar0^+_R\rangle^{(k)}
\nn
\end{align}
Note that we have suppressed the twist insertion dependence in the definitions of the left and right moving amplitudes for brevity, reintroducing it later when necessary.
Also note that for the factorization of the above amplitudes into left and right moving parts we have used (\ref{Ramond}) and (\ref{Ramond conj}) to factorize the vacuum state
\begin{align}
    |++\rangle^{(i)} = |0^{++}_R\rangle^{(i)} &=|0^+_R\rangle^{(i)}|\bar0^+_R\rangle^{(i)}
    \nn
    {}^{(i)}\langle++| = {}^{(i)}\langle0_{R,++}| &={}^{(i)}\langle0_{R,+}|{}^{(i)}\langle\bar0_{R,+}|
\end{align}
We also use the fact that in this particular twist configuration we can treat the action of the twist operators in a factorized way between left and right moving sectors and thus write
\begin{equation}
    \sigma^{\alpha\dot\alpha}(w_i,\bar w_i)=\sigma^{\alpha}(w_i)\sigma^{\dot\alpha}(\bar w_i)
\end{equation}
which is not always the case for generic configurations. \footnote{In general this action may not factorize which happens in the case of four twist operators \cite{Dixon:1986qv,Guo:2024edj} in the scenario where initial and final states are both in the singly wound sector. This is because the covering surface is no longer a sphere but a genus one torus which couples left and right movers.}
Let us make a comment about the notation which will slightly ease the proliferation of indices and labels. In the forthcoming computations we will label modes in the initial state with negative oscillator indices, defined with \text{ket} states. We will label modes in the final state with positive oscillator indices since they appear in \text{bra} states in the amplitudes. This way we don't have to add an additional label to specify whether the mode is in the initial state or final state a distinction which becomes important when computing the contractions between terms on the covering space.

\section{Covering space method}\label{covering space}
Twist operators introduce branch cuts in the plane as seen from the corresponding fields involved. These branch cuts leave the fields double valued and thus ill-defined. To solve this we will use the covering space method \cite{Lunin:2000yv,Lunin:2001pw} to derive the corresponding amplitudes. This includes mapping the system to a covering space in which the involved fields become single valued, thus yielding well defined quantities. In the next section we demonstrate this procedure in detail.
\subsection{Spectral flow on the cylinder}
To simplify the amplitude computation, it is convenient to first perform a spectral flow on the cylinder. Utilizing spectral flow relations in subsection (\ref{NS R}) and spectral flowing by $\a=-1$ yields the following changes of the fields and states which carry R-charge
 \bea
  \psi^{\pm A}(w)&\to& e^{\pm{1\over2} w }  \psi^{\pm A}(w)\cr
 \s^{\pm}_2(w_i)&\to& e^{\pm{w_i\over2}}\s^{\pm}_2(w_i),\quad i=1,2\cr
 G^{\pm}_{\dot A}(w)&\to&e^{\pm{w\over2}}G^{\pm}_{\dot A}(w)\cr
 |0^+_R\rangle^{(j)}&\to& |0_{NS}\rangle^{(j)},\quad j=1,2\cr
{}^{(j)} \langle 0_{R,+}|&\to& {}^{(j)}\langle 0_{NS}|,\quad j=1,2
 \eea 
and similarly for the right moving sector. The bosonic fields and modes are unchanged under spectral flow since they don't carry $R$-charge. Below we record how the corresponding modes defined before and after the twist operators change as a result. 
The modes before the twist operators, (\ref{w modes before}), become
 \bea\label{w modes before sf}
\a^{(j)}_{A\dot A,-m} &\to& {1\over 2\pi }\int_{\tau<\tau_i,\sigma=0}^{2\pi}dw e^{-mw}\partial X^{(j)}_{A\dot A}(w)\cr
d^{\pm A(j)}_{-r}&\to&{1\over 2\pi i}\int_{\tau<\tau_i,\sigma=0}^{2\pi}dw  e^{(-r\pm {1\over2})w}\psi^{\a A(j)}(w)
\eea
with the ones after, (\ref{w modes after twist}), becoming
 \bea\label{w modes after sf}
\a^{(j)}_{A\dot A,m} &\to& {1\over 2\pi }\int_{\tau>\tau_i,\sigma=0}^{2\pi}dw e^{mw}\partial X^{(j)}_{A\dot A}(w)\cr
d^{\pm A(j)}_{r}&\to&{1\over 2\pi i}\int_{\tau>\tau_i,\sigma=0}^{2\pi}dw  e^{(r \pm {1\over2})w}\psi^{\a A(j)}(w)
\eea
The deformation operators, (\ref{w def}), transform as
\bea\label{w def contours sf}
G^{-}_{\dot A,-{1\over2}}\s^+_2(w_1)\to G^{-}_{\dot A,-1}[e^{w_1\over2}\s^+_2(w_1)] &=& {1\over2\pi i}\oint_{w_1}dw G^{-}_{\dot A}(w)e^{-{1\over2}w}[e^{w_1\over2}\s^+_2(w_1)]\cr
G^{+}_{\dot A,{-{1\over2}}}\s^-_2(w_2)\to G^{+}_{\dot A,0}[e^{-{w_2\over2}}\s^-_2(w_2)] &=& {1\over2\pi i}\oint_{w_2}dw G^{+}_{\dot A}(w)e^{{1\over2}w}[e^{-{w_2\over2}}\s^-_2(w_2)]
\eea
We keep track of any Jacobian factors introduced by transformations of the twist operators, by collecting them within brackets. We will recompute this contribution later, separately, and combine it with the rest of the amplitude.
\subsection{Mapping to $z$-plane}
We now map our system, first to the $z$-plane using the map
\bea
z=e^{w}
\eea
The measure and fields transform as follows
\bea\label{z transf}
dw\partial X^{(j)}_{A\dot A}(w) &\to& dz\partial X^{(j)}_{A\dot A}(z),\qquad j=1,2\cr
dwG^{\a}_{\dot A}(w)&\to& dz\bigg({dz\over dw} \bigg)^{1\over2}G^{\a}_{\dot A}(z)=dzz^{1\over2}G^{\a}_{\dot A}(z)\cr
dw\psi^{\a A(j)}(w)&\to&dz\bigg({dz\over dw} \bigg)^{-{1\over2}} \psi^{\a A}(z) = dzz^{-{1\over2}} \psi^{\a A(j)}(z),\qquad j=1,2\cr
\s^{\pm}_2(w_i)&\to& \bigg({dz\over dw}\bigg)^{1\over2}\bigg|_{z=z_i}\s^{\pm}_2(z_i)=z_i^{1\over2}\s^{\pm}_2(z_i),\qquad i=1,2
\eea
where we note that the combined transformation of the Jacobian and the bosonic field together remains unchanged from the cylinder to the plane.
Applying the transformations (\ref{z transf}) to (\ref{w modes before sf}), (\ref{w modes after sf}) and (\ref{w def contours sf}),  the modes before the twist operators are now given by
\bea\label{z plane initial}
\a^{(j)}_{A\dot A,-m} &\to& {1\over 2\pi }\oint_{z=0}dz z^{-m}\partial X^{(j)}_{A\dot A}(z)\cr
d^{\pm A(j)}_{-r} &\to&{1\over 2\pi i}\oint_{z=0}dz  z^{(-r \pm {1\over2}-{1\over2})}\psi^{\a A,(j)}(z)
\eea
where we have freely deformed the $z$-plane contours corresponding to modes defined for $|z|<|z_i|$ to $z=0$.
The modes after the two twist operators become
\bea\label{z plane final}
\a^{(j)}_{A\dot A,m} &\to& {1\over 2\pi }\oint_{z=\infty}dz z^m\partial X^{(j)}_{A\dot A}(z)\cr
d^{\pm A(j)}_{r} &\to&{1\over 2\pi i}\oint_{z=\infty}dz  z^{(r \pm {1\over2}-{1\over2})}\psi^{\a A(j)}(z)
\eea
where we have freely deformed the $z$-plane contours corresponding to modes defined for $|z|>|z_i|$ to $z=\infty$.
The deformation operators become
\bea\label{z plane deform}
G^{-}_{\dot A,-{1\over2}}\s^+_2(w_1) &\to&  {1\over2\pi i}\oint_{z_1}dz G^{-}_{\dot A}(w)[z_1\s^+_2(z_1)]\cr
G^{+}_{\dot A,{-{1\over2}}}\s^-_2(w_2) &\to& {1\over2\pi i}\oint_{z_2}dz G^{+}_{\dot A}(z)z[\s^-_2(z_2)]
\eea

\subsection{Mapping to the $t$-plane}
Now we map our system to the covering space. To do so we must define the appropriate covering map.
Since we begin and end in the singly wound sector we use the following map, which was developed and first utilized in \cite{Carson:2015ohj},
\begin{equation}\label{map}
z={(t+a)(t+b)\over t}
\end{equation}
in which the double-valued fields at the respective $z$-plane locations are resolved by the following $t$-plane images
\begin{align}\label{copy images}
z=0&\implies t=-a ~ (\text{Copy 1}),~ t=-b ~(\text{Copy 1})\nn
z=\infty&\implies t=\infty ~ (\text{Copy 1}),~~ t=0 ~(\text{Copy 2})
\end{align}
We see that the map (\ref{map}) has resolved both the initial copies, which are defined at the same $z$-plane location, and the final copies, which are defined at the same $z$-plane location, into distinct $t$-plane images. This is the power of the covering map.
We have made an explicit choice of where to define the image of copy $1$ and copy $2$ initial and final. Since our system is copy symmetric we could equally have made the other choice.

In order to find the location of the twist operators in the $t$-plane we require that at those points the derivative of the covering map with respect to the covering space coordinate be zero, i.e.
\bea
{dz\over dt}= {t^2-ab\over t^2} = {(t-t_1)(t-t_2)\over t^2} = 0
\eea
Physically, this condition enables the two Riemann sheets to touch at a point, where the derivative is set zero, also enabling the expansion around the twist location to be second order in the covering space coordinate, i.e. $z-z_0\approx C(t-t_0)^2$ with $C$ a constant, which is the necessary behavior of a twist operator of order $2$. This gives
\bea
t_1&=&-\sqrt{ab}\cr
t_2&=&\sqrt{ab}
\eea
This is also a choice. We could have defined the t-plane images of the twist operators to be switched.
Based on the above choice, using (\ref{map}) the corresponding $z$-plane locations of the twist operators written in terms of the $t$-plane locations above are given by
\begin{align}\label{zt}
    z_1 &= {(t_1+a)(t_1+b)\over t_1} = a + b -2\sqrt{ab}\nn
    z_2 &= {(t_2+a)(t_2+b)\over t_2} = a + b +2\sqrt{ab}
\end{align}
From the map (\ref{map}) and image identifications (\ref{copy images}) the fields and measures transform in the following way
\bea
dz\partial X^{(j)}_{A\dot A}(z) &\to& dt\partial X^{(j)}_{A\dot A}(t),\qquad j=1,2\cr
dzG^{\a}_{\dot A}(z)&\to& dt\bigg({dz\over dt} \bigg)^{-{1\over2}}G^{\a}_{\dot A}(t)=dt{t\over(t-t_1)^{1\over2}(t-t_2)^{1\over2}}G^{\a}_{\dot A}(t)\cr
dz\psi^{\a A(j)}(z)&\to&dt\bigg({dz\over dt} \bigg)^{{1\over2}} \psi^{\a A(j)}(t) = dt{(t-t_1)^{1\over2}(t-t_2)^{1\over2}\over t} \psi^{\a A(j)}(t),\qquad j=1,2\cr
\s^{\pm}_2(z_i)&\to& C_iS^{\pm}(t_i),\qquad i=1,2
\eea
where $C_i$ is a constant.
From (\ref{copy images}), the following coordinates define the locations of copy $1$ initial, copy $2$ initial, copy $1$ final and copy $2$ final
\bea
&&\text{Copy 1 initial}: w=-\infty, ~z=0, ~t=-a
\cr\cr
&&\text{Copy 2 initial}: w=-\infty, ~z=0, ~t=-b
\cr\cr
&&\text{Copy 1 final}: w=\infty, ~z=\infty, ~t=\infty
\cr\cr
&&\text{Copy 2 final}: w=\infty, ~z=\infty, ~t=0
\eea
Applying the above changes to the modes defined before the twist operators, (\ref{z plane initial}), and after the twist operators, (\ref{z plane final}) we get 
\subsubsection*{Copy 1 initial}
\begin{align}\label{t plane init 1}
\a^{(1)}_{A\dot A,-m} &\to {1\over 2\pi }\oint_{t=-a}dt {(t+a)^{-m}(t+b)^{-m}\over t^{-m}}\partial X_{A\dot A}(t)\cr
d^{\pm A(1)}_{-r} &\to{1\over 2\pi i}\oint_{t=-a} dt {(t-t_1)^{1\over2}(t-t_2)^{1\over2}(t+a)^{(-r \pm {1\over2}-{1\over2})}(t+b)^{(-r \pm {1\over2}-{1\over2})}\over t^{(-r \pm {1\over2}+{1\over2})}} \psi^{\pm A}(t)
\end{align}
\subsubsection*{Copy 2 initial}
\begin{align}\label{t plane init 2}
\a^{(2)}_{A\dot A,-m} &\to {1\over 2\pi }\oint_{t=-b}dt {(t+a)^{-m}(t+b)^{-m}\over t^{-m}}\partial X_{A\dot A}(t)\cr
d^{\pm A(2)}_{-r} &\to{1\over 2\pi i}\oint_{t=-b} dt {(t-t_1)^{1\over2}(t-t_2)^{1\over2}(t+a)^{(-r \pm {1\over2}-{1\over2})}(t+b)^{(-r \pm {1\over2}-{1\over2})}\over t^{(-r \pm {1\over2}+{1\over2})}} \psi^{\pm A}(t)
\end{align}
\subsubsection*{Copy 1 final}
\begin{align}\label{t plane fin 1}
\a^{(1)}_{A\dot A,m} &\to {1\over 2\pi }\oint_{t=\infty}dt {(t+a)^m(t+b)^m\over t^m}\partial X_{A\dot A}(t)\cr
d^{\pm A(1)}_{r} &\to{1\over 2\pi i}\oint_{t=\infty} dt {(t-t_1)^{1\over2}(t-t_2)^{1\over2}(t+a)^{(r \pm {1\over2}-{1\over2})}(t+b)^{(r \pm {1\over2}-{1\over2})}\over t^{(r \pm {1\over2}+{1\over2})}} \psi^{\pm A}(t)
\end{align}
\subsubsection*{Copy 2 final}
\begin{align}\label{t plane fin 2}
\a^{(2)}_{A\dot A,m} &\to - {1\over 2\pi }\oint_{t=0}dt {(t+a)^m(t+b)^m\over t^m}\partial X_{A\dot A}(t)\cr
d^{\pm A(2)}_{r} &\to-{1\over 2\pi i}\oint_{t=0} dt {(t-t_1)^{1\over2}(t-t_2)^{1\over2}(t+a)^{(r \pm {1\over2}-{1\over2})}(t+b)^{(r \pm {1\over2}-{1\over2})}\over t^{(r \pm {1\over2}+{1\over2})}} \psi^{\pm A}(t)
\end{align}
The minus sign for the copy 2 final modes arises because the map goes as $z\sim 1/ t$ in which case the contour reverses direction.
The deformation operators, (\ref{z plane deform}), become
\bea\label{t plane deform}
G^{-}_{\dot A,-{1\over2}}\s^+_2(w_1) &\to& {1\over2\pi i}\oint_{t_1}dt {t\over (t-t_1)^{1\over2}(t-t_2)^{1\over2}} G^{-}_{\dot A}(t)[C_1z_1S^+(t_1)]\cr
G^{+}_{\dot A,-{1\over2}}\s^-_2(w_2) &\to& {1\over2\pi i}\oint_{t_2}dt{(t+a)(t+b)\over (t-t_1)^{1\over2}(t-t_2)^{1\over2}} G^{+}_{\dot A}(t)[C_2S^-(t_2)]
\eea
Recalling (\ref{bare plus sf}), we note that the $t$-plane retains the spin field insertions $S^{\pm}(t_i)$ at various points $t_i$ which we see explicitly in (\ref{t plane deform}) while the information about the bare twist operators becomes encoded in the covering map itself. We must therefore perform an appropriate number of spectral flows to remove these spin fields so that only the bosonic and fermionic modes remain allowing us to compute the $t$-plane amplitudes by standard Wick contraction.

Below we write out the spectral flow sequence and the transformations they impose on the corresponding fields which are affected, namely those which carry $R$-charge. Again recalling the spectral flow transformations (\ref{sf}) and (\ref{vacuum sf}) we have the following
\subsubsection*{$\a=-1$ at $t_1$}
\bea
S^{+}(t_1)&\to& 1\cr
\psi^{\pm A}(t)&\to& (t-t_1)^{\pm {1\over2}}\psi^{\pm A}(t)\cr
G^{\pm}_{\dot A}(t)&\to& (t-t_1)^{\pm {1\over2}} G^{\pm}_{\dot A}(t)
\eea
\subsubsection*{$\a=1$ at $t_2$}
\bea
S^{-}(t_2)&\to& 1\cr
\psi^{\pm A}(t)&\to& (t-t_2)^{\mp {1\over2}}\psi^{\pm A}(t)\cr
G^{\pm}_{\dot A}(t)&\to& (t-t_2)^{\mp {1\over2}} G^{\pm}_{\dot A}(t)
\eea
Our various modes change as follows
\subsubsection*{Copy 1 initial}
From (\ref{t plane init 1})
\bea\label{c1 init pr}
\a^{(1)}_{A\dot A,-m} &\to& \a'^{(1)}_{A\dot A,-m} \equiv{1\over 2\pi }\oint_{t=-a}dt{(t+a)^{-m}(t+b)^{-m}\over t^{-m}}\partial X_{A\dot A}(t)\cr
d^{+ A(1)}_{-r} &\to&d'^{+ A(1)}_{-r}\equiv{1\over 2\pi i}\oint_{t=-a} dt {(t-t_1)(t+a)^{-r}(t+b)^{-r}\over t^{-r+1}} \psi^{+ A}(t)\cr
d^{- A(1)}_{-r} &\to&d'^{- A(1)}_{-r}\equiv{1\over 2\pi i}\oint_{t=-a} dt {(t-t_2)(t+a)^{-r-1}(t+b)^{-r-1}\over t^{-r}} \psi^{- A}(t)
\eea

\subsubsection*{Copy 2 initial}
From (\ref{t plane init 2})
\bea\label{c2 init pr}
\a^{(2)}_{A\dot A,-m} &\to& \a'^{(2)}_{A\dot A,-m} \equiv {1\over 2\pi }\oint_{t=-b}dt {(t+a)^{-m}(t+b)^{-m}\over t^{-m}}\partial X_{A\dot A}(t)\cr
d^{+ A(2)}_{-r} &\to&d'^{+ A(2)}_{-r}\equiv{1\over 2\pi i}\oint_{t=-b} dt {(t-t_1)(t+a)^{-r}(t+b)^{-r}\over t^{-r+1}} \psi^{+ A}(t)\cr
d^{- A(2)}_{-r} &\to&d'^{- A(2)}_{-r}\equiv{1\over 2\pi i}\oint_{t=-b} dt {(t-t_2)(t+a)^{-r-1}(t+b)^{-r-1}\over t^{-r}} \psi^{- A}(t)
\eea

\subsubsection{Copy 1 final}
From (\ref{t plane fin 1})
\bea\label{c1 fin pr}
\a^{(1)}_{A\dot A,m} &\to&\a'^{(1)}_{A\dot A,m}\equiv {1\over 2\pi }\oint_{t=\infty}dt {(t+a)^m(t+b)^m\over t^m}\partial X_{A\dot A}(t)\cr
d^{+ A(1)}_{r} &\to&d'^{+ A(1)}_{r}\equiv{1\over 2\pi i}\oint_{t=\infty} dt {(t-t_1)(t+a)^{r}(t+b)^{r}\over t^{r+1}} \psi^{+ A}(t)\cr
d^{- A(1)}_{r} &\to&d'^{- A(1)}_{r}\equiv{1\over 2\pi i}\oint_{t=\infty} dt {(t-t_2)(t+a)^{r-1}(t+b)^{r-1}\over t^{r}} \psi^{- A}(t)
\eea
\subsubsection{Copy 2 final}
From (\ref{t plane fin 2})
\bea\label{c2 fin pr}
\a^{(2)}_{A\dot A,m} &\to&\a'^{(2)}_{A\dot A,m}\equiv -{1\over 2\pi }\oint_{t=0}dt {(t+a)^m(t+b)^m\over t^m}\partial X_{A\dot A}(t)\cr
d^{+ A(2)}_{r} &\to&d'^{+ A(2)}_{r}\equiv-{1\over 2\pi i}\oint_{t=0} dt {(t-t_1)(t+a)^{r}(t+b)^{r}\over t^{r+1}} \psi^{+ A}(t)\cr
d^{- A(2)}_{r} &\to&d'^{- A(2)}_{r}\equiv-{1\over 2\pi i}\oint_{t=0} dt {(t-t_2)(t+a)^{r-1}(t+b)^{r-1}\over t^{r}} \psi^{- A}(t)
\eea
where the primes, $'$, in the above mode definitions denote their expressions after all spectral flow and coordinate transformations have been performed.
The deformations, (\ref{t plane deform}), also become
\bea\label{t plane deform sf}
G^{-}_{\dot A,-{1\over2}}\s^+_2(w_1) &\to& {1\over2\pi i}\oint_{t_1}dt {t\over (t-t_1)} G^{-}_{\dot A}(t)[C'_1z_1]\cr
G^{+}_{\dot A,-{1\over2}}\s^-_2(w_2) &\to& {1\over2\pi i}\oint_{t_2}dt{(t+a)(t+b)\over (t-t_2)} G^{+}_{\dot A}(t)[C'_2]
\eea
Defining modes natural to the $t$-plane coordinate for the supercharge operator we have 
\bea
\tilde G^{\a,t_i}_{\dot A,r} = {1\over 2\pi i}\oint_{t_i} dt (t-t_i)^{r+{1\over2}}G^{\a}_{\dot A}(t)
\eea
which are centered around $t_i$.
We can write the deformation operators, in terms of these $t$-plane modes, as
\bea\label{GG after sf}
G^{-}_{\dot A,-{1\over2}}\s^+_2(w_1) &\to&t_1\tilde{G}^{-,t_1}_{\dot A,-{3\over2}} [C'_1z_1]\cr
G^{+}_{\dot A,-{1\over2}}\s^-_2(w_2) &\to&(t_2+a)(t_2+b)\tilde{G}^{-,t_2}_{\dot A,-{3\over2}} [C'_2]
\eea
where, since we have removed all spin fields, we have used the fact that locally, around the respective t-plane images of the twist locations, we have the following constraint from the locally defined vacuum
\bea \label{vacuum constraint}
\tilde G^{t_i}_{r\over2}|0_{NS}\rangle_{t_i} =0,\quad r\geq -1
\eea
This condition enables one to keep only the first term when expanding the integrands in (\ref{t plane deform sf}) around their respective location of integration, the $t_i$'s.
The constants in the brackets $C'_1,C'_2$ can be reabsorbed into the normalization of the twist operators and the factor of $z_1$ will be included in the formula that we obtain when we compute the vacuum amplitude with just the twist operators, which we do in the next subsection. We will need to multiply our final amplitudes by this contribution.

We can write the $G^{\pm,t_i}_{\dot A,-{3\over2}}$'s in terms of bosonic and fermionic modes through the expansion
\bea\label{Gt}
\tilde G^{\a,t_i}_{\dot A,r}=-i\sum_{k}\tilde d^{\a A,t_i}_{r-k}\tilde{\a}^{t_i}_{A\dot A,k}
\eea
which, utilizing (\ref{GG after sf}) and (\ref{Gt}), gives us
\bea\label{factors}
G^{-}_{\dot A,-{1\over2}}\s^+_2(w_1) &\to&-it_1\tilde{d}^{-A,t_1}_{-{1\over2}}\tilde{\a}^{t_1}_{A\dot A,-1}[C'_1z_1]\cr
G^{+}_{\dot A,-{1\over2}}\s^-_2(w_2) &\to&-i(t_2+a)(t_2+b)\tilde{d}^{+A,t_2}_{-{1\over2}}\tilde{\a}^{t_2}_{A\dot A,-1} [C'_2]
\eea
where we only keep the above modes due to the constraint (\ref{vacuum constraint}) being applied to the expansion (\ref{Gt}). We note that the bosonic and fermionic modes natural to the $t$-plane 
are defined around their respective $t$-plane locations $t_i$ as the following
\begin{align}\label{bos ferm ti}
\tilde{\a}^{t_i}_{A\dot{A},n}&={1\over 2\pi}\oint_{t_i} dt (t-t_i)^n\partial X_{A\dot{A}}(t)
\nn
\tilde d^{\a A,t_i}_{s} &= {1\over2\pi i}\oint_{t_i}dt (t-t_i)^{s-{1\over2}} \psi^{\a A}(t)
\end{align}

\subsection{Normalization coming from the twist operators}
Let us take a moment to systematically compute the normalization factor coming from just the twist operators themselves, part of which we collected in brackets throughout the sequence of Jacobian and spectral flow transformations, which will enter as a multiplicative factor to the $t$-plane amplitudes which we derive later. Starting on the cylinder, in the original configuration before applying any maps or spectral flows, we have the following left moving vacuum correlation function (the right moving contribution will be analogous)
\bea
U={}^{(1)}\langle 0_{R,+}|{}^{(2)}\langle 0_{R,+}|\s^-_2(w_2)\s^+_2(w_1)|0^+_R\rangle^{(1)}|0^+_R\rangle^{(2)}
\eea
Following the same chain of spectral flows and coordinate transformations as before (up until reaching the $z$-plane) let us first spectral flow by $\a=-1$ taking us to the NS sector. This gives
\bea
U= z_{1}^{{1\over2}}z_2^{-{1\over2}}{}^{(1)}\langle 0_{NS}|{}^{(2)}\langle 0_{NS}|\s^-_2(w_2)\s^+_2(w_1)|0_{NS}\rangle^{(1)}|0_{NS}\rangle^{(2)}\nn
\eea
Now let us map this to the $z$-plane using $z=e^w$. This gives
\bea
U&=&z_1{}^{(1)}\langle 0_{NS}|{}^{(2)}\langle 0_{NS}|\s^-_2(z_2)\s^+_2(z_1)|0_{NS}\rangle^{(1)}|0_{NS}\rangle^{(2)}
\eea
Since $\sigma^{\pm}_{2}(z_i)$ is a chiral primary operator of dimension and charge 
\bea 
h = j = {1\over2},\quad m=\pm {1\over2}
\eea 
the amplitude $U(z_1,z_2)$ is given by
\bea \label{U}
U(z_1,z_2) = {z_1\over z_2 - z_1}
\eea
and similarly for the right movers the expression is given by 
\bea \label{Ubar}
\bar U(\bar z_1,\bar z_2) = {\bar z_1\over \bar z_2 - \bar z_1}
\eea
\subsection{$t$-plane amplitudes}
In this subsection we demonstrate how to compute the $t$-plane amplitudes resulting from the various Jacobian and spectral flow transformations performed in the previous subsections. Let's first combine the various components we have obtained so far. Applying the deformation transformations (\ref{factors}), the twist amplitudes (\ref{U}), (\ref{Ubar}) and the mode transformations (\ref{c1 init pr}), (\ref{c2 init pr}), (\ref{c1 fin pr}), (\ref{c2 fin pr}) to the amplitude recorded in (\ref{subsub ms plus gr}) whose left and right moving components are given in (\ref{subsub ms plus gr lr}),
we collect together the following results
\subsubsection*{$|\Psi_0\rangle\to|\Psi_f\rangle$}
Summing over the copy contributions gives
\begin{align}
\mathcal{A}^{0\to f} &\equiv\mathcal{A}^{0\to f}_{1221} + \mathcal{A}^{0\to f}_{1212} + \mathcal{A}^{0\to f}_{2121} + \mathcal{A}^{0\to f}_{2112}
\end{align}
where
\begin{equation}\label{unint ms plus gr}
\mathcal{A}^{0\to f}_{ijkl} = {1\over 4mr}\e_{AB}\e^{\dot A_2\dot B_2}\e^{\dot A_1\dot B_1}\e_{A'B'}\big[\mathcal{A}^{0\to  f}_{ijkl}\big]^{AA'}_{\dot A_2\dot A_1}\big[\bar{\mathcal{A}}^{0\to  f}_{ijkl}\big]^{BB'}_{\dot B_2\dot B_1}
\end{equation}
\begin{align}\label{ms plus gr tp}
\big[\mathcal{A}^{0\to f}_{ijkl}\big]^{AA'}_{\dot A_2\dot A_1}
&\equiv-{z_1z_2\over z_2-z_1}t_1t_2\big[\tilde{\mathcal{A}}^{0\to f}_{ijkl}\big]^{AA'}_{\dot A_2\dot A_1}
\nn 
\big[\bar{\mathcal{A}}^{0\to f}_{ijkl}\big]^{BB'}_{\dot B_2\dot B_1}&\equiv -{\bar z_1\bar z_2\over \bar z_2-\bar z_1}\bar t_1\bar t_2\big[\tilde{\bar{\mathcal{A}}}^{0\to f}_{ijkl}\big]^{BB'}_{\dot B_2\dot B_1}
\end{align}
where the $t$-plane amplitudes are defined as
\begin{align}\label{t plane ms plus gr}
\big[\tilde{\mathcal{A}}^{0\to f}_{ijkl}\big]^{AA'}_{\dot A_2\dot A_1}&\equiv {}_t\langle0_{NS}|  d'^{+A(i)}_{p+q}\alpha'^{(j)}_{--,r}\tilde{d}^{+A_2,t_2}_{-{1\over2}}\tilde{\a}^{t_2}_{A_2\dot A_2,-1}\tilde{d}^{-A_1,t_1}_{-{1\over2}}\tilde{\a}^{t_1}_{A_1\dot A_1,-1}\a'^{(k)}_{++,-m}d'^{-A'(l)}_{-n} |0_{NS}\rangle_t
\nn
\big[\tilde{\bar{\mathcal{A}}}^{0\to f}_{ijkl}\big]^{BB'}_{\dot B_2\dot B_1}&\equiv
{}_{\bar{t}}\langle\bar 0_{NS}|\bar d'^{+B(i)}_p\bar{\alpha}'^{(j)}_{++,r}\tilde{\bar{d}}^{+B_2,\bar t_2}_{-{1\over2}}\tilde{\bar{\a}}^{\bar t_2}_{B_2\dot B_2,-1}\tilde{\bar{d}}^{-B_1,\bar t_1}_{-{1\over2}}\tilde{\bar{\a}}^{\bar t_1}_{B_1\dot B_1,-1}\bar\a'^{(k)}_{--,-m}\bar d'^{-B'(l)}_0|\bar0_{NS}\rangle_{\bar{t}}
\end{align}
Note for the above amplitudes that all of the spin fields have been removed leaving only the fermionic and bosonic modes. Now we can perform standard Wick contractions which we do next.

\subsection{Wick contractions}
Since we now have amplitudes comprised of only bosonic and fermionic modes, (\ref{t plane ms plus gr}), 
we can compute them by simply performing Wick contractions of the various terms. We note that most of the charge combinations have vanishing contributions and thus only write those which are nonzero.

We will denote Wick contractions by the shorthand notation $\langle\cdot\,,\cdot\rangle$. Summarizing the nonzero contributions for (\ref{t plane ms plus gr}), which record again below for clarity this time including the additional step of expanding contracted indices, we have $A_1,A_2,B_1,B_2$
\begin{align}
\big[\tilde{\mathcal{A}}^{0\to f}_{ijkl}\big]^{AA'}_{\dot A_2\dot A_1}&\equiv {}_t\langle0_{NS}|  d'^{+A(i)}_{p+q}\alpha'^{(j)}_{--,r}\tilde{d}^{+A_2,t_2}_{-{1\over2}}\tilde{\a}^{t_2}_{A_2\dot A_2,-1}\tilde{d}^{-A_1,t_1}_{-{1\over2}}\tilde{\a}^{t_1}_{A_1\dot A_1,-1}\a'^{(k)}_{++,-m}d'^{-A'(l)}_{-n} |0_{NS}\rangle_t
\nn
&= 
{}_t\langle0_{NS}|  d'^{+A(i)}_{p+q}\alpha'^{(j)}_{--,r}\tilde{d}^{++,t_2}_{-{1\over2}}\tilde{\a}^{t_2}_{+\dot A_2,-1}\tilde{d}^{--,t_1}_{-{1\over2}}\tilde{\a}^{t_1}_{-\dot A_1,-1}\a'^{(k)}_{++,-m}d'^{-A'(l)}_{-n} |0_{NS}\rangle_t
\nn
&+{}_t\langle0_{NS}|  d'^{+A(i)}_{p+q}\alpha'^{(j)}_{--,r}\tilde{d}^{+-,t_2}_{-{1\over2}}\tilde{\a}^{t_2}_{-\dot A_2,-1}\tilde{d}^{-+,t_1}_{-{1\over2}}\tilde{\a}^{t_1}_{+\dot A_1,-1}\a'^{(k)}_{++,-m}d'^{-A'(l)}_{-n} |0_{NS}\rangle_t 
\nn
\nn
\big[\tilde{\bar{\mathcal{A}}}^{0\to f}_{ijkl}\big]^{BB'}_{\dot B_2\dot B_1}&\equiv
{}_{\bar{t}}\langle\bar 0_{NS}|\bar d'^{+B(i)}_p\bar{\alpha}'^{(j)}_{++,r}\tilde{\bar{d}}^{+B_2,\bar t_2}_{-{1\over2}}\tilde{\bar{\a}}^{\bar t_2}_{B_2\dot B_2,-1}\tilde{\bar{d}}^{-B_1,\bar t_1}_{-{1\over2}}\tilde{\bar{\a}}^{\bar t_1}_{B_1\dot B_1,-1}\bar\a'^{(k)}_{--,-m}\bar d'^{-B'(l)}_0|\bar0_{NS}\rangle_{\bar{t}}
\nn
&=
{}_{\bar{t}}\langle\bar 0_{NS}|\bar d'^{+B(i)}_p\bar{\alpha}'^{(j)}_{++,r}\tilde{\bar{d}}^{++,\bar t_2}_{-{1\over2}}\tilde{\bar{\a}}^{\bar t_2}_{+\dot B_2,-1}\tilde{\bar{d}}^{--,\bar t_1}_{-{1\over2}}\tilde{\bar{\a}}^{\bar t_1}_{-\dot B_1,-1}\bar\a'^{(k)}_{--,-m}\bar d'^{-B'(l)}_0|\bar0_{NS}\rangle_{\bar{t}} 
\nn
&+{}_{\bar{t}}\langle\bar 0_{NS}|\bar d'^{+B(i)}_p\bar{\alpha}'^{(j)}_{++,r}\tilde{\bar{d}}^{+-,\bar t_2}_{-{1\over2}}\tilde{\bar{\a}}^{\bar t_2}_{-\dot B_2,-1}\tilde{\bar{d}}^{-+,\bar t_1}_{-{1\over2}}\tilde{\bar{\a}}^{\bar t_1}_{+\dot B_1,-1}\bar\a'^{(k)}_{--,-m}\bar d'^{-B'(l)}_0|\bar0_{NS}\rangle_{\bar{t}}
\end{align}
we obtain the following contributions
\begin{align}\label{ms plus gr t left}
\big[\tilde{\mathcal{A}}^{0\to f}_{ijkl}\big]^{+-}_{+-}
=&\big(\langle d'^{++(i)}_{p+q}d'^{--(l)}_{-n}\rangle\langle\tilde{d}^{++,t_2}_{-{1\over2}}\tilde{d}^{--,t_1}_{-{1\over2}}\rangle-\langle d'^{++(i)}_{p+q}\tilde{d}^{--,t_1}_{-{1\over2}}\rangle\langle\tilde{d}^{++,t_2}_{-{1\over2}}d'^{--(l)}_{-n}\rangle\big)
\nn
&\big(\langle\alpha'^{(j)}_{--,r}\tilde{\a}^{t_2}_{++,-1}\rangle\langle\tilde{\a}^{t_1}_{--,-1}\a'^{(k)}_{++,-m}\rangle
+\langle\alpha'^{(j)}_{--,r}\a'^{(k)}_{++,-m}\rangle\langle\tilde{\a}^{t_2}_{++,-1}\tilde{\a}^{t_1}_{--,-1}\rangle\big)
\nn 
&+\langle d'^{++(i)}_{p+q}d'^{--(l)}_{-n}\rangle\langle\tilde{d}^{+-,t_2}_{-{1\over2}}\tilde{d}^{-+,t_1}_{-{1\over2}}\rangle\langle\alpha'^{(j)}_{--,r}\a'^{(k)}_{++,-m}\rangle\langle\tilde{\a}^{t_2}_{-+,-1}\tilde{\a}^{t_1}_{+-,-1}\rangle
\nn 
\nn
\big[\tilde{\mathcal{A}}^{0\to f}_{ijkl}\big]^{-+}_{+-}
=&\langle d'^{+-(i)}_{p+q}d'^{-+(l)}_{-n}\rangle\langle\tilde{d}^{++,t_2}_{-{1\over2}}\tilde{d}^{--,t_1}_{-{1\over2}}\rangle
\nn
&\big(\langle\alpha'^{(j)}_{--,r}\tilde{\a}^{t_2}_{++,-1}\rangle\langle\tilde{\a}^{t_1}_{--,-1}\a'^{(k)}_{++,-m}\rangle
+\langle\alpha'^{(j)}_{--,r}\a'^{(k)}_{++,-m}\rangle\langle\tilde{\a}^{t_2}_{++,-1}\tilde{\a}^{t_1}_{--,-1}\rangle\big)
\nn 
&+\big(\langle d'^{+-(i)}_{p+q}d'^{-+(l)}_{-n}\rangle\langle\tilde{d}^{+-,t_2}_{-{1\over2}}\tilde{d}^{-+,t_1}_{-{1\over2}}\rangle-\langle d'^{+-(i)}_{p+q}\tilde{d}^{-+,t_1}_{-{1\over2}}\rangle\langle\tilde{d}^{+-,t_2}_{-{1\over2}}d'^{-+(l)}_{-n}\rangle\big)
\nn
&\langle\alpha'^{(j)}_{--,r}\a'^{(k)}_{++,-m}\rangle\langle\tilde{\a}^{t_2}_{-+,-1}\tilde{\a}^{t_1}_{+-,-1}\rangle
\nn 
\nn
\big[\tilde{\mathcal{A}}^{0\to f}_{ijkl}\big]^{+-}_{-+}
=&\big(\langle d'^{++(i)}_{p+q}d'^{--(l)}_{-n}\rangle\langle\tilde{d}^{++,t_2}_{-{1\over2}}\tilde{d}^{--,t_1}_{-{1\over2}}\rangle-\langle d'^{++(i)}_{p+q}\tilde{d}^{--,t_1}_{-{1\over2}}\rangle\langle\tilde{d}^{++,t_2}_{-{1\over2}}d'^{--(l)}_{-n}\rangle\big)
\nn
&\langle\alpha'^{(j)}_{--,r}\a'^{(k)}_{++,-m}\rangle\langle\tilde{\a}^{t_2}_{+-,-1}\tilde{\a}^{t_1}_{-+,-1}\rangle
\nn 
&+\langle d'^{++(i)}_{p+q}d'^{--(l)}_{-n}\rangle\langle\tilde{d}^{+-,t_2}_{-{1\over2}}\tilde{d}^{-+,t_1}_{-{1\over2}}\rangle
\nn
&\big(\langle\alpha'^{(j)}_{--,r}\tilde{\a}^{t_1}_{++,-1}\rangle\langle\tilde{\a}^{t_2}_{--,-1}\a'^{(k)}_{++,-m}\rangle
+\langle\alpha'^{(j)}_{--,r}\a'^{(k)}_{++,-m}\rangle\langle\tilde{\a}^{t_2}_{--,-1}\tilde{\a}^{t_1}_{++,-1}\rangle\big)
\nn 
\nn
\big[\tilde{\mathcal{A}}^{0\to f}_{ijkl}\big]^{-+}_{-+}
=&\langle d'^{+-(i)}_{p+q}d'^{-+(l)}_{-n}\rangle\langle\tilde{d}^{++,t_2}_{-{1\over2}}\tilde{d}^{--,t_1}_{-{1\over2}}\rangle
\langle\alpha'^{(j)}_{--,r}\a'^{(k)}_{++,-m}\rangle\langle\tilde{\a}^{t_2}_{+-,-1}\tilde{\a}^{t_1}_{-+,-1}\rangle
\nn 
&+\big(\langle d'^{+-(i)}_{p+q}d'^{-+(l)}_{-n}\rangle\langle\tilde{d}^{+-,t_2}_{-{1\over2}}\tilde{d}^{-+,t_1}_{-{1\over2}}\rangle-\langle d'^{+-(i)}_{p+q}\tilde{d}^{-+,t_1}_{-{1\over2}}\rangle\langle\tilde{d}^{+-,t_2}_{-{1\over2}}d'^{-+(l)}_{-n}\rangle\big)
\nn
&\big(\langle\alpha'^{(j)}_{--,r}\tilde{\a}^{t_1}_{++,-1}\rangle\langle\tilde{\a}^{t_2}_{--,-1}\a'^{(k)}_{++,-m}\rangle
+\langle\alpha'^{(j)}_{--,r}\a'^{(k)}_{++,-m}\rangle\langle\tilde{\a}^{t_2}_{--,-1}\tilde{\a}^{t_1}_{++,-1}\rangle\big)
\nn
\nn
\big[\tilde{\bar{\mathcal{A}}}^{0\to f}_{ijkl}\big]^{+-}_{+-}
=&\big(\langle\bar d'^{++(i)}_p\bar d'^{--(l)}_0\rangle\langle\tilde{\bar{d}}^{++,\bar t_2}_{-{1\over2}}\tilde{\bar{d}}^{--,\bar t_1}_{-{1\over2}}\rangle - \langle\bar d'^{++(i)}_p\tilde{\bar{d}}^{--,\bar t_1}_{-{1\over2}}\rangle\langle\tilde{\bar{d}}^{++,\bar t_2}_{-{1\over2}}\bar d'^{--(l)}_0\rangle \big)
\nn
&\big(
\langle\bar{\alpha}'^{(j)}_{++,r}\tilde{\bar{\a}}^{\bar t_1}_{--,-1}\rangle\langle\tilde{\bar{\a}}^{\bar t_2}_{++,-1}\bar\a'^{(k)}_{--,-m}\rangle+\langle\bar{\alpha}'^{(j)}_{++,r}\bar\a'^{(k)}_{--,-m}\rangle\langle\tilde{\bar{\a}}^{\bar t_2}_{++,-1}\tilde{\bar{\a}}^{\bar t_1}_{--,-1}\rangle\big)
\nn
+&\langle\bar d'^{++(i)}_p\bar d'^{--(l)}_0\rangle\langle\tilde{\bar{d}}^{+-,\bar t_2}_{-{1\over2}}\tilde{\bar{d}}^{-+,\bar t_1}_{-{1\over2}}\rangle 
\langle\bar{\alpha}'^{(j)}_{++,r}\bar\a'^{(k)}_{--,-m}\rangle\langle\tilde{\bar{\a}}^{\bar t_2}_{-+,-1}\tilde{\bar{\a}}^{\bar t_1}_{+-,-1}\rangle
\nn
\nn
\big[\tilde{\bar{\mathcal{A}}}^{0\to f}_{ijkl}\big]^{+-}_{-+}
=&\big(\langle\bar d'^{++(i)}_p\bar d'^{--(l)}_0\rangle\langle\tilde{\bar{d}}^{++,\bar t_2}_{-{1\over2}}\tilde{\bar{d}}^{--,\bar t_1}_{-{1\over2}}\rangle - \langle\bar d'^{++(i)}_p\tilde{\bar{d}}^{--,\bar t_1}_{-{1\over2}}\rangle\langle\tilde{\bar{d}}^{++,\bar t_2}_{-{1\over2}}\bar d'^{--(l)}_0\rangle \big)
\nn
&
\langle\bar{\alpha}'^{(j)}_{++,r}\bar\a'^{(k)}_{--,-m}\rangle\langle\tilde{\bar{\a}}^{\bar t_2}_{+-,-1}\tilde{\bar{\a}}^{\bar t_1}_{-+,-1}\rangle\nn
+&\langle\bar d'^{++(i)}_p\bar d'^{--(l)}_0\rangle\langle\tilde{\bar{d}}^{+-,\bar t_2}_{-{1\over2}}\tilde{\bar{d}}^{-+,\bar t_1}_{-{1\over2}}\rangle
\nn
&\big(\langle\bar{\alpha}'^{(j)}_{++,r}\tilde{\bar{\a}}^{\bar t_2}_{--,-1}\rangle\langle\tilde{\bar{\a}}^{\bar t_1}_{++,-1}\bar\a'^{(k)}_{--,-m}\rangle
+\langle\bar{\alpha}'^{(j)}_{++,r}\bar\a'^{(k)}_{--,-m}\rangle\langle\tilde{\bar{\a}}^{\bar t_2}_{--,-1}\tilde{\bar{\a}}^{\bar t_1}_{++,-1}\rangle\big)
\nn
\nn
\big[\tilde{\bar{\mathcal{A}}}^{0\to f}_{ijkl}\big]^{-+}_{+-}
=&\langle\bar d'^{+-(i)}_p\bar d'^{-+(l)}_0\rangle\langle\tilde{\bar{d}}^{++,\bar t_2}_{-{1\over2}}\tilde{\bar{d}}^{--,\bar t_1}_{-{1\over2}}\rangle
\nn
&\big(
\langle\bar{\alpha}'^{(j)}_{++,r}\tilde{\bar{\a}}^{\bar t_1}_{--,-1}\rangle\langle\tilde{\bar{\a}}^{\bar t_2}_{++,-1}\bar\a'^{(k)}_{--,-m}\rangle
+\langle\bar{\alpha}'^{(j)}_{++,r}\bar\a'^{(k)}_{--,-m}\rangle\langle\tilde{\bar{\a}}^{\bar t_2}_{++,-1}\tilde{\bar{\a}}^{\bar t_1}_{--,-1}\rangle\big)
\nn
+&\big(\langle\bar d'^{+-(i)}_p\bar d'^{-+(l)}_0\rangle\langle\tilde{\bar{d}}^{+-,\bar t_2}_{-{1\over2}}\tilde{\bar{d}}^{-+,\bar t_1}_{-{1\over2}}\rangle - \langle\bar d'^{+-(i)}_p\tilde{\bar{d}}^{-+,\bar t_1}_{-{1\over2}}\rangle\langle\tilde{\bar{d}}^{+-,\bar t_2}_{-{1\over2}}\bar d'^{-+(l)}_0\rangle \big)
\nn
&
\langle\bar{\alpha}'^{(j)}_{++,r}\bar\a'^{(k)}_{--,-m}\rangle\langle\tilde{\bar{\a}}^{\bar t_2}_{-+,-1}\tilde{\bar{\a}}^{\bar t_1}_{+-,-1}\rangle
\nn
\nn
\big[\tilde{\bar{\mathcal{A}}}^{0\to f}_{ijkl}\big]^{-+}_{-+}
=&\langle\bar d'^{+-(i)}_p\bar d'^{-+(l)}_0\rangle\langle\tilde{\bar{d}}^{++,\bar t_2}_{-{1\over2}}\tilde{\bar{d}}^{--,\bar t_1}_{-{1\over2}}\rangle
\langle\bar{\alpha}'^{(j)}_{++,r}\bar\a'^{(k)}_{--,-m}\rangle\langle\tilde{\bar{\a}}^{\bar t_2}_{+-,-1}\tilde{\bar{\a}}^{\bar t_1}_{-+,-1}\rangle
\nn
+&\big(\langle\bar d'^{+-(i)}_p\bar d'^{-+(l)}_0\rangle\langle\tilde{\bar{d}}^{+-,\bar t_2}_{-{1\over2}}\tilde{\bar{d}}^{-+,\bar t_1}_{-{1\over2}}\rangle - \langle\bar d'^{+-(i)}_p\tilde{\bar{d}}^{-+,\bar t_1}_{-{1\over2}}\rangle\langle\tilde{\bar{d}}^{+-,\bar t_2}_{-{1\over2}}\bar d'^{-+(l)}_0\rangle \big)
\nn
&\big(\langle\bar{\alpha}'^{(j)}_{++,r}\tilde{\bar{\a}}^{\bar t_2}_{--,-1}\rangle\langle\tilde{\bar{\a}}^{\bar t_1}_{++,-1}\bar\a'^{(k)}_{--,-m}\rangle 
+\langle\bar{\alpha}'^{(j)}_{++,r}\bar\a'^{(k)}_{--,-m}\rangle\langle\tilde{\bar{\a}}^{\bar t_2}_{--,-1}\tilde{\bar{\a}}^{\bar t_1}_{++,-1}\rangle\big)
\end{align}
To compute the Wick contraction terms, which enter as building blocks of the above amplitudes, explicitly using the transformed bosonic and fermionic modes (\ref{c1 init pr}), (\ref{c2 init pr}), (\ref{c1 fin pr}) and (\ref{c2 fin pr}) and supercharge modes (\ref{factors}), one must expand these into a linear combination of modes which are defined naturally around the appropriate $t$-plane image of the cover. Since this step is straightforward but lengthy we relegate the details to appendix \ref{wick contractions} where we display two examples explicitly and record the final expressions for the remaining ones.
\subsection{Unintegrated amplitude}
Now that we have written down the nonzero Wick contracted $t$-plane amplitude contributions we can write out the form of the full unintegrated amplitude, (\ref{unint ms plus gr}), by expanding the charge combinations using the Levi-Civita tensor, where $\e_{+-}=1,\e^{+-}=-1$, which gives
\bea 
\mathcal{A}^{0\to f} &=&\mathcal{A}^{0\to f}_{1221} + \mathcal{A}^{0\to f}_{1212} + \mathcal{A}^{0\to f}_{2121} + \mathcal{A}^{0\to f}_{2112}
\eea
with
\begin{align}\label{ms plus gr unint}
\mathcal{A}^{0\to f}_{ijkl}&={1\over4mr}\bigg|{z_1z_2\over z_2-z_1}t_1t_2\bigg|^2\nn
&\bigg(\big[\tilde{\mathcal{A}}^{0\to f}_{ijkl}\big]^{-+}_{ + -}\big[\tilde{\bar{\mathcal{A}}}^{0\to f}_{ijkl}\big]^{+-}_{-+}
+\big[\tilde{\mathcal{A}}^{0\to f}_{ijkl}\big]^{-+}_{-+}\big[\tilde{\bar{\mathcal{A}}}^{0\to f}_{ijkl}\big]^{+-}_{+-}
\nn
&+\big[\tilde{\mathcal{A}}^{0\to f}_{ijkl}\big]^{+-}_{+-}\big[\tilde{\bar{\mathcal{A}}}^{0\to f}_{ijkl}\big]^{-+}_{-+}
+\big[\tilde{\mathcal{A}}^{0\to f}_{ijkl}\big]^{+-}_{-+}\big[\tilde{\bar{\mathcal{A}}}^{0\to f}_{ijkl}\big]^{-+}_{+-}\bigg)
\end{align}
with the tilded amplitudes written explicitly in (\ref{ms plus gr t left}) for left-moving
and right moving modes.

\section{Amplitude integration}\label{integration}
In this section we show how to integrate the unintegrated amplitudes, computed in the previous section, over the twist locations on the cylinder. To better understand what our results will look like let us record some useful information about the twist locations. 
\subsection{Coordinate relations}
The relation between twist locations on the $w$-cylinder and the $z$-plane were given in (\ref{wz}) as
\begin{align}\label{zw pr}
z_1 = e^{w_1}=e^{s-{\Delta w\over2}},\quad z_2 = e^{w_2}=e^{s+{\Delta w\over2}}
\end{align}
where $s$ and $\Delta w$ are convenient center of mass and difference coordinates written in terms of the individual twist locations $w_i$ recorded in (\ref{s dw}).
The relation between twist locations on the $z$-plane and their images on the $t$-plane, where the spin fields were mapped to before spectral flowing them away and where the supercharge operators were mapped to, were given in (\ref{zt})
\begin{align}\label{zt pr}
    z_1 &= {(t_1+a)(t_1+b)\over t_1} = a + b +2t_1,\quad t_1=-\sqrt{ab}\nn
    z_2 &= {(t_2+a)(t_2+b)\over t_2} = a + b +2t_2,\quad t_2=\sqrt{ab}
\end{align}
The Wick contraction terms, recorded in appendix \ref{wick contractions}, all contain the $t$-plane parameters, $a,b,t_1,t_2$. Therefore, in order to obtain explicit results which correspond to the physical process we are interested in, which is defined on the cylinder, we need to know how the $t$-plane coordinates are written in terms of the cylinder coordinates. Thus, combining (\ref{zw pr}) and (\ref{zt pr}) 
we find the convenient relations 
\begin{align}\label{a b t1 t2}
    a&=e^s\cosh^2\bigg({\Delta w\over4}\bigg),& b&=e^s\sinh^2\bigg({\Delta w\over4}\bigg)
    \nn
    t_1&=-e^s\cosh\bigg({\Delta w\over4}\bigg)\sinh\bigg({\Delta w\over4}\bigg),&t_2&=e^s\cosh\bigg({\Delta w\over4}\bigg)\sinh\bigg({\Delta w\over4}\bigg)
\end{align}
Furthermore, the front factor in the amplitudes (\ref{ms plus gr unint}), using (\ref{zw pr}) and (\ref{zt pr}), can be written in terms of $s,\Delta w$ as 
\begin{equation}
 \bigg|{z_1z_2\over z_2-z_1}t_1t_2\bigg|^2 = \bigg({z_1z_2t_1t_2\over z_1-z_2}\bigg)\bigg({\bar z_1\bar z_2\bar t_1\bar t_2\over \bar z_1-\bar z_2}\bigg) = \bigg({1\over8}e^{3s}\sinh\bigg({\Delta w\over2}\bigg)\bigg)\bigg({1\over8}e^{3\bar s}\sinh\bigg({\Delta \bar w\over2}\bigg)\bigg)
\end{equation}
The Wick contraction terms don't have closed form solutions in terms of elementary functions but are instead composed of finite sums of terms which include the $t$-plane parameters $a,b,t_1,t_2$. However, since these parameters can be written in terms of $s,\Delta w$, as seen in (\ref{a b t1 t2}) the amplitude (\ref{ms plus gr unint}) for a given choice of initial and final mode numbers can be expanded as a sum of terms polynomial in $e^{\Delta w},e^{\Delta \bar w}$ multiplying numerical coefficients with the center of mass dependence $e^s,e^{\bar s}$ eventually dropping out of the expressions when energy and momentum conservation are imposed. Therefore, we write below a general form that our solutions will take in terms of a double sum, one over powers of $e^{\Delta w}$ and the other over powers of $e^{\Delta \bar w}$:
\bea\label{amp sum}
\mathcal{A}^{0 \to f}(w_1,w_2,\bar w_1,\bar w_2) &=&\sum_{k=-2(m+n) }^{2(m+n)}\sum_{\bar k =-2m}^{2m}B^{0\to f}_{k,\bar k}( m,n; p,q,r)e^{{k\Delta w\over2} + {\bar k\Delta \bar w\over2}}
\eea
The mode numbers characterizing the initial state are $m,n$ and the modes characterizing the final state are $p,q,r$. We see that the minimum and maximum values of the left-moving and right-moving sums, are governed by left and right moving energies respectively (where we've imposed energy conservation).
We also note a symmetry we observe for the coefficients over the sum coming from explicit computations
\be\label{B symmetry}
B^{0 \to f}_{k,k}(m,n; p,q,r) = B^{0\to f}_{- k,- k}(m,n; p,q,r)
\ee
Next, using the above form, we can easily integrate the amplitudes to obtain the final results of the transition process.

\subsection{Integrating the amplitude}
Let's show how to integrate the amplitudes to obtain the final result. Imposing energy and momentum conservation from the outset
\begin{align}
    h_f +\bar h_f = h_0 + \bar h_0,\quad h_f-\bar h_f = h_0-\bar h_0
\end{align}
which for the amplitudes (\ref{Asec5}), (\ref{subsub ms plus gr}), (\ref{subsub ms plus gr lr}) become
\begin{align}\label{em cons}
    2(r + p) +q &=2m + n ,\quad q=n\implies r+p=m
\end{align} 
the amplitudes we compute take the general form (\ref{amp sum}) and thus the integration over the twist insertions is simply over powers of exponential factors. Thus inserting the general form (\ref{amp sum}) into (\ref{gen int}) and restoring the twist insertion dependence we have
\bea\label{int amp one}
A^{0\to f}(\tau)&=&{1\over2}\lambda^2\int d^2w_2d^2w_1\mathcal{A}^{0 \to f}(w_1,w_2,\bar w_1,\bar w_2)\cr
&=&{1\over2}\lambda^2\int d^2w_2d^2w_1\sum_{k=-2(m+n) }^{2(m+n)}\sum_{\bar k =-2m}^{2m}B^{0\to f}_{k,\bar k}(m,n; p,q,r)e^{{k \Delta w\over2}+{\bar k\Delta \bar w\over2}}\cr
&=&\lambda^2\sum_{k=-2(m+n) }^{2(m+n)}\sum_{\bar k =-2m}^{2m}B^{0 \to f}_{k,\bar k}(m,n; p,q,r)I_{k,\bar k}
\eea
Where the integral $I_{k,\bar k}$ is defined as
\bea\label{I}
I_{k,\bar k} \equiv{1\over2} \int d^2w_2d^2w_1e^{{k \Delta w\over2}+{\bar k \Delta \bar w\over2}}
\eea
Recalling $\Delta w$ in (\ref{s dw}) and the cylinder coordinate (\ref{w}) we have
\bea
\Delta w &=& w_2 - w_1=\tau_2 - \tau_1 + i(\sigma_2 - \sigma_1)\cr
\Delta \bar w &=& \bar w_2 - \bar w_1 =\tau_2 - \tau_1 - i(\sigma_2 - \sigma_1)
\eea
To obtain physical results we must Wick rotate back to the physical time coordinate, $\tau_j\to it_j$. This gives 
\bea\label{dw mink}
\Delta w &=& w_2 - w_1=i(t_2 - t_1 + \sigma_2 - \sigma_1)\cr
\Delta \bar w &=& \bar w_2 - \bar w_1 =i(t_2 - t_1 - (\sigma_2 - \sigma_1))
\eea
The integration region is given by
\bea
-{ t\over2}\leq t_i \leq {t\over2},~~~~\quad 0\leq \sigma_i< 2\pi
\eea
where $t$ is the physical time which measures the duration of the deformation operators \footnote{Note that there is a factor of the radius of the physical $y$ circle, $R_y$, which is present in the theory. However, we consider our coordinates to be dimensionless quantities which is like taking $R_y=1$. It can be restored if necessary.}. Therefore (\ref{int amp one}) becomes
\bea
A^{0\to f}(t)&=&\lambda^2\sum_{k=-2(m+n) }^{2(m+n)}\sum_{\bar k =-2m}^{2m}B^{0\to f}_{k,\bar k}(m,n;p,q,r)I_{k,\bar k}(t)
\label{amplitude}
\eea
with the integration term a function of time $t$. Let us now evaluate the integral. Inserting (\ref{dw mink}) into (\ref{I}) and noting that we pick a particular ordering of the two twist operators, $t_1\leq t_2$, to demonstrate the integration explicitly, we have\footnote{The factor of ${1\over2}$ in the first line of (\ref{integral}) goes away in the second line, when defining the integration region, because the full integration includes an additional contribution which arises when the coordinate $t_1\geq t_2$. Therefore we simply multiply the contribution coming from the integration for $t_1\leq t_2$ by $2$.}
\bea
I_{k,\bar k}(t)&=&\frac{1}{2}\int d^2w_2\,d^2w_1\, e^{{k \Delta w\over2}+{\bar k \Delta \bar w\over2}}\cr
&=&\int_{-{t\over2}}^{{t\over2}}dt_2\int_{-{t\over2}}^{t_2}dt_1\int_{\s_2=0}^{2\pi} d\s_2\int_{\s_1=0}^{2\pi} d\s_1 e^{{i(k+\bar k)\over2}(t_2-t_1)}e^{{i(k-\bar k)\over2}(\s_2-\s_1)}
\label{integral}
\eea
We note that our amplitudes yield expansions in which $k,\bar k$ are always even, thus implying that $k-\bar k$ is always even further implying that ${k-\bar k\over 2}$ is an integer. Furthermore the integral over $\sigma_i$ enforces momentum conservation in the intermediate states.
There are two nonzero contributions coming from integrating over the twist locations. Those where $k=\bar k\neq0$ and those where $k=\bar k=0$.
The first contribution is when 
\bea
k=\bar k \neq 0
\eea
which yields the integral
\bea\label{kneqkbar}
I_{k=\bar k\neq 0}(t)
&=&\frac{1}{2}\int d^2w_2\,d^2w_1\, e^{{k \Delta w\over2}+{\bar k \Delta \bar w\over2}}\cr
&=&\int_{-{t\over2}}^{{t\over2}}dt_2\int_{-{t\over2}}^{t_2}dt_1\int_{\s=0}^{2\pi}d\s_2\int_{\s=0}^{2\pi}d\s_1\, e^{i k (t_2-t_1)}\cr
&=& {4i \pi^2 \over k ^2} \bigg(kt  -  2e^{i {kt\over2}}\sin\bigg(  {kt\over 2}\bigg)  \bigg)
\label{integral three}
\eea
The second contribution is when 
\bea
k=\bar k = 0
\eea
which gives
\bea
I_{k=\bar k= 0}(t) &=&\frac{1}{2}\int d^2w_2\,d^2w_1\, e^{{k \Delta w\over2}+{\bar k \Delta \bar w\over2}}\cr
&=&\int_{-{t\over2}}^{{t\over2}}dt_2\int_{-{t\over2}}^{t_2}dt_1\int_{\s=0}^{2\pi}d\s_2\int_{\s=0}^{2\pi}d\s_1\cr
&=&2\pi^2t^2
\eea
We note that since the left and right moving sectors differ in mode number, the sums over $k$ and $\bar k$ may have different ranges. Thus, in the case where $k=\bar k\neq 0$, the sum is defined over the variable with the smaller range which will typically be $\bar k$. 
We also recall the symmetry amongst coefficients
(\ref{B symmetry})
\bea\label{A0tof}
B^{0 \to f}_{ k, k}(m,n; p,q,r) = B^{0\to f}_{- k,- k}(m,n; p,q,r)
\eea
Thus for the contribution, $k=\bar k\neq 0 $ (\ref{kneqkbar}), the terms which are linear in $t$ will cancel in the sum. Thus we can write the final integrated amplitude in the form 
\bea\label{At}
A^{0 \to f}(t)&=&\lambda^2 2\pi^2t^2B^{0 \to f}_{0,0}(m,n; p,q,r)  + \lambda^2\sum_{  k = 1}^{ 2m }B^{0 \to f}_{ k, k}(m,n; p,q,r){16\pi^2\over k^2}\sin^2\bigg({ kt\over2}\bigg)\nn
&\equiv&A^{{0 \to f},scr}(t) + A^{ {0 \to f},osc}(t)
\label{int amplitude}
\eea 
where
\begin{align}\label{Ascr Aosc}
    A^{{0 \to f},scr}(t)&\equiv\lambda^2 2\pi^2t^2B^{0 \to f}_{0,0}(m,n; p,q,r)\nn
     A^{ {0 \to f},osc}(t)&\equiv \lambda^2\sum_{  k = 1}^{ 2m }B^{0 \to f}_{ k, k}(m,n; p,q,r){16\pi^2\over k^2}\sin^2\bigg({ kt\over2}\bigg)
\end{align}
We note that there are two contributions, one which oscillates in $t$, $A^{ {0 \to f},osc}(t)$, and one which grows as $t^2$, $A^{ {0 \to f},scr}(t)$, where $scr$ stands for scrambling and $osc$ stands of oscillating. The $A^{ {0 \to f},osc}(t)$ term represents a particle traveling from the boundary into the bulk into the superstratum geometry. A similar such term was analyzed when investigating the CFT dual of a graviton propagating in AdS where the oscillatory behavior represented the particle propagating from the boundary freely into the geometry and back to the boundary. See \cite{Guo:2021ybz} for more details.
However, the scrambling term, $A^{{0 \to f},scr}(t)$, which grows as $t^2$, represents the interaction of the particle with superstratum geometry. For large time scales the $t^2$ contribution will begin to dominate over the oscillatory contribution giving a growing transition amplitude and thus probability suggesting that the in-falling particle and the initial superstratum geometry are beginning to back-react into a new geometry. However, we must keep in mind that this computation is perturbative and will break down when the coupling becomes of order $\sim1$.
\subsection{Simple example}
For the following quantum numbers
\begin{align}
\lbrace m=2, n=1,p=1,q=1,r=1\rbrace
\end{align}
we record a simple example of (\ref{Ascr Aosc}) explicitly
\begin{align}
A^{0\to f}&=\lambda^2\pi^2\bigg( {27  t^2\over8192} + {83  \sin^2(t)^2\over16384} + {
 45 \sin^2(2 t)\over65536} \bigg)
\end{align}
where 
\begin{align}
 A^{{0 \to f},scr}(t) &= \lambda^2\pi^2{27  t^2\over8192}
 \nn
 A^{ {0 \to f},osc}(t)&=\lambda^2\pi^2\bigg({83  \sin^2(t)\over16384} + {
 45 \sin^2(2 t)\over65536}\bigg)
\end{align}

\section{Numerical analysis}\label{analysis}
In this section we numerically analyze the amplitude corresponding to the transition of a superstratum state and graviton in the initial state into a microstratum state and graviton in the final state, (\ref{At}). We would like to understand, at least from the CFT perspective, what the preferred transitions are.
Of course, there are many possible final states which span the Hilbert space, including those which correspond to more stringy excitations. However, for ease of computation we considered the final state utilized in this paper because it is one of the simplest to compute but yet able to offer some insight into transition dynamics serving as a step in understanding scrambling and thermalization in a more comprehensive way in the future.
\begin{figure}
\centering
        \includegraphics[width=14cm]{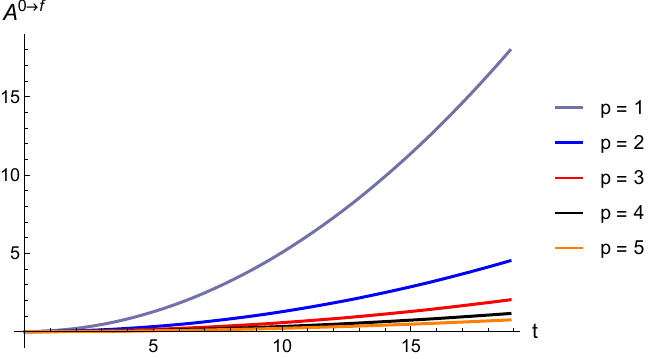}
\caption{$A^{0\to f}_{m,n}(t)$  vs. $t$  for $m=18,n=1$ and $p=1,2,3,4,5$ for $0\leq t\leq 6\pi$ with $\lambda^2=1$.}\label{Avstm18n1p}
\end{figure} 
\begin{figure}
\centering
        \includegraphics[width=14cm]{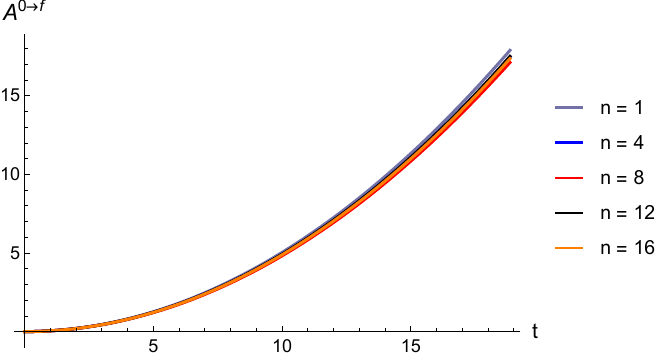}
\caption{$A^{0\to f}(t)$  vs. $t$  for $m=12,p=1$ and $n=1,4,8,12,16$ for $0\leq t\leq 6\pi$ with $\lambda^2=1$.}\label{Avstm12np1}
\end{figure} 
\begin{figure}
\centering
        \includegraphics[width=14cm]{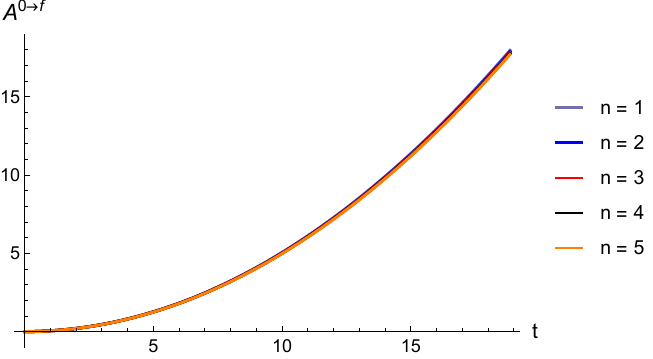}
\caption{$A^{0\to f}(t)$  vs. $t$  for $m=18,p=1$ and $n=1,2,3,4,5$ for $0\leq t\leq 6\pi$ with $\lambda^2=1$.}\label{Avstm18np1}
\end{figure} 
\begin{figure}
\centering
        \includegraphics[width=14cm]{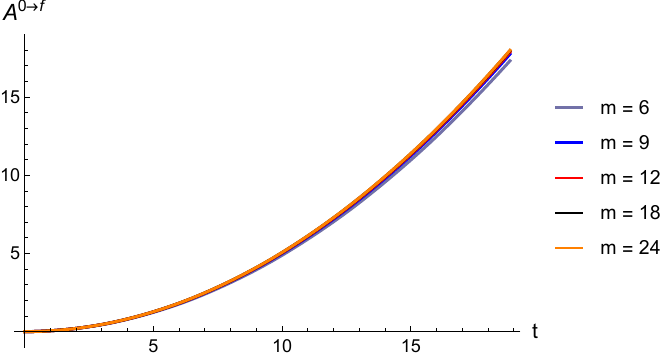}
\caption{$A^{0\to f}(t)$  vs. $t$  for $p=1,n=1$ and $m=6,9,12,18,24$ for $0\leq t\leq 6\pi$ with $\lambda^2=1$.}\label{Avstmn1p1}
\end{figure} 
In figure \ref{Avstm18n1p} we plot (\ref{At}) as a function of $t$ for an initial state where the graviton mode has quantum numbers characterized by $m=18$ and the superstratum state has momentum $n=1$ for different values of $p$, the quantum number of the microstratum state which characterizes how non-BPS it is. Due to energy and momentum conservation, (\ref{em cons}), the final graviton mode has quantum numbers $r=m-p$. We observe that the largest transition value occurs when $p=1$, the lowest value it can take for this microstratum state. The initial graviton mode would rather deposit a small amount of its energy, in a given two-deformation interaction, to the superstratum state converting it into a microstratum state that is just slightly non-BPS. As the amount of energy transferred from the initial graviton mode to the original superstratum state increases, the transition amplitude decreases. The minimum amplitude for this scenario occurs when $p=m-1$ in which the final graviton mode has quantum numbers characterized by $m=1$ the lowest value it can have (in our scenario there are no bosonic zero modes). 
The amplitude $A^{0\to f}(t)$ appears to become independent of $n$ as $n$ increases which is shown in figure \ref{Avstm12np1} for $m=12$ and $p=1$ and in figure \ref{Avstm18np1} for $m=18$ and $p=1$. As $m$ increases it seems that changes in $n$ cause the amplitude to vary less and less as it approaches a profile which goes to a constant w.r.t to $n$. In figure \ref{Avstmn1p1} $A^{0\to f}(t)$ appears to become independent of $m$ for $n=1$ and $p=1$ as $m$ increases. Combining these behaviors seem to suggest that for sufficiently large $m,n$ for $p=1$, the amplitude approaches a limiting profile becoming independent of $m,n$. We can understand this behavior in the following way. 

By imposing energy and momentum conservation for initial and final states, (\ref{em cons}), which requires $q=n$ and $r+p=m$, we have three independent quantum numbers $m,n,p$ with $m-1\geq p$ since there are no bosonic zero mode contributions in our scenario. Furthermore to ensure that the final state includes a microstratum state we must have $p\geq1$. Most importantly we found that for $p=1$ for a given $m,n$, the amplitude achieved its highest value, with increasing values of $p$ yielding a smaller and smaller amplitude, with $p=m-1$ giving the smallest amplitude. The amplitude is sensitive only to relative changes between initial and final states characterized by the value $p$. The main conclusion is that the amplitude, $A^{0\to f}(t)$ has the fastest growth for $n=p=1$ for a given $m$ which seems to increase to a profile independent of $m$ as $m$ increases.
\subsection*{Superstratum limit}
The limit where the microstratum state becomes a superstratum state occurs by taking $p=0$ in the final state (\ref{ms plus gr ket}), (\ref{ms+gr}) and thus in the amplitudes (\ref{Asec5}), (\ref{subsub ms plus gr}) (\ref{subsub ms plus gr lr}). This corresponds to the transition
\begin{equation}
    |\Psi_0\rangle\to|\Psi_f\rangle_{p=0}
\end{equation}
Subsequent application of energy and momentum conservation (\ref{em cons}) then requires that the initial and final states be identical. This yields an unintegrated amplitude which is actually an expectation value and not a transition, which we record below
\begin{align}\label{amp exp}
\mathcal{A}^{\text{exp}}(w_1,w_2,\bar w_1,\bar w_2) &\equiv (\mathcal{A}^{0\to f}(w_1,w_2,\bar w_1,\bar w_2))_{p=0,q=n,r=m}\nn
&=(\langle\Psi_f|DD|\Psi_0\rangle)_{p=0,q=n,r=m}\nn
&= \langle\Psi_0|DD|\Psi_0\rangle\nn
&\equiv\mathcal{A}^{\text{exp}}_{1221} + \mathcal{A}^{\text{exp}}_{1212} + \mathcal{A}^{\text{exp}}_{2121} + \mathcal{A}^{\text{exp}}_{2112}
\end{align}
where the individual amplitudes are now defined as  
\begin{equation}\label{exp}
\mathcal{A}^{\text{exp}}_{ijkl} \equiv \langle\Psi_0|DD|\Psi_0\rangle={1\over 4m^2}\e_{AB}\e^{\dot A_2\dot B_2}\e^{\dot A_1\dot B_1}\e_{A'B'}\big[\mathcal{A}^{\text{exp}}_{ijkl}\big]^{AA'}_{\dot A_2\dot A_1}\big[\bar{\mathcal{A}}^{\text{exp}}_{ijkl}\big]^{BB'}_{\dot B_2\dot B_1}
\end{equation}
where
\begin{align}\label{exp lr}
&\big[\mathcal{A}^{\text{exp}}_{ijkl}\big]^{AA'}_{\dot A_2\dot A_1}\nn
&\equiv{}^{(j)}\langle0_{R,+}|{}^{(i)}\langle0_{R,+}| d^{+A(i)}_{n}\alpha^{(j)}_{--,m}G^+_{-{1\over2},\dot A_2}\sigma_{2}^{-}(w_2)G^-_{-{1\over2},\dot A_1}\sigma_{2}^{+}(w_1)\a^{(k)}_{++,-m}d^{-A'(l)}_{-n}|0^+_R\rangle^{(l)}|0^+_R\rangle^{(k)}
\nn
\nn
&\big[\bar{\mathcal{A}}^{\text{exp}}_{ijkl}\big]^{BB'}_{\dot B_2\dot B_1}\nn
&\equiv{}^{(j)}\langle\bar0_{R,+}|{}^{(i)}\langle\bar0_{R,+}|\bar d^{+B(i)}_0 \bar\alpha^{(j)}_{++,m}\bar G^+_{-{1\over2},\dot B_2}\sigma_{2}^{-}(\bar w_2)\bar G^-_{-{1\over2},\dot B_1}\sigma_{2}^{+}(\bar w_1)\bar\a^{(k)}_{--,-m}\bar d^{-B'(l)}_0|\bar0^+_R\rangle^{(l)}|\bar0^+_R\rangle^{(k)}
\nn
\end{align}
As such, this introduces divergences as the deformation operators are brought together. This is due to the fact that since the  left and right moving conformal dimensions, $(h,\bar h)$, of each deformation, $D$, is $(1,1)$ then schematically as two deformations $D(z_2,\bar z_2),D(z_1,\bar z_1)$ are brought close together then to leading order, the amplitude involving the transition of some initial state $|i\rangle$ to some final state $|f\rangle$ is
\begin{equation}
    \langle f | D(z_2,\bar z_2)D(z_1,\bar z_1)|i\rangle\sim {\langle f |i\rangle\over(z_2-z_1)^2(\bar z_2-\bar z_1)^2}
\end{equation}
If $|f\rangle \neq |i\rangle$ the above amplitude vanishes but if $|f\rangle = |i\rangle$ then the amplitude diverges as the deformation operators are brought together. The vanishing of the above amplitude for $|f\rangle\neq |i\rangle$, being orthogonal states, explains why for $p>0$ we have no such divergence in the amplitude computed in this paper.
As expected, for $p=0$, we were able to verify that the amplitude, (\ref{amp exp}), does diverge in the limit where the deformation operators are brought together. Obtaining a finite result then requires regularization of the amplitude by subtracting off the divergent part. This seems to become nontrivial given the form in which the amplitude is expressed, a sum over many terms. We do not address this limit in this paper as it corresponds to a different type of process than the one in which we are currently interested. This is left for a potential future work.
For the amplitude we've considered so far, the total number of strands is $N=N_{00}+N_{++}=N_1+N_{++}=2$ since $N_{00}=N_1=1,N_{++}=1$. In this regime the gravity dual would be highly stringy, far from the supergravity parameter regime which requires $N$ to be large. Thus the results above really represent very stringy transitions. In the next section we discuss the extrapolation to large $N$.

\section{Large $N$ limit}\label{large N}
In this section we focus on the large $N$ limit of the amplitude we have computed for $N=2$ in previous sections. Recall that our initial state, for arbitrary $N$, is given by  
\bea\label{N init}
|\Psi_0;N\rangle &=& \bigg({1\over n!}(L_{-1}-J^3_{-1})^n|00\rangle\bigg)^{N_{00}}\bigg(|++\rangle\bigg)^{N_{++}-1}\bigg({1\over m}\alpha_{++,-m}\alpha_{--,-m}\big|++\rangle\bigg)
\label{ss state}
\eea
corresponding to an initial graviton mode moving in a superstratum background in the gravity dual. For the moment we have omitted the copy indices. Two deformation operators can twist and untwist the graviton mode with either the $00$ strand or $++$ strands. In the case of twisting with only $++$ strands the amplitude was found to be periodic with no terms which grew as $t^2$, representing a graviton moving from the boundary into the bulk and back to the boundary \cite{Guo:2021ybz}. However, in \cite{Guo:2021gqd} it was shown that when twisting a graviton mode with a $00$ strand, the amplitude contained a $t^2$ term in addition to an oscillatory term in $t$, representing an interaction of the probe graviton with the superstratum geometry in the gravity dual, becoming tidally excited into stringy states. In that case however, the superstratum background was considered fixed, being represented in the CFT as a transition in which both the initial and final superstata states were the same. Here, we go a step further by allowing the superstratum state in the CFT itself to change due to its interaction with the graviton which is no longer seen as a probe but now involved in the back reaction of the geometry itself. 

In this paper we have considered a final state corresponding to a change in the background geometry in the gravity dual. We consider one of the simplest nontrivial scenarios where the initial background transitions to a final state containing a microstratum state and graviton suggesting the dual to be a microstratum background with a graviton 
in the bulk. 
To incorporate large $N$ we consider the transition
\begin{equation}
|\Psi_0;N\rangle \to |\Psi_f;N\rangle
\end{equation}
where
\begin{align}\label{N fin}
    |\Psi_{ f};N\rangle &=
    \bigg({1\over n!}(L_{-1}-J^{3}_{-1})^{n}{1\over\sqrt2}\e_{AB}d^{-A}_{0}\bar d^{-B}_{0}|++\rangle\bigg)^{N_{00}-1}\big(|++\rangle\big)^{N_{++}-1}\nn
    &\bigg({1\over(m+n)!m!}(L_{-1}-J^{3}_{-1})^{m+n}(\bar L^{}_{-1}-\bar J^{3}_{-1})^m{1\over\sqrt2}\e_{AB}d^{-A}_{0}\bar d^{-B}_{0}|++\rangle\bigg)\nn
    &\bigg({1\over r}\alpha_{++,-r}\bar\alpha_{--,-r}|++\rangle\bigg)
\end{align}
We remind the reader that we have taken the number of microstrata states to be $N_1=1$ with the number of $00$ strands we start with to be $N_{00}$. For $N_{00}=N_{++}=1$ and thus $N=2$, the amplitude will be given by the expression (\ref{int amplitude}). In this regime the bulk dual is highly stringy. 
Recall that $N_{00}$ counts the number of supergravitons in the initial superstratum state. We define the amplitude for arbitrary $N$ as
\begin{equation}\label{AN}
\tilde A^{0 \to f,scr}_{N}(t) \equiv 
A^{0 \to f,scr}(t) = \lambda^2 2\pi^2t^2B^{0 \to f}_{0,0}(m,n; p,q,r) = {1\over N} g^2 2\pi^2t^2B^{0 \to f,scr}_{0,0}(m,n; p,q,r)
\end{equation}
where we have only included the term which grows for large $t$, $A^{0 \to f,scr}(t)$ and where we have used the relation (\ref{g}) to write the amplitude in terms of the t'Hooft like coupling $g$ and total copy number $N=N_1N_5=N_{00}+N_{++}$. A sufficient condition on the value of $t$ for which the above approximation is justified can be derived through the inequalities below
\begin{align}\label{ineq}
A^{ {0 \to f},scr}(t)&\gg \lambda^2\sum_{k=1}^{2m}{16\pi^2\over k^2}\nn
&> \lambda^2\sum_{k=1}^{2m}{16\pi^2\over k^2}\sin^2\bigg({ kt\over2}\bigg)
\nn
&>\lambda^2\sum_{k=1}^{2m}B^{0 \to f}_{ k, k}(m,n; p,q,r){16\pi^2\over k^2}\sin^2\bigg({ kt\over2}\bigg)
\nn
&=A^{ {0 \to f},osc}(t)
\end{align}
where we have used the empirical fact that so far it seems
\begin{equation}
  1>  |B^{0 \to f}_{ k, k}(m,n; p,q,r)|
\end{equation}
Thus (\ref{ineq}), which implies using (\ref{Ascr Aosc}) that
\begin{equation}\label{t ineq}
    t\gg \sqrt{\sum_{k=1}^{2m}{8\over k^2}}
\end{equation}
certainly satisfies the requirement 
\begin{equation}\label{Ascr Aosc ineq}
    A^{ {0 \to f},scr}(t)\gg A^{ {0 \to f},osc}(t)
\end{equation}
So if $t$ satisfies (\ref{ineq}) then certainly (\ref{Ascr Aosc ineq}) is satisfied.

While the second equality in (\ref{AN}) is the equation imported from the computation for $N=2$ performed in previous sections, the last equality is where the amplitude is upgraded to the case for arbitrary $N$ through the coupling relation (\ref{g}).

Upon inspection, for fixed large $t$ in the large $N$ limit we see that the amplitude is small
\begin{equation}
\tilde A^{0 \to f,scr}_{N}(t)  = {g^2\over N} 2\pi^2t^2B^{0 \to f,scr}_{0,0}(m,n; p,q,r)\ll 1
\end{equation}
since we are in the regime where $g\ll\sqrt N$.

Thus, since we've only computed transitions of one 00 strand out of $N_{00}$ 00 strands in the initial state into one microstratum state and $N_{00}-1$ 00 strands in the final state, we see no enhancement effect. Said another way, the transition to one microstratum state, $N_1=1$, and $N_{00}-1$ $00$ strands in the final state breaks the final state symmetry. There would have been an enhancement had all the final states been microstrata states, i.e, $N_1=N_{00}$ but this is a highly nontrivial scenario to obtain. As a result, we expect the system to take significantly longer to back-react, i.e. for the contribution of the above term to become of order one and hence for the perturbative expansion to break down, which occurs on a timescale of 
\begin{equation}\label{t N}
t\sim {\sqrt{N}\over g}{1\over \sqrt2\pi\sqrt{B^{0 \to f,scr}_{0,0}(m,n;p,q,r)}}
\end{equation}
where we have recorded explicit values of $B^{0 \to f,scr}_{0,0}(m,n;p,q,r)$ for various quantum numbers in table \ref{table}.\footnote{The original values of $B^{0 \to f,scr}_{0,0}(m,n;p,q,r)$ are exact, rational numbers, however in order to save space we have recorded their numerical value up to a certain level of precision.}
\begin{table}
\centering
\label{table}
\begin{tabular}{ | c | c | c | } 
\hline
  Coefficient  & Numerical Value\\
  \hline
  $B^{0\to f}_{0,0}(m=6,n=1;p=1,q=1,r=5)$&   0.0024697\\ 
  \hline
  $B^{0\to f}_{0,0}(m=9,n=1;p=1,q=1,r=8)$&  0.00253053 \\ 
  \hline
  $B^{0\to f}_{0,0}(m=12,n=1;p=1,q=1,r=11)$ & 0.00254984 \\ 
  \hline
  $B^{0\to f}_{0,0}(m=18,n=1;p=1,q=1,r=17)$ & 0.00256181 \\ 
  \hline
  $B^{0\to f}_{0,0}(m=24,n=1;p=1,q=1,r=23)$ &  0.00256514 \\ 
  \hline
\end{tabular}
\caption{Values of coefficient $B^{0\to f}_{0,0}(m,n;p,q,r)$ for various quantum numbers $m,n,p,q,r$.  }
\end{table}
It is understood that (\ref{t N}) is also in the regime which at least satisfies (\ref{t ineq}).

 A physical interpretation of the long timescale may be that the graviton must scatter many times (for long timescales) off the supergravitons which are present, eventually converting one of them into a microstratum state. However, if we were to consider many gravitons interacting with many of the superstrata states, converting them to microstrata states, we expect that there may be an enhancement effect giving a significantly shorter time scale for back-reaction to occur. This would require computing transition amplitudes including many deformation operators such that each copy is affected by the deformation (twist and supercharge) interactions. This is something to consider in the future.

\section{Discussion and Conclusion}\label{discussion conclusion}
In this paper, utilizing the D1D5 CFT, we initiated a study of time-dependent transitions between black hole microstates by applying marginal deformations moving us away from the free theory. Each marginal deformation contains a twist operator and a supercharge operator. The twist operator joins and splits copies of the CFT. The supercharge operator turns a boson into a fermion and fermion into a boson. 

More specifically, the CFT transition which we studied, from perspective of the gravity dual, corresponds to the back-reaction of certain bulk three-charge black hole microstate geometries known as superstrata. The CFT dual of these geometries are identified with coherent sums of supergraviton states which we call superstrata states. Furthermore, we considered this coherent sum to be highly peaked around an average value, the $(1,0,n)$ superstratum state, corresponding to the well studied $(1,0,n)$ superstratum geometry. We then studied, dynamically, how this particular superstratum state, interacting with a graviton mode, could transition into final a state in which the superstratum state itself was modified. We chose the final state to be a graviton mode and a microstratum state, the CFT dual of a microstratum geometry, the non-BPS analogue of a superstratum geometry. This process, in principle, should correspond to back-reaction in the bulk in which the original superstratum geometry, upon infall of a graviton, would dynamically evolve into a microstratum geometry with a different graviton \footnote{At even order in the deformation operator, this process requires there to be at least a graviton in the final state, in addition to the microstratum state. Given the initial state, the absence of anything except the microstratum state in the final state is prohibited by charge conservation}. 

To do this we considered the action of two deformation operators to compute transitions starting from two singly wound copies of the CFT, one containing a 00 strand, the other containing a $++$ strand with the combination thereof defining the superstratum state on which the graviton was acting. 
The first deformation operator twisted together the state on the initial two copies mentioned above, into a state on a doubly wound copy with the second deformation splitting this state back into one with two singly copies, a graviton acting on a microstratum state. We first computed the unintegrated amplitudes of the above described transition. To do this we utilized the covering space method, a technology which has been developed to compute correlation functions which include twist operators.

To obtain the full transition amplitude we integrated the unintegrated amplitude over the twist locations. 
We found that this amplitude contained both 1) an oscillatory term in $t$, characteristic of a mode whose dual corresponds to the infall of particle into the dual geometry from the boundary, and 2) a  $t^2$ term, indicating a growing probability for the initial state to transition into the final state for long time scales. In the gravity dual, this suggests that the interaction of the in-falling mode with the original geometry itself is causing a transition into a new one. We were able to plot, numerically, the amplitudes and observe their growth behavior for various explicit quantum numbers.

After imposing energy and momentum conservation the transition amplitude
is characterized by three quantum numbers, $m,n,p$ where $m$ characterizes the energy of the initial graviton, $n$ the initial momentum of the superstratum mode and $p$ the measure of the non-BPS nature of the microstratum state i.e. how much left and right moving momentum has been added on top of a superstratum state characterized by $q=n$ with the final graviton mode subject to the constraint $r=m-p$.
We find that the amplitude only significantly feels the difference between initial and final states with this difference characterized by $p$. Thus the largest transition probability for superstratum $+$ graviton $\to$ microstratum $+$ graviton' occurs when the initial graviton transfers a minimum amount of energy i.e. $p=1$ and with $n=1$ for sufficiently large $m$ with the amplitude approaching a profile which seems to be independent of both $n$ and $m$ as $n$ and $m$ are increased. The microstratum state with $p=1$ corresponds to a perturbatively non-BPS microstratum state \cite{Ganchev:2021ewa}.
When extrapolating the result to the large $N$ limit, we found no enhancement factor which could counteract the dilution by $N$. Thus the timescale for back-reaction of the transition considered here, i.e. when the growth term of the transition amplitude becomes of order one and where the perturbative expansion starts to break down, is significantly long ($\sim \sqrt{N}/g>>1$). Even though this is the case, the interaction computed here could be a building block of a larger set of processes which, when combined together, could give a significant effect. This would likely require many such twist operators in order to convert most of the $N_{00}$ $00$ strands into microstrata states counted by $N_1$. Here we only considered the conversion of one superstratum state into one microstratum state each with an accompanying graviton. 

In \cite{Guo:2021gqd} a superstratum state with an initial graviton was studied using two deformation operators. There the focus was on extracting the CFT dual of tidal force dynamics into stringy excitations of the initial probe graviton coming from interactions with the superstratum geometry which was studied on the gravitational side of the correspondence in \cite{Martinec:2020cml,Ceplak:2021kgl} for radial infall and \cite{Guo:2024pvv} for circular motion. Here the aim was to consider the CFT dual of how this interaction could change the background itself. Another interesting final state to consider, in the context of the current work, is to then replace the graviton mode with a stringy mode, corresponding to a higher number of bosonic or fermionic modes in the final state. This would allow the study of transitions in which the final state could include a superstratum state because the final energy could be distributed amongst more bosonic modes, evading the constraint required in the graviton case considered here. However preliminary results, which are not reported here, seem to show that stringy states have a smaller amplitude than their graviton counterparts. This is sensible since stringy states in the CFT correspond to strings which have a mass, arising from their vibrational modes. Thus transitions including such states should be suppressed more than the massless graviton counterparts. In \cite{Marolf:2016nwu} the authors consider transitions of two charge microstates in supergravity by investigating instabilities predicted by \cite{Eperon_2016}. Time-dependent microstrata were constructed in \cite{Houppe:2024hyj}. It would be interesting to understand the connection of the CFT to these results. In general dynamics are challenging to study and furthermore, less is generally known about how heavy black hole microstates back-react into other heavy black hole microstates.

The computations in this paper aim to address, holographically, how three charge microstates evolve in their allowed phase space when they back-react due to their interaction with an incoming excitation, a central question of the fuzzball paradigm. While there is still much to be done, we have presented here at least a computation, albeit in a highly simplified setting, which tries to address such a question, though in a regime that is perturbative in the CFT, extrapolated to strong coupling. Ideally, one would need to compute transitions to all orders in the deformation operator and sum their contributions to obtain a result for strongly coupled dynamics. Thus it would be interesting to explore higher orders in the deformation for this reason. See \cite{Guo:2024pvv} for some progress in this direction.  Furthermore, we've considered singly wound twist sectors in both the initial and final states. However, in the supergravity regime, the mass gap of superstrata, which is inversely proportional to the length of the AdS$_2\times S^1$ region, is proportional to $ 1/ N_1N_5$ where $N_1$ and $N_5$ are the number of D1 and D5 branes wrapping the compact directions. This corresponds to the maximally twisted sector in the CFT.  It would therefore be nice to explore back-reaction dynamics which evolves the singly wound sectors into multi-wound sectors. Additionally, in this paper, we considered a graviton and superstratum initial state in the CFT. One could also consider a superstratum, antisuperstratum state and how their interaction evolves dynamically, corresponding to a superstratum, antisuperstratum collision in the bulk. There are many questions one can explore as there are many final states to transition into with many different twist sectors. We hope to explore these various directions in the future. 

\section*{Acknowledgements}
SDH would like to thank Nejc \v{C}eplak for helpful discussions during the completion of this work. SDH would also like to thank the hospitality of IBS in Daejeon during the completion of this work. The work of SDH is supported by KIAS Grant PG096301.

\appendix

\section{Field content and $\mathcal N = 4$ superconformal algebra}\label{algebra}
In this section we record the field content and the $\mathcal N=(4,4)$ superconformal algebra. We explicitly only write the left moving sector, noting that the right moving sector has an analogous form obtained by replacing the holomorphic coordinates with antiholomorphic coordinates, $z\to\bar z$, and the holomorphic quantities with the antiholomorphic quantities.
\subsection{Fields}
The D1D5 CFT contains four left moving fermions and four right moving fermions. Only writing the left movers we have
 $\psi_1, \psi_2, \psi_3, \psi_4$ which can be organized as doublets  $\psi^{\alpha A}$:
\be
\begin{pmatrix}
\psi^{++} \cr \psi^{-+}
\end{pmatrix}
={1\over\sqrt{2}}
\begin{pmatrix}
\psi_1+i\psi_2 \cr \psi_3+i\psi_4
\end{pmatrix}
\ee
\be
\begin{pmatrix}
\psi^{+-} \cr \psi^{--}
\end{pmatrix}
={1\over\sqrt{2}}
\begin{pmatrix}
\psi_3-i\psi_4 \cr -(\psi_1-i\psi_2)
\end{pmatrix}.
\ee
where $\alpha=+,-$ corresponds to the $SU(2)_L$ subgroup of $S^3$ rotations and $A=+,-$ corresponds to the $SU(2)_1$ subgroup of $T^4$ rotations. The OPE between fermionic fields is given by
\be
\langle\psi^{\alpha A}(z)\psi^{\beta B}(w)\rangle=-\epsilon^{\alpha\beta}\epsilon^{AB}{1\over z-w}
\ee
where the epsilon symbol is defined as
\be
\epsilon_{+-}=1, ~~~\epsilon^{+-}=-1
\ee
The theory also contains four real bosons $X_1, X_2, X_3, X_4$ which can also be grouped into doublets 
\be
X_{A\dot A}= {1\over\sqrt{2}} X_i \sigma_i
={1\over\sqrt{2}}
\begin{pmatrix}
X_3+iX_4 & X_1-iX_2 \cr X_1+iX_2&-X_3+iX_4
\end{pmatrix}
\ee
where
$\sigma_j=(\sigma_a, iI)$ for $a=1,2,3$ with $\sigma_a$ the Pauli matrices. The OPE for the left moving bosonic sector is
\be
\langle\partial X_{A\dot A}(z) \partial X_{B\dot B}(w)\rangle=\epsilon_{AB}\epsilon_{\dot A\dot B}{1\over (z-w)^2} \,.
\ee
with a similar relation for the right-moving sector. Written in terms of the bosonic and fermionic fields, the $R$-currents, supercurrents, and stress-energy tensor
\bea
J^a&=& {1\over 4}\e_{\a \g}\e_{AC}\psi^{\g C} (\sigma^{Ta})^\alpha{}_\beta \psi^{\beta A},\qquad a=1,2,3
\cr\cr
G^\alpha_{\dot A}&=& \psi^{\alpha A} \partial X_{A\dot A},\qquad \a = +,-
\cr\cr
T&=& {1\over 2} \e^{AB}\e^{\dot{A}\dot{B}}\partial X_{B\dot B}\partial X_{A\dot A} + {1\over 2} \e_{\a\beta}\e_{AB}\psi^{\beta B} \partial \psi^{\alpha A}
\eea
generate the chiral algebra.

\subsection{OPE algebra}
Here we record the chiral OPE algebra derived from the currents above.

\subsection{OPE's of currents with $\partial X_{A\dot{A}}(z)$ and $\psi^{\a A}(z)$ }
First recording the boson-current and fermion-current OPE's we have
\bea
T(z)\partial X_{A\dot{A}}(w) &\sim& {\partial X_{A \dot A}(w)\over (z-w)^2} + { \partial^2 X_{A \dot A}(w)\over z-w} \cr
T(z)\psi^{\a A}(w) &\sim&  {{1\over 2}\psi^{\a A}(w)\over (z-w)^2}+{\partial \psi^{\a A}(w)\over z-w} \cr
G^{\a}_{\dot{A}}(z)\psi^{\beta B}(w) &\sim& \e^{\a\beta}\e^{BA} {\partial X_{A\dot{A}}(w) \over z-w}\cr
G^{\a}_{\dot{A}}(z)\partial X_{ B\dot{B}}(w)&\sim& \e_{AB}\e_{\dot{A}\dot{B}}{ \psi^{\a A}(w)\over  (z-w)^2} + \e_{AB}\e_{\dot{A}\dot{B}}{ \partial \psi^{\a A}(w)\over  z-w}\cr
J^a(z)\psi^{\a A}(w) &\sim&  {1\over 2} {1\over z-w}(\s^{Ta})^{\a}_{\beta}\psi^{\beta A}(w)\cr
J^{+}(z)\psi^{+ A}(w) &=& 0,\qquad J^{-}(z)\psi^{+ A}(w) =   {\psi^{- A}(w)\over z-w} \cr
J^{+}(z)\psi^{- A}(w) &=&   {\psi^{+ A}(w)\over z-w} ,\qquad  J^{-}(z)\psi^{- A}(w) = 0
\eea

\subsection{OPE's of currents with currents }
The current-current OPE's are
\bea
T(z)T(w)&\sim&{{c\over2}\over (z-w)^4} + {2T(w)\over (z-w)^2}  + {\partial T(w)\over z-w}
\cr
J^a(z)J^b(w)&\sim&  {{c\over 12}\delta^{ab}\over(z-w)^2} + {i\e^{ab}_{\,\,\,\,c}J^c(w)\over z-w}
\cr
G^{\a}_{\dot{A}}(z)G^{\beta}_{\dot{B}}(w) &\sim& -\e_{\dot{A}\dot{B}}\bigg[\e^{\beta\a}{{c\over3}\over (z-w)^3}  + \e^{\beta\g}(\s^{aT})^{\a}_{\g}\bigg({2J^a(w)\over (z-w)^2} + {\partial J^a(w)\over z-w}\bigg)  + \e^{\beta\a}{1\over z-w}T(w)   \bigg]
\cr
J^a(z)G^{\a}_{\dot{A}}(w)&\sim& {1\over z-w}{1\over2}(\s^{aT})^{\a}_{\beta}G^{\beta}_{\dot{A}}(w)
\cr
T(z)J^a(w) &\sim&  {J^a(w)\over (z-w)^2} + {\partial J^a(w)\over z-w}
\cr
T(z)G^{\a}_{\dot{A}}(w)&\sim& {{3\over2}G^{\a}_{\dot{A}}(w)\over (z-w)^2} + { \partial G^{\a}_{\dot{A}}(w)\over z-w}
\eea
Organizing the $R$-current components $J^1,J^2$ into currents $J^+,J^-$ carrying charge under $SU_L(2)$ we have
\bea
J^+ &=& J^1 + i J^2\cr
J^-&=& J^1 - i J^2
\eea
with the current-current algebra being rewritten as 
\begin{align}
J^{+}(z)J^{-}(w)&\sim  {{c\over6}\over (z-w)^2} +  {2J^3(w)\over z-w},& J^{-}(z)J^{+}(w)&\sim  {{c\over6}\over (z-w)^2} -  {2J^3(w)\over z-w}\cr
J^3(z)J^{+}(w)&\sim  {J^{+}(w)\over z-w}, & J^3(z)J^{-}(w)&\sim- {J^{-}(w)\over z-w}\cr
J^+(z)J^3(w)&\sim- {J^+(w)\over z-w}, &J^-(z)J^3(w)&\sim {J^-(w)\over z-w}\cr
T(z)J^{+}(w)&\sim{J^{+}(w)\over (z-w)^2} + {\partial J^{+}(w)\over z-w},&T(z)J^{-}(w)&\sim {J^{-}(w)\over (z-w)^2} + {\partial J^{-}(w)\over z-w}\cr
J^{+}(z)G^{-}_{\dot{A}}(w) &\sim {G^{+}_{\dot{A}}(w)\over z-w},& J^{-}(z)G^{+}_{\dot{A}}(w)& \sim {G^{-}_{\dot{A}}(w)\over z-w}
\end{align}

\subsection{Mode and contour definitions of the fields} 

We relate the fields recorded previously in terms of modes through contour integrals
\bea
L_m&=&\oint {dz\over 2\pi i}z^{m+1}T(z)\cr
J^a_m&=&\oint {dz\over 2\pi i} z^{m}J^a(z)\cr
G^{\a}_{\dot{A},r}&=&\oint {dz\over 2\pi i} z^{r+{1\over2}}G^{\a}_{\dot{A}}(z)\cr
\a_{A\dot{A},m}&=&i\oint {dz\over 2\pi i} z^{m}\partial X_{A\dot{A}}(z)\cr
d^{\a A}_r&=& \oint {dz\over 2\pi i} z^{r-{1\over2}}\psi^{\a A}(z)
\eea 
which can be inverted to give the relations
\bea
T(z) &=& \sum_{m}z^{-m-2}L_m\cr
J^a(z) &=& \sum_{m}z^{-m-1}J^a_m\cr
G^{\a}_{\dot{A}}(z) &=& \sum_{r}z^{-r-{3\over2}}G^{\a}_{\dot{A},r}\cr
\partial X_{A\dot{A}}(z) &=& -i\sum_m z^{-m-1}\a_{A\dot{A},m}\cr
\psi^{\a A}(z) &=& \sum_m z^{-m-{1\over2}}d^ {\a A}_m
\eea

\subsection{Commutators of currents with currents}
Here we record current - current commutation relations
\bea\label{commutations_ii}
[L_m,L_n] &=& {c\over12}m(m^2-1)\delta_{m+n,0}+ (m-n)L_{m+n}\cr
[J^a_{m},J^b_{n}] &=&{c\over12}m\delta^{ab}\delta_{m+n,0} +  i\e^{ab}_{\,\,\,\,c}J^c_{m+n}\cr
\lbrace G^{\a}_{\dot{A},r} , G^{\beta}_{\dot{B},s} \rbrace&=&  \e_{\dot{A}\dot{B}}\bigg[\e^{\a\beta}{c\over6}(r^2-{1\over4})\delta_{r+s,0}  + (\s^{aT})^{\a}_{\g}\e^{\g\beta}(r-s)J^a_{r+s}  + \e^{\a\beta}L_{r+s}  \bigg]\cr
[J^a_{m},G^{\a}_{\dot{A},r}] &=&{1\over2}(\s^{aT})^{\a}_{\beta} G^{\beta}_{\dot{A},m+r}\cr
[L_{m},J^a_n]&=& -nJ^a_{m+n}\cr
[L_{m},G^{\a}_{\dot{A},r}] &=& ({m\over2}  -r)G^{\a}_{\dot{A},m+r}\cr
[J^+_{m},J^-_{n}]&=&{c\over6}m\delta_{m+n,0} + 2J^3_{m+n}\cr
[L_m,J^{+}_n] &=& -nJ^{+}_{m+n},\qquad ~[L_m,J^{-}_n] ~=~ -nJ^{-}_{m+n}\cr
[J^{+}_{m},G^{+}_{\dot{A},r}]  &=& 0 ,\qquad\qquad ~~~[J^{-}_{m},G^{+}_{\dot{A},r}]  ~=~ G^{-}_{\dot{A},m+r}\cr
[J^{+}_{m},G^{-}_{\dot{A},r}]  &=&G^{+}_{\dot{A},m+r},\qquad ~[J^{-}_{m},G^{-}_{\dot{A},r}]  ~=~ 0 \cr
[J^3_m , J^{+}_n] &=& J^{+}_{m+n},\qquad\qquad [J^3_m , J^{-}_n] ~=~ -J^{-}_{m+n}
\eea

\subsection{Commutators of $\alpha_{A\dot{A},m}$ and $d^{\a A}_r$}
The bosonic and fermionic mode relations are
\bea
[\a_{A\dot{A},m},\a_{B\dot{B},n}] &=& -m\e_{A\dot{A}}\e_{B\dot{B}}\delta_{m+n,0}\cr
\lbrace d^{\alpha A}_r , d^{\beta B}_s\rbrace  &=&-\e^{\alpha\beta}\e^{AB}\delta_{r+s,0}
\eea
\subsection{Commutators of currents with $\alpha_{A\dot{A},m}$ and $d^{\a A}_r$}
The bosonic mode - current and fermionic mode - current relations are given by
\bea\label{commutations}
[L_m,\a_{A\dot{A},n}] &=&-n\a_{A\dot{A},m+n} \cr
[L_m ,d^{\a A}_r] &=&-({m\over2}+r)d^{\a A}_{m+r}\cr
\lbrace G^{\a}_{\dot{A},r} ,  d^{\beta B}_{s} \rbrace&=&i\e^{\a\beta}\e^{AB}\a_{A\dot{A},r+s}\cr
[G^{\a}_{\dot{A},r} , \a_{B \dot{B},m}]&=&  -im\e_{AB}\e_{\dot{A}\dot{B}}d^{\a A}_{r+m}\cr
[J^a_m,d^{\a A}_r] &=&{1\over 2}(\s^{Ta})^{\a}_{\beta}d^{\beta A}_{m+r}\cr
[J^{+}_m,d^{+ A}_r] &=& 0,\qquad~~~~~ [J^{-}_m,d^{+ A}_r] ~=~ d^{-A}_{m+r}\cr
[J^{-}_m,d^{+ A}_r] &=& d^{-A}_{m+r},\qquad [J^{+}_m,d^{+ A}_r] ~=~ 0
\eea

\subsection{Current modes written in terms of $\alpha_{A\dot{A},m}$ and $d^{\a A}_r$}
Finally, using the free field realization, we write the current operators in terms of bosonic and fermionic modes.
\bea
J^a_m &=& {1\over 4k}\sum_{r}\epsilon_{AB}d^ {\g B}_r\epsilon_{\alpha\gamma}(\s^{aT})^{\a}_{\beta}d^ {\beta A}_{m-r},\qquad a=1,2,3\cr
J^3_m &=&  - {1\over 2k}\sum_{r} d^ {+ +}_{r}d^ {- -}_{m-r} - {1\over 2k}\sum_{r}d^ {- +}_r d^ {+ -}_{m-r}\cr
J^{+}_m&=&{1\over k}\sum_{r}d^ {+ +}_rd^ {+ -}_{m-r} ,\qquad J^{-}_m={1\over k}\sum_{r}d^ {--}_rd^ {- +}_{m-r}\cr
G^{\a}_{\dot{A},r} &=& -{i\over k}\sum_{n}d^ {\a A}_{r-n} \a_{A\dot{A},n}\cr
L_m&=& -{1\over 2k}\sum_{n} \e^{AB}\e^{\dot A \dot B}\a_{A\dot{A},n}\a_{B\dot{B},m-n}- {1\over 2k}\sum_{r}(m-r+{1\over2})\epsilon_{\alpha\beta}\epsilon_{AB}d^ {\a A}_r d^ {\beta B}_{m-r}\nn
\eea

\section{Wick contraction terms}\label{wick contractions}
In this appendix we record all of the possible Wick contraction terms which enter the expressions for the various amplitudes computed in this paper. For brevity we show an explicit computation for one bosonic contraction and one fermionic contraction noting that the others can be computed in a similar way. Let's first tabulate the the $t$-plane images of the copies and supercharges
\bea
\text{Copy 1 initial}&:& t=-a\cr
\text{Copy 2 initial}  &:& t=-b\cr
\text{Copy 1 final} &:& t=\infty\cr
\text{Copy 2 final} &:& t=0\cr
\text{Supercharge 1} &:& t=t_1\cr
\text{Supercharge 2} &:& t=t_2
\eea
Next we record all pairs of $t$-plane locations labeling the corresponding copy for which we have a Wick contraction between pairs of modes, each at the respective locations.
\begin{align}
\text{Copy 1 final, Copy 1 final}~&:~\infty,\infty\cr
\text{Copy 1 final, Supercharge $i$}&~:~\infty,t_i\cr
\text{Copy 1 final, Copy 1 initial}&~:~\infty,-a\cr
\text{Copy 1 final, Copy 2 initial}&~:~\infty,-b\cr
\cr
\text{Copy 2 final, Copy 2 final}&~:~0,0\cr
\text{Copy 2 final, Supercharge $i$}&~:~0,t_i\cr
\text{Copy 2 final, Copy 1 initial}&~:~0,-a\cr
\text{Copy 2 final, Copy 2 initial}&~:~0,-b\cr
\cr
\text{Copy 1 initial, Copy 1 initial}&~:~-a,-a\cr
\text{Copy 2 initial, Copy 2 initial}&~:~-b,-b\cr
\text{Copy 1 initial, Copy 2 initial}&~:~-a,-b
\cr
\cr
\text{Supercharge $i$, Copy 1 initial}&~:~t_i,-a\cr
\text{Supercharge $i$, Copy 2 initial}&~:~t_i,-b\cr
\cr
\text{Supercharge $1$, Supercharge $2$}&~:~t_1,t_2
\end{align}
where $i=1,2$ in the above list. We note that we don't explicitly compute or record all possible combinations of contractions identified above since for this paper some of them are not needed.
Let us record into one place the various boson and fermion $t$-plane contours below which we obtain from (\ref{c1 init pr}), (\ref{c2 init pr}), (\ref{c1 fin pr}) (\ref{c2 fin pr}) and (\ref{bos ferm ti})
\bea\label{modes}
\a'^{(1)}_{A\dot{A},m} &=& {1\over 2\pi}\oint_{t=\infty} dt {(t+a)^m(t+b)^m\over t^m}\partial X_{A\dot{A}}(t),\qquad \cr
\a'^{(2)}_{A\dot{A},m} &=& -{1\over 2\pi}\oint_{t=0} dt {(t+a)^m(t+b)^m\over t^m}\partial X_{A\dot{A}}(t)\cr
d'^{+ A(1)}_{r} &=&{1\over 2\pi i}\oint_{t=\infty} dt {(t-t_1)(t+a)^{r}(t+b)^{r}\over t^{r+1}} \psi^{+ A}(t)\cr
d'^{- A(1)}_{r} &=&{1\over 2\pi i}\oint_{t=\infty} dt {(t-t_2)(t+a)^{r-1}(t+b)^{r-1}\over t^{r}} \psi^{- A}(t)\cr
d'^{+ A(2)}_{r} &=&-{1\over 2\pi i}\oint_{t=0} dt {(t-t_1)(t+a)^{r}(t+b)^{r}\over t^{r+1}} \psi^{+ A}(t)\cr
d'^{- A(2)}_{r} &=&-{1\over 2\pi i}\oint_{t=0} dt {(t-t_2)(t+a)^{r-1}(t+b)^{r-1}\over t^{r}} \psi^{- A}(t)\cr
d'^{+ A(1)}_{-r} &=&{1\over 2\pi i}\oint_{t=-a} dt {(t-t_1)(t+a)^{-r}(t+b)^{-r}\over t^{-r+1}} \psi^{+ A}(t)\cr
d'^{- A(1)}_{-r} &=&{1\over 2\pi i}\oint_{t=-a} dt {(t-t_2)(t+a)^{-r-1}(t+b)^{-r-1}\over t^{-r}} \psi^{- A}(t)\cr
d'^{+ A(2)}_{-r} &=&{1\over 2\pi i}\oint_{t=-b} dt {(t-t_1)(t+a)^{-r}(t+b)^{-r}\over t^{-r+1}} \psi^{+ A}(t)\cr
d'^{- A(2)}_{-r} &=&{1\over 2\pi i}\oint_{t=-b} dt {(t-t_2)(t+a)^{-r-1}(t+b)^{-r-1}\over t^{-r}} \psi^{- A}(t)\cr
\a'^{(1)}_{A\dot{A},-n}&=&  {1\over 2\pi}\oint_{t=-a} dt {(t+a)^{-n}(t+b)^{-n}\over  t^{-n} }\partial X_{A\dot{A}}(t)\cr
\a'^{(2)}_{A\dot{A},-n}&=&  {1\over 2\pi}\oint_{t=-b} dt {(t+a)^{-n}(t+b)^{-n}\over t^{-n}}\partial X_{A\dot{A}}(t)\cr
\tilde{\a}^{t_i}_{A\dot{A},-1}&=&i \partial X_{A\dot{A}}(t_i)\cr
\tilde{d}^{\a A,t_i}_{-{1\over2}} &=&\psi^{\a A}(t_i)
\eea
where the minus sign appearing in front of all copy two final modes arises due to the fact that locally at $t=0$ the map behaves as $z\sim 1 / t$ therefore reversing the direction of the contour.

To compute the contraction terms one must expand the integrand of the mode in consideration around its appropriate $t$-plane image. This yields a linear combination of modes defined with respect to the vacuum at that point. If the two modes are on the same copy, meaning they are located at the same $t$-plane image, then one can straightforwardly compute the contraction between those expansions, using local commutation relations. 

However, if the modes are on different copies, meaning they are located at different $t$-plane images, one must first expand each mode around their respective locations. But because they are at separate points they cannot be contracted together. To solve this, one must then pick a mode, which has already been expanded around its respective image giving a linear combination of modes, and expand each member defined at that image into an additional linear combination of modes defined at the image of the other mode. Therefore, one of the modes is expanded twice, once around its own image, and again around the other's image. One can then perform Wick contractions between the pairs of modes since they are both written as expansions around the same $t$-plane image. Since this is, in some cases, the most intricate type of Wick contraction we will choose one from the the set of bosonic contractions and one from the set fermionic contractions, respectively, of this type, that is used in this paper, to demonstrate explicitly. The other contractions can be computed using similar techniques.

\subsection{Contraction between bosons on copy $2$ final ($t=0$) and copy $1$ initial ($t=-a$)}
Here we demonstrate an explicit example by computing the contraction between a bosonic mode on copy $2$ final located at the image $t=0$ and copy $1$ initial located at the image at $t=-a$. We choose this particular contraction to show the steps explicitly because it captures the more intricate details of the contraction computation. Any other bosonic contraction will either contain a similar level of intricacy or less.
To do so we need to first expand the modes involved in a linear combination around their respective $t$-plane images. For copy $2$, from the list (\ref{modes}), we have the bosonic mode
\bea\label{copy 2 fin}
\a'^{(2)}_{A\dot{A},p} &=& - {1\over 2\pi}\oint_{t=0} dt {(t+a)^p(t+b)^p\over t^p}\partial X_{A\dot{A}}(t)\cr
&=&-\sum_{j,j'\geq 0}{}^pC_j{}^pC_{j'} a^{p-j}b^{p-j'}\tilde{\a}^0_{A\dot{A},j+j'-p}
\eea
where we have expanded the integrand around $t=0$ and have used the following definition of modes natural to the $t$-plane at that location
\bea\label{contour at 0}
\tilde{\a}^0_{A\dot{A},n}={1\over 2\pi}\oint_{t=0}dt t^n\partial X_{A\dot{A}}(t)
\eea
The coefficient ${}^{p}C_{q}$ is the binomial coefficient
\begin{equation}
{}^{p}C_{q}\equiv {p!\over q!(p-q)!}
\end{equation}
As a reminder, the minus in (\ref{copy 2 fin}) accounts for the reversal of the contour direction when expanding around $t=0$, because the map goes as $z\sim 1/ t$.
For Copy 1 initial, at the image $t=-a$, we have the expansion
\bea\label{copy 1 init}
\a'^{(1)}_{A\dot{A},-n}&=&  {1\over 2\pi}\oint_{t=-a} dt {t^n\over   (t+a)^n(t+b)^n}\partial X_{A\dot{A}}(t)\cr
&=&\sum_{j,j'\geq 0}{}^nC_{j}{}^{-n}C_{j'}(-a)^{n-j}(b-a)^{-n-j'}\tilde{\a}^{-a}_{A\dot{A},-n+j+j'}
\eea
where modes natural to the $t$-plane at the image $t=-a$ are defined as
\bea
\tilde{\a}^{-a}_{A\dot{A},m}={1\over 2\pi}\oint_{t=-a} dt(t+a)^m\partial X_{A\dot{A}}(t)
\eea
In order for the action on the local vacuum to not vanish, i.e.
\begin{equation}
    \tilde{\a}^{-a}_{A\dot{A},m}|0\rangle^{-a}_t\neq0,\quad m<0
\end{equation}
the mode indices for the $t$-plane modes in (\ref{copy 1 init}) should obey
\bea
-n+j+j'<0&\to& j'<n-j\cr
j'\geq0&\to& j<n
\eea
This gives
\bea
\a'^{(1)}_{A\dot{A},-n}&=&\sum_{j= 0}^{n-1}\sum_{j'= 0}^{n-j-1}{}^nC_{j}{}^{-n}C_{j'}(-a)^{n-j}(b-a)^{-n-j'}\tilde{\a}^{-a}_{A\dot{A},-n+j+j'}
\label{copy one initial b}
\eea
So far, both modes have been expanded around their respective $t$-plane images. However, in order to contract them they must be expanded around the same $t$-plane image. Proceeding forward we choose to deform the contour circling $t=0$, into a contour circling $t=-a$, situated on the outside of the contour already circling $t=-a$. Thus utilizing the definition (\ref{contour at 0}) we record the $t$-plane mode from the expansion (\ref{copy 2 fin})
\bea\label{contour at 0 pr}
\tilde{\a}^0_{A\dot{A},k+k'-p} &=& {1\over 2\pi}\oint_{t=0} dt t^{k+k'-p}\partial X_{A\dot{A}}(t)
\eea
Expanding the integrand around $t=-a$ yields 
\bea
 t^{k+k'-p}=\sum_{l\geq0}{}^{k+k'-p}C_l(-a)^{k+k'-p-l}(t+a)^l
\eea
Inserting this into the contour in (\ref{contour at 0 pr}) yields an expansion of the modes defined at $t=0$ in terms of modes defined at $t=-a$
\bea\label{0 exp at minus a}
\tilde{\a}^0_{A\dot{A},k+k'-p}  &=&-\sum_{l\geq0}{}^{k+k'-p}C_l(-a)^{k+k'-p-l}\tilde{\a}^{-a}_{A\dot{A},l}
\eea
where the minus sign comes from reversing the direction of the contour when deforming it from around the point $t=0$ to the point $t=-a$.
Therefore inserting (\ref{0 exp at minus a}) into (\ref{copy 2 fin}) we obtain
\bea\label{copy 2 final pr}
\a'^{(2)}_{A\dot{A},p}&=&\sum_{k=0}^{p-1}\sum_{k'=0}^{p-k-1}\sum_{l\geq0}{}^pC_k{}^pC_{k'} {}^{k+k'-p}C_l (-a)^{k+k'-p-l}a^{p-k}b^{p-k'}\tilde{\a}^{-a}_{A\dot{A},l}
\eea
Now that copy 2 final has been expanded around copy 1 initial we can now perform the Wick contraction between them. We first record the definition of commutation relations for modes defined at the $t$-plane image $t=-a$, which given by
\begin{equation}\label{comm rel at minus a}
    [\tilde\alpha^{-a}_{A\dot A,m},\tilde\alpha^{-a}_{B\dot B,n}]=-m\e_{AB}\e_{\dot A\dot B}\d_{m+n,0}
\end{equation}
Thus contracting together (\ref{copy 2 final pr}) and (\ref{copy 1 init}) and utilizing the relation (\ref{comm rel at minus a}) we obtain
\bea\label{bos fin two bos init one}
\langle\a'^{(2)}_{B\dot{B},p}\a'^{(1)}_{A\dot{A},-n}\rangle &=&\sum_{k=0}^{p-1}\sum_{k'=0}^{p-k-1}\sum_{j= 0}^{n-1}\sum_{j'= 0}^{n-j-1}\sum_{l\geq0}{}^pC_k{}^pC_{k'} {}^{k+k'-p}C_l {}^nC_{j}{}^{-n}C_{j'}\cr
&&(-a)^{n-j+k+k'-p-l}a^{p-k}(b-a)^{-n-j'}b^{p-k'}[\tilde{\a}^{-a}_{B\dot{B},l},\tilde{\a}^{-a}_{A\dot{A},-n+j+j'}]
\cr
\cr
&=&\sum_{k=0}^{p-1}\sum_{k'=0}^{p-k-1}\sum_{j= 0}^{n-1}\sum_{j'= 0}^{n-j-1}\sum_{l\geq0}{}^pC_k{}^pC_{k'}{}^{k+k'-p}C_l {}^nC_{j}{}^{-n}C_{j'} \cr
&&(-a)^{n-j+k+k'-p-l}a^{p-k}(b-a)^{-n-j'}b^{p-k'}(-l\e_{BA}\e_{\dot{B}\dot{A}}\d_{l-n+j+j',0})
\cr
\cr
&=&-\e_{BA}\e_{\dot{B}\dot{A}}\sum_{k=0}^{p-1}\sum_{k'=0}^{p-k-1}\sum_{j= 0}^{n-1}\sum_{j'= 0}^{n-j-1}(n-j-j'){}^pC_k{}^pC_{k'} {}^{k+k'-p}C_{n-j-j'} \cr
&&{}^nC_{j}{}^{-n}C_{j'}(-a)^{j'+k+k'-p}a^{p-k}b^{p-k'}(b-a)^{-n-j'}
\cr
\cr
&=&-\e_{BA}\e_{\dot{B}\dot{A}}\sum_{k=0}^{p-1}\sum_{k'=0}^{p-k-1}\sum_{j= 0}^{n-1}\sum_{j'= 0}^{n-j-1}(n-j-j'){}^pC_k{}^pC_{k'} {}^{k+k'-p}C_{n-j-j'} \cr
&&{}^nC_{j}{}^{-n}C_{j'}(-1)^{p+j'+k+k'}a^{j'+k'}b^{p-k'}(b-a)^{-n-j'}
\eea
To obtain the contraction where the initial state is on copy 2 we switch $a$ and $b$ in the above expression giving
\bea\label{bos fin two bos init two}
\langle\a'^{(2)}_{B\dot{B},p}\a'^{(2)}_{A\dot{A},-n}\rangle &=& -\e_{BA}\e_{\dot{B}\dot{A}}\sum_{k=0}^{p-1}\sum_{k'=0}^{p-k-1}\sum_{j= 0}^{n-1}\sum_{j'= 0}^{n-j-1}(n-j-j'){}^pC_k{}^pC_{k'} {}^{k+k'-p}C_{n-j-j'} \cr
&&{}^nC_{j}{}^{-n}C_{j'}(-1)^{p+j'+k+k'}b^{j'+k'}a^{p-k'}(a-b)^{-n-j'}
\eea

\subsection{Contraction between fermions with $+R$ - charge on copy $1$ final $(t=\infty)$ and $-R$ - charge on copy $1$ initial $(t=-a)$.}
For the fermionic case we will demonstrate explicitly how to compute the contraction between a mode on copy $1$ final with $R$-charge $+$ located at the image $t=\infty$ and copy $1$ initial with $R$-charge $-$ located at the image $t=-a$. We start by recording the relevant copy $1$ final and copy $2$ initial modes from the list (\ref{modes}) in which we have
\bea
d'^{+ A(1)}_{r} &=&{1\over 2\pi i}\oint_{t=\infty} dt {(t-t_1)(t+a)^{r}(t+b)^{r}\over t^{r+1}} \psi^{+ A}(t)\cr
d'^{- A(1)}_{-r} &=&{1\over 2\pi i}\oint_{t=-a} dt {(t-t_2)(t+a)^{-r-1}(t+b)^{-r-1}\over t^{-r}} \psi^{- A}(t)
\eea
Focusing first on copy $1$ final we expand the integrand of $d'^{(1),+ A}_r$ around the image $t=\infty$. This gives the expansion
\bea\label{ferm copy one plus inf}
d'^{+ A(1)}_r&=&\sum_{k,k'\geq 0}{}^{r}C_k{}^{r}C_{k'}a^kb^{k'}\tilde{d}^{+A,\infty}_{r-k-k'+{1\over2}}+\sum_{k,k'\geq 0}{}^{r}C_k{}^{r}C_{k'}a^kb^{k'}t_2\tilde{d}^{+A,\infty}_{r-k-k'-{1\over2}}\nn
\eea
where the fermionic modes natural to the $t$-plane at the image $t=\infty$ are given by
\bea\label{modes inf}
\tilde d^{\a A,\infty}_{s} &=& {1\over2\pi i}\oint_{t=\infty}dt t^{s-{1\over2}} \psi^{\a A}(t)
\eea
Since we are at $t=\infty$ we have the following condition arising from the definition of the local vacuum there
\bea
\langle 0_{NS}|\tilde d^{\a A,\infty}_s=0,\quad  s<0
\eea
This condition places the following constraints on the modes in each term of (\ref{ferm copy one plus inf}) in order to yield a nonzero contraction
\bea
&&\text{Term 1}: r-k-k'+{1\over2}>0\implies r-k+{1\over2} > k'\cr
&&\qquad\qquad k'\geq 0 \implies r+{1\over2}>k
\cr
&&\text{Term 2}: r-k-k'-{1\over2}>0\implies r-k-{1\over2} > k'\cr
&&\qquad\qquad k'\geq 0 \implies r-{1\over2}>k
\eea
Thus this gives
\bea\label{copy one fin}
d'^{+ A(1)}_r&=&\sum_{k= 0}^{r}\sum_{k'=0}^{r-k}{}^{r}C_k{}^{r}C_{k'}a^kb^{k'}\tilde{d}^{+A,\infty}_{r-k-k'+{1\over2}}+\sum_{k= 0}^{r-1}\sum_{k'=0}^{r-k-1}{}^{r}C_k{}^{r}C_{k'}a^kb^{k'}t_2\tilde{d}^{+A,\infty}_{r-k-k'-{1\over2}}\nn
\eea
note that we have used the relation that $t_1=-t_2$. Let us now turn to the mode on copy $1$ final which we write again for continuity
\begin{equation}
    d'^{- A(1)}_{-r} ={1\over 2\pi i}\oint_{t=-a} dt {(t-t_2)(t+a)^{-r-1}(t+b)^{-r-1}\over t^{-r}} \psi^{- A}(t)
\end{equation}
This mode is defined at the image $t=-a$ and we should expand the integrand around that point. Since there are several factors in the integrand let us show the expansion more explicitly for each factor. First we consider the factor
\bea
(t+b)^{-r-1} &=& (t+a + b-a)^{-r-1} \cr
&=&(b-a)^{-r-1}\bigg(1+{t+a\over b-a}\bigg)^{-r-1}\cr
&=&\sum_{k\geq0}{}^{-r-1}C_k(b-a)^{-r-k-1}(t+a)^k
\eea
Taking $b=0$ in above equation and taking $-r-1\to r$ immediately gives the expansion
\bea
t^{r} &=&\sum_{k\geq0}{}^{r}C_k(-a)^{r-k}(t+a)^k
\eea
Defining modes natural to the $t$-plane around $t=-a$ we have
\bea
\tilde d^{\a A,-a}_{s} &=& {1\over2\pi i}\oint_{t=-a}dt (t+a)^{s-{1\over2}} \psi^{\a A}(t)
\eea
Therefore
\bea\label{ferm copy one minus minus a}
d'^{- A(1)}_{-r} &=&\sum_{k,k'\geq 0}{}^{-r-1}C_k{}^{r}C_{k'}(b-a)^{-r-k-1}(-a)^{r-k'}\tilde d^{- A,-a}_{k+k'-r+{1\over2}}\cr
&&\quad - \sum_{k,k'\geq 0}{}^{-r-1}C_k{}^{r}C_{k'}(b-a)^{-r-k-1}(-a)^{r-k'}(a+t_2)\tilde d^{- A,-a}_{k+k'-r-{1\over2}}
\eea
Since there is nothing sitting inside of this contour at $t=-a$ we have the following constraints coming from the vacuum defined at that point
\bea
\tilde d^{\a A}_{s}|0_{NS}\rangle^{-a}_t =0,\quad s>0
\eea
This condition gives the following constraints on the mode indices in the sums of the two terms in (\ref{ferm copy one minus minus a})
\bea
\text{Term 1}: &&k+ k' -r + {1\over2} < 0 \implies k' < r - k  - {1\over2}\cr
&&k'\geq 0\implies k  < r - {1\over2}
\cr
\cr
\text{Term 2 } &&k+ k' -r - {1\over2} < 0 \implies k' < r - k  +{1\over2}\cr
&&k'\geq 0\implies k  < r + {1\over2}
\eea
Applying these constraints to the sums in (\ref{ferm copy one minus minus a}) 
\bea\label{ferm copy one minus minus a p}
d'^{- A(1)}_{-r} &=&\sum_{k=0}^{r-1}\sum_{k'=0}^{r-k-1}{}^{-r-1}C_k{}^{r}C_{k'}(b-a)^{-r-k-1}(-a)^{r-k'}\tilde d^{- A,-a}_{k+k'-r+{1\over2}}\cr
&&\quad - \sum_{k = 0}^{r}\sum_{k'=0}^{r-k}{}^{-r-1}C_k{}^{r}C_{k'}(b-a)^{-r-k-1}(-a)^{r-k'}(t_2+a)\tilde d^{- A,-a}_{k+k'-r-{1\over2}}
\eea
Now we deform the contour circling $t=-a$ into a contour circling $t=\infty$ enclosed by the contour that is already there. This corresponds to expanding the modes currently defined at $t=-a$, given above, to modes at $t=\infty$. Recalling that modes expanded around $t=-a$ are given by 
\bea
\tilde d^{\a A,-a}_{s'} &=& {1\over2\pi i}\oint_{t=-a}dt (t+a)^{s'-{1\over2}} \psi^{\a A}(t)
\eea
Expanding the integrand now around $t=\infty$ yields 
\bea
\tilde d^{\a A,-a}_{s'} &=& {1\over2\pi i}\oint_{t=-a}dt (t+a)^{s'-{1\over2}} \psi^{\a A}(t)
\eea
To do that we write
\bea
(t+a)^{s'-{1\over2}} = \sum_{k''\geq 0} {}^{s'-{1\over2}}C_{k''}a^{k''}t^{s'-k''-{1\over2}}
\eea
\bea\label{copy one ini to infty}
\tilde d^{\a A,-a}_{s'} &=&\sum_{k''\geq 0} {}^{s'-{1\over2}}C_{k''}a^{k''} \tilde d^{\a A,\infty}_{s'-k''}
\eea
where we have utilized the definition of modes natural to the $t$-plane defined at the image $t=\infty$ given in (\ref{modes inf}). Therefore inserting expansion (\ref{copy one ini to infty}) into (\ref{ferm copy one minus minus a p}) yields
\begin{align}\label{exp copy one init}
d'^{- A(1)}_{-r} &=\sum_{k=0}^{r-1}\sum_{k'=0}^{r-k-1}\sum_{k''\geq0}{}^{-r-1}C_k{}^{r}C_{k'}{}^{k+k'-r}C_{k''}(b-a)^{-r-k-1}(-a)^{r-k'}a^{k''}\tilde d^{- A,\infty}_{k+k'-k''-r+{1\over2}}\cr
&- \sum_{k = 0}^{r}\sum_{k'=0}^{r-k}\sum_{k''\geq0}{}^{-r-1}C_k{}^{r}C_{k'}{}^{k+k'-r-1}C_{k''}(b-a)^{-r-k-1}(-a)^{r-k'}a^{k''}(t_2+a)\tilde d^{- A,\infty}_{k+k'-k''-r-{1\over2}}\nn
\end{align}
Now that both modes involved in the contraction are expanded around the same $t$-plane image $t=\infty$ we can compute the Wick contraction. We first write the local fermionic commutation relations that we'll need
\begin{equation}\label{ferm comm infty}
    \lbrace \tilde d^{\a A,\infty}_{r},\tilde d^{\b B,\infty}_{s} \rbrace = -\e^{\alpha\beta}\e^{AB}\d_{r+s,0}
\end{equation}
Utilizing (\ref{copy one fin}), (\ref{exp copy one init}) and the above commutation relations, we write the Wick contraction terms as
\begin{align}
&\langle d'^{+A(1)}_rd^{-B(1)}_{-s}\rangle\cr
&=\left\langle\bigg( \sum_{j= 0}^{r}\sum_{j'=0}^{r-j}{}^{r}C_j{}^{r}C_{j'}a^jb^{j'}\tilde{d}^{+A,\infty}_{r-j-j'+{1\over2}}+\sum_{j= 0}^{r-1}\sum_{j'=0}^{r-k-1}{}^{r}C_j{}^{r}C_{j'}a^jb^{j'}t_2\tilde{d}^{+A,\infty}_{r-j-j'-{1\over2}}\bigg)\right.\cr
 &\left. \bigg(\sum_{k=0}^{s-1}\sum_{k'=0}^{s-k-1}\sum_{k''\geq0}{}^{-s-1}C_k{}^{s}C_{k'}{}^{k+k'-s}C_{k''}(b-a)^{-s-k-1}(-a)^{s-k'}a^{k''}\tilde d^{- B,\infty}_{k+k'-k''-s+{1\over2}}\right.\cr
&\left. - \sum_{k = 0}^{s}\sum_{k'=0}^{s-k}\sum_{k''\geq0}{}^{-s-1}C_k{}^{s}C_{k'}{}^{k+k'-s-1}C_{k''}(b-a)^{-s-k-1}(-a)^{s-k'}a^{k''}(t_2+a)\tilde d^{- B,\infty}_{k+k'-k''-s-{1\over2}}  \bigg)\right\rangle
\cr
&= \sum_{j= 0}^{r}\sum_{j'=0}^{r-j}{}^{r}C_j{}^{r}C_{j'}a^jb^{j'} \sum_{k=0}^{s-1}\sum_{k'=0}^{s-k-1}\sum_{k''\geq0}{}^{-s-1}C_k{}^{s}C_{k'}{}^{k+k'-s}C_{k''}(b-a)^{-s-k-1}(-a)^{s-k'}a^{k''}\cr
&\quad\lbrace \tilde{d}^{+A,\infty}_{r-j-j'+{1\over2}} , \tilde d^{- B,\infty}_{k+k'-k''-s+{1\over2}} \rbrace
\cr
&+\sum_{j= 0}^{r-1}\sum_{j'=0}^{r-j-1}{}^{r}C_j{}^{r}C_{j'}a^jb^{j'}t_2 \sum_{k=0}^{s-1}\sum_{k'=0}^{s-k-1}\sum_{k''\geq0}{}^{-s-1}C_k{}^{s}C_{k'}{}^{k+k'-s}C_{k''}(b-a)^{-s-k-1}(-a)^{s-k'}a^{k''}\cr
&\quad \lbrace \tilde{d}^{+A,\infty}_{r-j-j'-{1\over2}}, \tilde d^{- B,\infty}_{k+k'-k''-s+{1\over2}} \rbrace
\cr
&- \sum_{j= 0}^{r}\sum_{j'=0}^{r-j}{}^{r}C_j{}^{r}C_{j'}a^jb^{j'}   \sum_{k = 0}^{s}\sum_{k'=0}^{s-k}\sum_{k''\geq0}{}^{-s-1}C_k{}^{s}C_{k'}{}^{k+k'-s-1}C_{k''}(b-a)^{-s-k-1}(-a)^{s-k'}a^{k''}\cr
&\qquad(t_2+a)\lbrace \tilde{d}^{+A,\infty}_{r-j-j'+{1\over2}}, \tilde d^{- B,\infty}_{k+k'-k''-s-{1\over2}}  \rbrace
\cr
&- \sum_{j= 0}^{r-1}\sum_{j'=0}^{r-j-1}{}^{r}C_j{}^{r}C_{j'}a^jb^{j'}t_2 \sum_{k = 0}^{s}\sum_{k'=0}^{s-k}\sum_{k''\geq0}{}^{-s-1}C_k{}^{s}C_{k'}{}^{k+k'-s-1}C_{k''}(b-a)^{-s-k-1}(-a)^{s-k'}a^{k''}\cr
&\quad(t_2+a)\lbrace \tilde{d}^{+A,\infty}_{r-j-j'-{1\over2}} , \tilde d^{- B,\infty}_{k+k'-k''-s-{1\over2}} \rbrace
\end{align}
Utilizing commutation relations (\ref{ferm comm infty}) we obtain
\begin{align}\label{copy one fin copy two init contraction}
&\langle d'^{+A(1)}_rd^{-B(1)}_{-s}\rangle\cr
&= \sum_{j= 0}^{r}\sum_{j'=0}^{r-j}{}^{r}C_j{}^{r}C_{j'}a^jb^{j'} \sum_{k=0}^{s-1}\sum_{k'=0}^{s-k-1}\sum_{k''\geq0}{}^{-s-1}C_k{}^{s}C_{k'}{}^{k+k'-s}C_{k''}(b-a)^{-s-k-1}(-a)^{s-k'}a^{k''}\cr
&\quad(-\e^{+-}\e^{AB}\d_{r-j-j'+k+k'-k''-s+1,0})
\cr
&+\sum_{j= 0}^{r-1}\sum_{j'=0}^{r-j-1}{}^{r}C_j{}^{r}C_{j'}a^jb^{j'}t_2 \sum_{k=0}^{s-1}\sum_{k'=0}^{s-k-1}\sum_{k''\geq0}{}^{-s-1}C_k{}^{s}C_{k'}{}^{k+k'-s}C_{k''}(b-a)^{-s-k-1}(-a)^{s-k'}a^{k''}\cr
&\quad(-\e^{+-}\e^{AB}\d_{r-j-j'+k+k'-k''-s,0})
\cr
&- \sum_{j= 0}^{r}\sum_{j'=0}^{r-j}{}^{r}C_j{}^{r}C_{j'}a^jb^{j'}   \sum_{k = 0}^{s}\sum_{k'=0}^{s-k}\sum_{k''\geq0}{}^{-s-1}C_k{}^{s}C_{k'}{}^{k+k'-s-1}C_{k''}(b-a)^{-s-k-1}(-a)^{s-k'}a^{k''}\cr
&\quad(t_2+a)(-\e^{+-}\e^{AB}\d_{r-j-j'+k+k'-k''-s,0})
\cr
&- \sum_{j= 0}^{r-1}\sum_{j'=0}^{r-j-1}{}^{r}C_j{}^{r}C_{j'}a^jb^{j'}t_2 \sum_{k = 0}^{s}\sum_{k'=0}^{s-k}\sum_{k''\geq0}{}^{-s-1}C_k{}^{s}C_{k'}{}^{k+k'-s-1}C_{k''}(b-a)^{-s-k-1}(-a)^{s-k'}a^{k''}\cr
&\quad(t_2+a)(-\e^{+-}\e^{AB}\d_{r-j-j'+k+k'-k''-s-1,0})
\end{align}
Enforcing the kronecker delta and positivity constraints we obtain the following limits on the summation indices of each of the four terms in the above expression
\bea
&&\text{Term 1}:\d_{r-j-j'+k+k'-k''-s + 1,0},\implies k''=r-j-j'+k+k'-s+1
\cr
\cr
&&\qquad\qquad k''\geq 0\implies k'\geq s -k-1  - (r-j-j')
\cr
\cr
&&\text{Term 2}:\d_{r-j-j'+k+k'-k''-s,0},\implies k''=r-j-j'+k+k'-s
\cr
\cr
&&\qquad\qquad k''\geq 0\implies k'\geq s -k  - (r-j-j')
\cr
\cr
&&\text{Term 3}:\d_{r-j-j'+k+k'-k''-s,0},\implies k''=r-j-j'+k+k'-s
\cr
\cr
&&\qquad\qquad k''\geq 0\implies k'\geq s -k  - (r-j-j')
\cr
\cr
&&\text{Term 4}:\d_{r-j-j'+k+k'-k''-s-1,0},\implies k''=r-j-j'+k+k'-s-1
\cr
\cr
&&\qquad\qquad k''\geq 0\implies k'\geq s -k+1  - (r-j-j') 
\eea
Inserting these constraints into the contraction (\ref{copy one fin copy two init contraction}) we obtain
\begin{align}
&\langle d'^{+A(1)}_rd^{-B(1)}_{-s}\rangle\cr
&= \e^{AB}\sum_{j= 0}^{r}\sum_{j'=0}^{r-j} \sum_{k=0}^{s-1}\sum_{k'=\max [0,s-k-1-(r-j-j')]}^{s-k-1}{}^{r}C_j{}^{r}C_{j'}{}^{-s-1}C_k{}^{s}C_{k'}{}^{k+k'-s}C_{r-j-j'+k+k'-s+1}\cr
&\qquad(b-a)^{-s-k-1}(-1)^{s-k'}a^{r-j'+k+1}b^{j'}
\cr
&+\e^{AB}\sum_{j= 0}^{r-1}\sum_{j'=0}^{r-j-1} \sum_{k=0}^{s-1}\sum_{k'=\max[0,s-k-(r-j-j')]}^{s-k-1}{}^{r}C_j{}^{r}C_{j'}{}^{-s-1}C_k{}^{s}C_{k'}{}^{k+k'-s}C_{r-j-j'+k+k'-s}\cr
&\qquad(b-a)^{-s-k-1}(-1)^{s-k'}a^{r-j'+k}b^{j'}t_2
\cr
&- \e^{AB}\sum_{j= 0}^{r}\sum_{j'=0}^{r-j} \sum_{k = 0}^{s}\sum_{k'=\max[0,s-k-(r-j-j')]}^{s-k}{}^{r}C_j{}^{r}C_{j'}  {}^{-s-1}C_k{}^{s}C_{k'}{}^{k+k'-s-1}C_{r-j-j'+k+k'-s}\cr
&\qquad(b-a)^{-s-k-1}(-1)^{s-k'}a^{r-j'+k}(t_2+a)b^{j'}
\cr
&- \e^{AB}\sum_{j= 0}^{r-1}\sum_{j'=0}^{r-j-1}\sum_{k = 0}^{s}\sum_{k'=\max[0,s-k+1-(r-j-j')]}^{s-k}{}^{r}C_j{}^{r}C_{j'} {}^{-s-1}C_k{}^{s}C_{k'}{}^{k+k'-s-1}C_{r-j-j'+k+k'-s-1}\cr
&\qquad(b-a)^{-s-k-1}(-1)^{s-k'}a^{r-j'+k-1}b^{j'}t_2(t_2+a)
\end{align}
For the contraction of a fermion mode on copy $1$ final with a fermion mode on copy $2$ initial we simply make the switch $a\leftrightarrow b$ in the above expressions. This yields
\begin{align}\label{ferm fin one plus ferm init one minus}
&\langle d'^{+A(1)}_rd^{-B(1)}_{-s}\rangle\cr
&= \e^{AB}\sum_{j= 0}^{r}\sum_{j'=0}^{r-j} \sum_{k=0}^{s-1}\sum_{k'=\max [0,s-k-1-(r-j-j')]}^{s-k-1}{}^{r}C_j{}^{r}C_{j'}{}^{-s-1}C_k{}^{s}C_{k'}{}^{k+k'-s}C_{r-j-j'+k+k'-s+1}\cr
&\qquad(a-b)^{-s-k-1}(-1)^{s-k'}b^{r-j'+k+1}a^{j'}
\cr
&+\e^{AB}\sum_{j= 0}^{r-1}\sum_{j'=0}^{r-j-1} \sum_{k=0}^{s-1}\sum_{k'=\max[0,s-k-(r-j-j')]}^{s-k-1}{}^{r}C_j{}^{r}C_{j'}{}^{-s-1}C_k{}^{s}C_{k'}{}^{k+k'-s}C_{r-j-j'+k+k'-s}\cr
&\qquad(a-b)^{-s-k-1}(-1)^{s-k'}b^{r-j'+k}a^{j'}t_2
\cr
&- \e^{AB}\sum_{j= 0}^{r}\sum_{j'=0}^{r-j} \sum_{k = 0}^{s}\sum_{k'=\max[0,s-k-(r-j-j')]}^{s-k}{}^{r}C_j{}^{r}C_{j'}  {}^{-s-1}C_k{}^{s}C_{k'}{}^{k+k'-s-1}C_{r-j-j'+k+k'-s}\cr
&\qquad(a-b)^{-s-k-1}(-1)^{s-k'}b^{r-j'+k}(t_2+b)a^{j'}
\cr
&- \e^{AB}\sum_{j= 0}^{r-1}\sum_{j'=0}^{r-j-1}\sum_{k = 0}^{s}\sum_{k'=\max[0,s-k+1-(r-j-j')]}^{s-k}{}^{r}C_j{}^{r}C_{j'} {}^{-s-1}C_k{}^{s}C_{k'}{}^{k+k'-s-1}C_{r-j-j'+k+k'-s-1}\cr
&\qquad(a-b)^{-s-k-1}(-1)^{s-k'}b^{r-j'+k-1}a^{j'}t_2(t_2+b)
\end{align}

\begin{align}\label{ferm fin one plus ferm init two minus}
&\langle d'^{+A(1)}_rd^{-B(1)}_{-s}\rangle\cr
&= \e^{AB}\sum_{j= 0}^{r}\sum_{j'=0}^{r-j} \sum_{k=0}^{s-1}\sum_{k'=\max [0,s-k-1-(r-j-j')]}^{s-k-1}{}^{r}C_j{}^{r}C_{j'}{}^{-s-1}C_k{}^{s}C_{k'}{}^{k+k'-s}C_{r-j-j'+k+k'-s+1}\cr
&\qquad(a-b)^{-s-k-1}(-1)^{s-k'}b^{r-j'+k+1}a^{j'}
\cr
&+\e^{AB}\sum_{j= 0}^{r-1}\sum_{j'=0}^{r-j-1} \sum_{k=0}^{s-1}\sum_{k'=\max[0,s-k-(r-j-j')]}^{s-k-1}{}^{r}C_j{}^{r}C_{j'}{}^{-s-1}C_k{}^{s}C_{k'}{}^{k+k'-s}C_{r-j-j'+k+k'-s}\cr
&\qquad(a-b)^{-s-k-1}(-1)^{s-k'}b^{r-j'+k}a^{j'}t_2
\cr
&- \e^{AB}\sum_{j= 0}^{r}\sum_{j'=0}^{r-j} \sum_{k = 0}^{s}\sum_{k'=\max[0,s-k-(r-j-j')]}^{s-k}{}^{r}C_j{}^{r}C_{j'}  {}^{-s-1}C_k{}^{s}C_{k'}{}^{k+k'-s-1}C_{r-j-j'+k+k'-s}\cr
&\qquad(a-b)^{-s-k-1}(-1)^{s-k'}b^{r-j'+k}(t_2+b)a^{j'}
\cr
&- \e^{AB}\sum_{j= 0}^{r-1}\sum_{j'=0}^{r-j-1}\sum_{k = 0}^{s}\sum_{k'=\max[0,s-k+1-(r-j-j')]}^{s-k}{}^{r}C_j{}^{r}C_{j'} {}^{-s-1}C_k{}^{s}C_{k'}{}^{k+k'-s-1}C_{r-j-j'+k+k'-s-1}\cr
&\qquad(a-b)^{-s-k-1}(-1)^{s-k'}b^{r-j'+k-1}a^{j'}t_2(t_2+b)
\end{align}

\subsection{Contractions tabulated}\label{all contractions}
All bosonic and fermionic Wick contractions required for the amplitudes computed in 
(\ref{ms plus gr unint})
including both the contractions that were worked out explicitly in (\ref{bos fin two bos init one}), (\ref{bos fin two bos init two}), (\ref{ferm fin one plus ferm init one minus}) and (\ref{ferm fin one plus ferm init two minus}) above as well as all of the remaining contractions whose derivations were not recorded explicitly but can be obtained by performing similar computations, are tabulated below with a reminder that the coefficient ${}^{p}C_{q}$ is the binomial coefficient
\begin{equation}
{}^{p}C_{q}\equiv {p!\over q!(p-q)!}
\end{equation}
\subsubsection{Bosons}\label{boson app}
\begin{align}
\langle\a'^{(1)}_{B\dot{B},p}\tilde{\a}^{t_i}_{C\dot{C},-1}\rangle&=-\e_{BC}\e_{\dot{B}\dot{C }}\sum_{k=0}^{p-1}\sum_{k'=0}^{p-k-1}(p-k-k'){}^{p}C_{k}{}^{p}C_{k'}a^kb^{k'}t_i^{p-k-k'-1}
\cr
\cr
\langle\a'^{(1)}_{B\dot{B},p}\a'^{(1)}_{A\dot{A},-n}\rangle&=- \e_{BA}\e_{\dot{B}\dot{A}}\sum_{j= 0}^{n-1}\sum_{j'= \max{[0,n-p -j]}}^{n-j-1}\sum_{k=0}^{p-n +j+j'}\sum_{k'=0}^{p-n +j+j'-k}(p-k-k'){}^{p}C_{k}{}^{p}C_{k'}\cr
&\qquad\quad{}^nC_{j}{}^{-n}C_{j'}{}^{-n+j+j'}C_{p-k-k' -n+j+j'}(-1)^{n-j}a^{p-k' +j'}b^{k'}(b-a)^{-n-j'}
\cr
\cr
\langle\a'^{(2)}_{B\dot{B},p}\tilde{\a}^{t_i}_{A\dot{A},-1}\rangle &=-\e_{\dot{B}\dot{A}}\e_{BA}\sum_{k=0}^{p-1}\sum_{k'=0}^{p-k-1}{}^pC_k{}^pC_{k'} {}^{k+k'-p}C_1a^{p-k}b^{p-k'}t_i^{k+k'-p-1}
\cr\cr
\langle\a'^{(2)}_{B\dot{B},p}\a'^{(2)}_{A\dot{A},-n}\rangle &=-\e_{BA}\e_{\dot{B}\dot{A}}\sum_{k=0}^{p-1}\sum_{k'=0}^{p-k-1}\sum_{j= 0}^{n-1}\sum_{j'= 0}^{n-j-1}(n-j-j'){}^pC_k{}^pC_{k'} {}^{k+k'-p}C_{n-j-j'} \cr
&\qquad\qquad{}^nC_{j}{}^{-n}C_{j'}(-1)^{p+j'+k+k'}b^{j'+k'}a^{p-k'}(a-b)^{-n-j'}
\cr
\cr
\langle\a'^{(2)}_{B\dot{B},p}\a'^{(1)}_{A\dot{A},-n}\rangle&=-\e_{BA}\e_{\dot{B}\dot{A}}\sum_{k=0}^{p-1}\sum_{k'=0}^{p-k-1}\sum_{j= 0}^{n-1}\sum_{j'= 0}^{n-j-1}(n-j-j'){}^pC_k{}^pC_{k'} {}^{k+k'-p}C_{n-j-j'} \cr
&\qquad\qquad{}^nC_{j}{}^{-n}C_{j'}(-1)^{p+j'+k+k'}a^{j'+k'}b^{p-k'}(b-a)^{-n-j'}
\cr\cr
\langle\a'^{(1)}_{B\dot{B},p}\a'^{(2)}_{A\dot{A},-n}\rangle&=  - \e_{BA}\e_{\dot{B}\dot{A}}\sum_{j= 0}^{n-1}\sum_{j'= \max{[0,n-p -j]}}^{n-j-1}\sum_{k=0}^{p-n +j+j'}\sum_{k'=0}^{p-n +j+j'-k}(p-k-k'){}^{p}C_{k}{}^{p}C_{k'}\cr
&\qquad\qquad{}^nC_{j}{}^{-n}C_{j'}{}^{-n+j+j'}C_{p-k-k' -n+j+j'}(-1)^{n-j}b^{p-k' +j'}a^{k'}\cr
&\qquad\qquad(a-b)^{-n-j'}
\cr\cr
\langle\tilde{\a}^{t_i}_{C\dot{C},-1}\a'^{(1)}_{A\dot{A},-n}\rangle &=\e_{CA}\e_{\dot{C}\dot{A}}\sum_{j= 0}^{n-1}\sum_{j'= 0}^{n-j-1}{}^nC_{j}{}^{-n}C_{j'}{}^{-n+j+j'}C_1(-a)^{n-j}(b-a)^{-n-j'}\cr
&\qquad\quad(t_i+a)^{-n + j + j'-1}
\cr
\cr
\langle\tilde{\a}^{t_i}_{C\dot{C},-1}\a'^{(2)}_{A\dot{A},-n}\rangle&=\e_{CA}\e_{\dot{C}\dot{A}}\sum_{j= 0}^{n-1}\sum_{j'= 0}^{n-j-1}{}^nC_{j}{}^{-n}C_{j'}{}^{-n+j+j'}C_1(-b)^{n-j}(a-b)^{-n-j'}\cr
&\qquad\quad(t_i+b)^{-n + j + j'-1}
\cr
\cr
\langle\tilde{\a}^{t_2}_{C\dot{C},-1}\tilde{\a}^{t_1}_{A\dot{A},-1}\rangle&=-\e_{CA}\e_{\dot{C}\dot{A}}{1\over (t_2-t_1)^2}
\end{align}

\subsubsection{Fermions}\label{ferm app}
The fermion wick contractions are given by

\begin{align}
\langle d'^{- A(1)}_q\tilde{d}^{+B,t_2}_{-{1\over2}} \rangle&=-\e^{AB}\sum_{j = 0}^{ q-1}\sum_{j' = 0}^{ q-j-1}{}^{q-1}C_j{}^{q-1}C_{j'}a^jb^{j'}t_2^{q-j-j'-1}\nn
&\quad+\e^{AB}\sum_{j = 0}^{q-2}\sum_{j' = 0}^{ q-j-2}{}^{q-1}C_j{}^{q-1}C_{j'}a^jb^{j'}t_2^{q-j-j'-1}\nn
\cr
\langle d'^{+A(1)}_q\tilde{d}^{-B,t_1}_{-{1\over2}}\rangle&=\e^{AB}\sum_{j = 0}^{ q}\sum_{j' = 0}^{q-j}{}^{q}C_j{}^{q}C_{j'}a^jb^{j'}t_1^{q-j-j'}  -  \e^{AB}\sum_{j = 0}^{ q-1}\sum_{j' = 0}^{ q-j-1}{}^{q}C_j{}^{q}C_{j'}a^jb^{j'}t_1^{q-j-j'}
\cr
\cr
\langle d'^{+ A(2)}_r\tilde{d}^{-B,t_1}_{-{1\over2}} \rangle&=\e^{AB}\sum_{j = 0}^{ r-1}\sum_{j' = 0}^{ r-j-1}{}^{r}C_j{}^{r}C_{j'}a^{r-j}b^{r-j'}t_1^{j+j'-r} -\e^{AB}\sum_{j = 0}^{r}\sum_{j' = 0}^{ r-j}{}^{r}C_j{}^{r}C_{j'}a^{r-j}b^{r-j'}t_1^{j+j'-r}
\nn
\cr
\langle d'^{- +(2)}_r \tilde{d}^{+-,t_2}_{-{1\over2}} \rangle &= -\e^{AB}\sum_{j = 0}^{r-2}\sum_{j' = 0}^{ r-j-2}{}^{r-1}C_j{}^{r-1}C_{j'}a^{r-j-1}b^{r-j'-1}t_2^{j+j'-r+1}\cr
&\quad +\e^{AB}\sum_{j = 0}^{ r-1}\sum_{j' = 0}^{ r-j-1}{}^{r-1}C_j{}^{r-1}C_{j'}a^{r-j-1}b^{r-j'-1}t_2^{j+j'-r+1}\nn
\cr
\langle d'^{+A(1)}_rd^{-B(1)}_{-s}\rangle&= \e^{AB}\sum_{j= 0}^{r}\sum_{j'=0}^{r-j} \sum_{k=0}^{s-1}\sum_{k'=\max [0,s-k-1-(r-j-j')]}^{s-k-1}{}^{r}C_j{}^{r}C_{j'}{}^{-s-1}C_k{}^{s}C_{k'}\cr
&\qquad{}^{k+k'-s}C_{r-j-j'+k+k'-s+1}(b-a)^{-s-k-1}(-1)^{s-k'}a^{r-j'+k+1}b^{j'}
\cr
&+\e^{AB}\sum_{j= 0}^{r-1}\sum_{j'=0}^{r-j-1} \sum_{k=0}^{s-1}\sum_{k'=\max[0,s-k-(r-j-j')]}^{s-k-1}C_{j'}{}^{-s-1}C_k{}^{s}C_{k'}\cr
&\qquad{}^{k+k'-s}C_{r-j-j'+k+k'-s}(b-a)^{-s-k-1}(-1)^{s-k'}a^{r-j'+k}b^{j'}t_2
\cr
&- \e^{AB}\sum_{j= 0}^{r}\sum_{j'=0}^{r-j} \sum_{k = 0}^{s}\sum_{k'=\max[0,s-k-(r-j-j')]}^{s-k}{}^{r}C_j{}^{r}C_{j'}  {}^{-s-1}C_k{}^{s}C_{k'}\cr
&\qquad{}^{k+k'-s-1}C_{r-j-j'+k+k'-s}(b-a)^{-s-k-1}(-1)^{s-k'}a^{r-j'+k}(t_2+a)b^{j'}
\cr
&- \e^{AB}\sum_{j= 0}^{r-1}\sum_{j'=0}^{r-j-1}\sum_{k = 0}^{s}\sum_{k'=\max[0,s-k+1-(r-j-j')]}^{s-k}{}^{r}C_j{}^{r}C_{j'} {}^{-s-1}C_k{}^{s}C_{k'}\cr
&{}^{k+k'-s-1}C_{r-j-j'+k+k'-s-1}(b-a)^{-s-k-1}(-1)^{s-k'}a^{r-j'+k-1}b^{j'}t_2(t_2+a)
\cr
\cr
\langle d'^{+ A(2)}_{r}d'^{- B(1)}_{-s}\rangle &= \e^{AB}\sum_{j=0}^{r-j-1}\sum_{j'=0}^{r-j-1} \sum_{k=0}^{s-1}\sum_{k'=0}^{s-k-1}{}^rC_j{}^rC_{j'} {}^{j+j'-r}C_{s-k-k'-1} {}^{-s-1}C_k{}^{s}C_{k'}\cr
&\quad (-1)^{j+j'-r+k+1}b^{r-j'}a^{k+j'+1}(b-a)^{-s-k-1}
\cr
&- \e^{AB}\sum_{j=0}^{r-j-1}\sum_{j'=0}^{r-j-1} \sum_{k = 0}^{s}\sum_{k'=0}^{s-k}{}^rC_j{}^rC_{j'} {}^{j+j'-r}C_{s-k-k'}{}^{-s-1}C_k{}^{s}C_{k'}\cr
&\quad (-1)^{j+j'-r+k}a^{k+j'}b^{r-j'}(b-a)^{-s-k-1}(t_2+a)
\cr
&- \e^{AB} \sum_{j=0}^{r-j}\sum_{j'=0}^{r-j}\sum_{k=0}^{s-1}\sum_{k'=0}^{s-k-1}{}^rC_j{}^rC_{j'}{}^{j+j'-r-1}C_{s-k-k'-1} {}^{-s-1}C_k{}^{s}C_{k'}\cr
&\quad (-1)^{j+j'-r+k}a^{j'+k}b^{r-j'}t_1(b-a)^{-s-k-1}
\cr
&+  \e^{AB}\sum_{j=0}^{r-j}\sum_{j'=0}^{r-j}\sum_{k = 0}^{s}\sum_{k'=0}^{s-k}{}^rC_j{}^rC_{j'}{}^{j+j'-r-1}C_{s-k-k'} {}^{-s-1}C_k{}^{s}C_{k'}\cr
&\quad (-1)^{j+j'-r+k-1}a^{j'+k-1}b^{r-j'}t_1(b-a)^{-s-k-1}(t_2+a)
\nn
\cr
\langle \tilde d^{+ A,t_i}_{-{1\over2}}d'^{- B(1)}_{-r}\rangle&=\e^{AB} \sum_{k=0}^{r-1}\sum_{k'=0}^{r-k-1}{}^{-r-1}C_k{}^{r}C_{k'}(t_i+a)^{k+k'-r} (b-a)^{-r-k-1}(-a)^{r-k'}
\cr
&  - \e^{AB}\sum_{k = 0}^{r}\sum_{k'=0}^{r-k}{}^{-r-1}C_k{}^{r}C_{k'}(t_i+a)^{k+k'-r-1}(b-a)^{-r-k-1}(-a)^{r-k'}(t_2+a)\nn
\cr
\langle\tilde{d}^{+C,t_2}_{-{1\over2}}\tilde{d}^{-A,t_1}_{-{1\over2}}\rangle &=\e^{CA}{1\over t_2-t_1}
\end{align}
\subsubsection{Coordinates}
Below we record the cylinder coordinates 
\begin{align}\label{zw pr app}
s={w_1+w_2\over2},\quad \Delta w=w_2 -w_1
\end{align}
and the their relation to the $t$-plane parameters
\begin{align}\label{a b t1 t2 app}
    a&=e^s\cosh^2\bigg({\Delta w\over4}\bigg),& b&=e^s\sinh^2\bigg({\Delta w\over4}\bigg)
    \nn
    t_1&=-e^s\cosh\bigg({\Delta w\over4}\bigg)\sinh\bigg({\Delta w\over4}\bigg),&t_2&=e^s\cosh\bigg({\Delta w\over4}\bigg)\sinh\bigg({\Delta w\over4}\bigg)
\end{align}
which are utilized in the various Wick contraction terms tabulated in in subsubsections \ref{boson app} and \ref{ferm app}.
To obtain the antiholomorphic bosonic and fermionic counterparts recorded therein, we simply make the replacements
$w_1\to \bar w_1,w_2\to \bar w_2$ in the coordinates (\ref{zw pr app}) and (\ref{a b t1 t2 app}) which then enter (\ref{a b t1 t2 app}). We note that bar notation should be performed while still Euclidean signature. Wick rotation of $\tau_j\to it_j$ for $j=1,2$ should be performed afterwards.

\bibliographystyle{JHEP}
\bibliography{bibliography.bib}

\end{document}